\documentclass[11pt,oneside,openright,a4paper]{thesis}
\usepackage{acronym}
\usepackage{algorithm}
\usepackage{algorithmicx}
\usepackage[noend]{algpseudocode}
\usepackage{amsfonts}
\usepackage{amsmath}
\usepackage{amsthm}
\usepackage[inline]{enumitem}
\usepackage{hyperref}
\usepackage{mathtools}
\usepackage{pifont}
\usepackage{pgfplots}
\usepackage{pgfplotstable}
\usepackage[strings]{underscore}
\usepackage[disable, colorinlistoftodos, backgroundcolor=red!40, linecolor=red!40]{todonotes}
\usepackage{xspace}
\usepackage{multicol}
\usepackage{makecell}
\usepackage{tabularx}
\usepackage{csquotes}

\theoremstyle{plain}

\usepackage[noend]{algpseudocode}
\makeatletter
\def\nobreakhline{%
  \noalign{\ifnum0=`}\fi
    \penalty\@M
    \futurelet\@let@token\LT@@nobreakhline}
\def\LT@@nobreakhline{%
  \ifx\@let@token\hline
    \global\let\@gtempa\@gobble
    \gdef\LT@sep{\penalty\@M\vskip\doublerulesep}
  \else
    \global\let\@gtempa\@empty
    \gdef\LT@sep{\penalty\@M\vskip-\arrayrulewidth}
  \fi
  \ifnum0=`{\fi}%
  \multispan\LT@cols
     \unskip\leaders\hrule\@height\arrayrulewidth\hfill\cr
  \noalign{\LT@sep}%
  \multispan\LT@cols
     \unskip\leaders\hrule\@height\arrayrulewidth\hfill\cr
  \noalign{\penalty\@M}%
  \@gtempa}
\makeatother

\pgfplotsset{compat=1.16}

\DeclareGraphicsExtensions{.jpg,.png}

\newcolumntype{M}[1]{>{\centering\arraybackslash}m{#1}}
\newcolumntype{R}[1]{>{\raggedleft\arraybackslash}p{#1}}
\newcolumntype{Y}{>{\centering\arraybackslash}X}

\hypersetup{
    pdftitle={The impossibility of automating ambiguity: a relational perspective},
    pdfauthor={Abeba Birhane},
    pdfsubject={Phd Thesis},
    pdfcreator={University College Dublin},
    pdfproducer={LaTeX},
    pdfkeywords={Phd Thesis},
}

\setlist[itemize]{itemsep=0pt,topsep=0pt}

\usepackage{pgf-pie}

\usepgfplotslibrary{groupplots}
\pgfplotsset{
        show sum on top/.style={
            /pgfplots/scatter/@post marker code/.append code={%
                \node[
                    at={(normalized axis cs:%
                            \pgfkeysvalueof{/data point/x},%
                            \pgfkeysvalueof{/data point/y})%
                    },
                    anchor=south,
                ]
                {\pgfmathprintnumber{\pgfkeysvalueof{/data point/y}}};
            },
        },
        show sum on top left/.style={
            /pgfplots/scatter/@post marker code/.append code={%
                \node[
                    at={(normalized axis cs:%
                            \pgfkeysvalueof{/data point/x} -0.2,%
                            \pgfkeysvalueof{/data point/y})%
                    },
                    anchor=south,
                ]
                {\pgfmathprintnumber{\pgfkeysvalueof{/data point/y}}};
            },
        }
    }
\usepgfplotslibrary{statistics}
\pgfplotsset{only if/.style args={entry of #1 is #2}{
/pgfplots/boxplot/data filter/.code={
\edef\tempa{\thisrow{#1}}
\edef\tempb{#2}
\ifx\tempa\tempb
\else
\def\pgfmathresult{}
\fi
}
}
}

\pgfplotsset{
discard if not/.style 2 args={
x filter/.append code={
\edef\tempa{\thisrow{#1}}
\edef\tempb{#2}
\ifx\tempa\tempb
\else
\def\pgfmathresult{inf}
\fi
}
}
}

\pgfplotsset{
discardbp if not/.style 2 args={
/pgfplots/boxplot/data filter/.append code={
\edef\tempa{\thisrow{#1}}
\edef\tempb{#2}
\ifx\tempa\tempb
\else
\def\pgfmathresult{inf}
\fi
}
}
}

\definecolor{gray1}{gray}{0.9}

\definecolor{gray2}{gray}{0.85}

\definecolor{gray3}{gray}{0.8}

\definecolor{gray4}{gray}{0.75}

\definecolor{gray5}{gray}{0.7}

\newcounter{researchquestionCount}

\title{
\sc{Automating Ambiguity: Challenges and Pitfalls of Artificial Intelligence}
}
 
\author{
    \it{Abeba Birhane}
}
\studentid{Student ID: 14202685}
\collegeordept{{School of Computer Science}}
\university{{University College Dublin}}

\renewcommand{\submittedtext}{
    The thesis is submitted to University College Dublin in fulfilment of the requirements for the degree of
}
\degree{Doctor of Philosophy}

\headschool{Head of School:\\Associate Prof. Chris Bleakley}
\supervisors{Under the supervision of:\\ Assistant Prof. Anthony Ventresque}
\dsp{Doctoral Study Panel:\\ Associate Prof. Brian Mac Namee \\ Prof. Maria Baghramian \\ Assistant Prof. Brendan Rooney}
\degreedate{October 2021}

\begin{document}

    \setcounter{page}{0}
    \maketitle

    \frontmatter
    \newpage
    \tableofcontents

    \chapter*{}

\vspace*{-5cm}

\begin{abstracts}

Machine learning (ML) and artificial intelligence (AI) tools increasingly permeate every possible social, political, and economic sphere; sorting, taxonomizing and predicting complex human behaviour and social phenomena. However, from fallacious and naive groundings regarding complex adaptive systems to datasets underlying models, these systems are beset by problems, challenges, and limitations. They remain opaque and unreliable, and fail to consider societal and structural oppressive systems, disproportionately negatively impacting those at the margins of society while benefiting the most powerful. 

The various challenges, problems and pitfalls of these systems are a hot topic of research in various areas, such as critical data/algorithm studies, science and technology studies (STS), embodied and enactive cognitive science, complexity science, Afro-feminism, and the broadly construed emerging field of Fairness, Accountability, and Transparency (FAccT). Yet, these fields of enquiry often proceed in silos. This thesis weaves together seemingly disparate fields of enquiry to examine core scientific and ethical challenges, pitfalls, and problems of AI.

In this thesis I, a) review the historical and cultural ecology from which AI research emerges, b) examine the shaky scientific grounds of machine prediction of complex behaviour illustrating how predicting complex behaviour with precision is impossible in principle, c) audit large scale datasets behind current AI demonstrating how they embed societal historical and structural injustices, d) study the seemingly neutral values of ML research and put forward 67 prominent values underlying ML research, e) examine some of the insidious and worrying applications of computer vision research, and f) put forward a framework for approaching challenges, failures and problems surrounding ML systems as well as alternative ways forward.

\end{abstracts}
\chapter*{}
\addcontentsline{toc}{chapter}{Declaration}

\begin{center}
    \vspace*{0cm}
\par\end{center}

\begin{minipage}[t]{260pt}
    \line(1,0){260}

    \noindent{
        \large
        \emph{
            I hereby certify that the submitted work is my own work, was completed while registered as a candidate for the degree stated on the Title Page, and I have not obtained a degree elsewhere on the basis of the research presented in this submitted work.
        }
    }
    {\large \par}

    \begin{center}
        \vspace*{1cm}
    \par\end{center}

    \noindent {
        \large Signed: \textit{Abeba Birhane}\\
    }
    {\large \par}

    \vspace*{1cm}
    \noindent {
        \large Student ID: \textit{14202685}
    }
    {\large \par}

    \vspace*{1cm}
    \noindent {
        \large Date: October 2021
    }
    {\large \par}

\end{minipage}
\chapter*{}
\addcontentsline{toc}{chapter}{Acknowledgements}

\begin{center}
    \vspace*{-6cm}
\par\end{center}

\begin{acknowledgements}

As the well-know African proverb 
goes, ``it takes a village to raise a child''. This PhD thesis is just that: an effort nurtured by a network of support. My intellectual journey from early days to getting through this PhD has been one that is marked by constant support, love, and encouragement both from academic peers as well as friends, family and networks of support. Black women scholars have especially been a source of inspiration and courage in navigating uncharted territories and asking seemingly unorthodox questions. As an academic that often struggles to fit into a neat category, my academic journey has been filled with frustration and various attempts to leave academia altogether. However, it is thanks to the network of support, love, and encouragement, often from unexpected places, that this PhD thesis came to completion. I am forever grateful to everyone that has uplifted my work, proofed and commented on my drafts, took a chance on me and offered me opportunities. Twitter family, your support and swift response to my rants, various questions and requests through out the year has been invaluable. I am grateful for my circle of friends within the Ethiopian community in Ireland who made me ``thesis writing {\selectlanguage{ethiop}'agalgl}'' (specially prepared Ethiopian food to fuel thesis writing), cheered me up when I felt down, and provided me with a vital social support. You know who you are. I surely am guaranteed to leave some people out if I attempt to name all, so I won't. It is my sincere hope and commitment to pay it forward. 

I would like to thank my supervisor Dr Anthony Ventresque for supporting, trusting and having faith in me to explore seemingly risky topics. Thank you also to Prof Maria Baghramian, Prof Brian Mac Namee, Prof Brendan Rooney, and Prof Fred Cummins for your counsel.

My deepest gratitude goes to all my collaborators from a multitude of disciplines for putting up with me throughout the years. I have benefited from your vast knowledge and research expertise. Thank you for inspiring me to continually aim for an ever higher research standard and for teaching me that kindness, sympathy, and appreciation for other triumphs while working on a collaborative project.  

Very special gratitude goes out to my lab-mate and dear friend Thomas Laurent, whom I constantly run to for last minute proof of my drafts at short notice, all my coding problems, and general rants. It has been a privilege going for a coffee to Pi every morning with you for the past number of years. I am grateful to have shared the lab and whole PhD experience with you.

\end{acknowledgements}

\chapter{Publications}
\label{chapter:publications}

\section*{Submitted}
{
\small
\renewcommand{\arraystretch}{0.75}
\begin{tabular}{l m{0.9\textwidth}}
  \cite{birhane2021debunking} & \bibentry{birhane2021debunking}\\
    &\\
  \cite{birhane22} & \bibentry{birhane22}\\
    &\\
  \cite{birhane2021multimodal} &\bibentry{birhane2021multimodal}\\
    &\\
  \cite{birhaneparticipatory} &\bibentry{birhaneparticipatory}\\
    &\\
  \cite{Birhanepostmodern} &\bibentry{Birhanepostmodern}\\
\end{tabular}
}
\normalsize

\section*{Published}
{
\small
\renewcommand{\arraystretch}{0.75}
\begin{tabular}{l m{0.9\textwidth}}

    \cite{abebe2021narratives} & \bibentry{abebe2021narratives}\\ 
    &\\
    \cite{BirhaneGuest2021} & \bibentry{BirhaneGuest2021}\\
    &\\
    \cite{birhane2021large} & \bibentry{birhane2021large} \textbf{Recipient of the Venture Beat AI Innovation Award in computer vision}\\
    &\\
    \cite{birhane2020robot} & \bibentry{birhane2020robot}\\
    &\\
    \cite{birhane2020algorithmic} & \bibentry{birhane2020algorithmic}\\
    &\\
    \cite{birhane2021impossibility} & \bibentry{birhane2021impossibility}\\
    &\\
    \cite{birhane2021algorithmic} & \bibentry{birhane2021algorithmic}\\
    &\\
    \cite{birhane2019algorithmic} & \bibentry{birhane2019algorithmic} \textbf{Recipient of the Black in AI Best Paper Award at NeurIPS2019}\\
    &\\
    \cite{ruane2019conversational} & \bibentry{ruane2019conversational}\\
      &\\
    \cite{saliency22} & \bibentry{saliency22}\\
      &\\
      \cite{birhane2021values} & \bibentry{birhane2021values}\\
    &\\
  \cite{birhaneforgotten} &\bibentry{birhaneforgotten}
\end{tabular}
}
\normalsize
\newpage
\listoffigures
\addcontentsline{toc}{chapter}{List of Figures}

\newpage
\listoftables
\addcontentsline{toc}{chapter}{List of Tables}

\chapter{List of Acronyms}

\begin{acronym}[XXXXXXXX]
    \setlength{\parskip}{0ex}
    \setlength{\itemsep}{1ex}

    \acro{AGI}{Artificial General Intelligence}
    \acro{AI}{Artificial Intelligence}
    \acro{CAS}{Complex Adaptive Systems}
    \acro{CNN}{Convolutional Neural Network}
    \acro{CV}{Computer Vision}
    \acro{DL}{Deep Learning}
    \acro{DS}{Data Science}
    \acro{GAN}{Generative Adversarial Network}
    \acro{LSVD}{Large Scale Vision Dataset}
    \acro{ML}{Machine Learning}
    \acro{NLP}{Natural Language Processing}
    \acro{SOTA}{State-of-the-art}
    \acro{STS}{Science and Technology Studies}
    \acro{STSS}{Science, Technology, and Society Studies}
\end{acronym}

    \mainmatter
    \renewcommand\baselinestretch{1.5}
    \baselineskip=16pt plus1pt
    \setlength{\parskip}{8pt plus1pt}
    \setcounter{page}{1}

    \part{}
    \label{part:i}
    \chapter{Introduction}
\label{chapter:introduction}

The integration of machine learning (ML) and artificial intelligence (AI) tools into every possible social, political, and economic sphere means that these tools increasingly sort, classify, categorize, and predict human behaviour and social phenomena. However, from fallacious scientific groundings, questionable assumptions to problematic models and underlying datasets, ML systems are beset by problems and limitations. They are, for the most part, opaque, unreliable, and embed historical and social stereotypes, disproportionately harming those at the margin of society when applied in the real world. 

In this thesis I, 1) interrogate the shaky scientific grounds of predicting behaviours of a complex system, informed by post-Cartesian approaches to cognition and social systems, 2) empirically examine large datasets behind current AI, demonstrate their limitations and downstream negative impacts and illustrate the underlying values of ML research through a systematic study of the most influential papers, and 3) put forward a framework for approaching failures, limitations, and problems surrounding ML/AI systems.

\section{Motivation}
\label{introduction:motivation}

It has become trivial to point out how ML/AI systems are increasingly an integral part of decision-making processes in various social, political, and economic spheres. With considerable progress in various domains from computer vision, natural language processing (NLP), and audio processing, the field can be said to be currently experiencing an ``AI spring''~\cite{mitchell2021ai}. Both inside the field as well as within the general public discourse, AI is largely marked by over-enthusiasm, blind trust, and little critical scrutiny~\cite{mitchell2019artificial,broussard2018artificial} . 

ML tools are spread ubiquitously, impacting lives in all directions, including whether one receives medical care or a mortgage loan, is hired or fired, is recognized 
as a 
pedestrian, or is considered an immigrant and receives aid. Thus, critical study of these systems is not only a matter of academic endeavour, but also has the potential to alter these tools, drive regulations and policy, as well as raise awareness around tools that often operate in the dark.     

More and more of ML research finds applications in various areas of life impacting individuals and communities. This means that the study and critical examination of the field requires more than engineers and computer scientists but also academics from various fields, policy makers, and regulators as well as civil society. Currently, the need for involvement of various stakeholders and multiple approaches is generally well recognized. Generally, there is a consensus that AI can and should benefit humanity. However, without further analysis of the cost and benefit of AI on different groups, the general sentiment remains vacuous. When algorithmic systems fail (and they often do), individuals and communities at the margins of society pay the highest price. Those that create these systems (predominantly big tech corporations and elite universities) receive the highest benefits. 

The nature of human behaviour and social systems, the limitations and problems of historical and web sourced data for training and validating AI, and the disproportionate negative impact of AI on minoritized communities are ongoing questions. However, these questions, for the most part, are advanced in silos in their respective fields. In the case of examining the nature of complex human behaviour and social systems with the aim of exposing the limitation of ML prediction, the work remains underdeveloped and this thesis marks one of the steps advancing in that direction.

Approaches such as embodied and enactive cognitive science, and complex systems provide invaluable foundations for examining the nature of complex human behaviour and social systems in light of pervasive predictive systems. As the ``success'' behind current AI systems is due to the ``availability'' of massive amounts of data, dataset audits give us a glimpse into AI systems and their downstream impacts. Systematic analysis of the most influential ML papers reveal underlying values of the field, where the field might be heading, and who the powerful actors are. Critical data and Black feminist approaches provide insights that propel us to look at algorithmic bias, harms and threats from the perspective of the most impacted. All these perspectives, approaches, and methods overlap and each lends something that the others are lacking. This thesis weaves together these various domains, methods, and approaches in a unique and novel way with the hope of producing as full a narrative as possible in approaching the following problems. 
\section{Thesis Approach}
\label{introduction:thesis_approach}

Unlike traditional theses, this thesis is not confined within a specific approach, domain, or method. Rather, it brings together the most relevant aspects of three main broad traditions and utilizes the most fitting methods in order to examine implicit theoretical assumptions, underlying values, historical trajectories, the AI pipeline (from data sourcing to application), deeply ingrained social, cultural, and historical injustice and power asymmetries embedded in current AI systems.   

To that end, the thesis embarks on \textit{theoretical explorations} surrounding questions of human behaviour and social systems, \textit{empirical} (both qualitative and quantitative) analysis of the data, history, and underlying values of current AI, and \textit{critical analysis} of oppressive social structures and power asymmetries from which AI systems are designed, built, and integrated into society.    

Reductive and simplistic approaches towards cognition, intelligence, human behaviour and social systems remain the default, especially in Western sciences. Implicit assumptions underlying ML, statistics, AI, and traditional cognitive science are no different. Complex, messy, contextual, dynamic, multivalent, and ambiguous phenomena are often reduced down to their abstract ``essence'' or single representation. Part of the problem with AI systems that sort, classify, and predict human behaviour and social phenomena, springs from this simplistic and limited understanding. To that end, this thesis leans on insights from post-Cartesian approaches to characterizing cognition, human behaviour, and social systems with the aim of revealing both the scientific and ethical limitations of machine classification and prediction.   




\section{Contributions}
\label{introduction:contributions}
The main contributions of this thesis are: 
\begin{itemize}
    \item~\textbf{Theoretical contribution.} Post-Cartesian perspectives provide important foundations for understanding and modelling complex systems. Yet, such insights remain largely missing from the point of view of machine classification and prediction of complex adaptive systems. Post-Cartesian approaches also remain largely 
    unaware of 
    structural oppression and power asymmetries that permeate social systems. Using insights from post-Cartesian perspectives and Black feminist epistemology, I put forward a theoretical framework that encompasses both for the messy, fluctuating, non-linear, open-ended and dynamic nature of complex systems as well as the oppressive and unjust social structures in the context of sorting, predicting and modeling behaviours and social systems.   
    \item~\textbf{Empirical contributions.} In order to have a better grasp of AI, both the \textit{dataset} behind it as well as understanding the wider \textit{ecology, history, and underlying values of ML research} are crucial elements. To this end, this thesis presents dataset audits for two large image datasets: ImageNet and Tiny Images. The contribution in this regard include rigorous audits and recommendation for dataset curators, in the latter case resulting in the withdrawal of Tiny Images from use, as described below. Furthermore, this thesis puts forward a deeper understanding the wider ecology, history, and underlying values of ML research through the analysis of the most highly cited papers from NeurIPS and ICML from the years 2008, 2009, 2018, and 2019 (see Chapter~\ref{chp:values} for further details on this).         
    \item~\textbf{Methodological contributions.} In order to gain an in-depth understanding of ML research, one of the contribution of this thesis has been to embark on a detailed analysis of 100 most cited ML papers from two premier AI conferences: NeurIPS and ICML. To that end, we develop and open source a fine-grained annotation scheme for the detection of values in research papers, including identifying a list of 67 values significant in machine learning research. To our knowledge, this annotation scheme is the first of its kind, and opens the door to further qualitative and quantitative analyses of research.\footnote{Template and all annotations as supplementary material can be found at \url{https://github.com/wagnew3/The-Values-Encoded-in-Machine-Learning-Research} with a CC BY-NC-SA license.} 
    
\end{itemize}


\section{Real-world Impact}

ML research increasingly finds applications in the real world, meaning its impacts are felt outside the field itself, requiring engagement from various stakeholders (from engineers, scientists, policy makers, auditors, regulators, to data subjects). In this regard, the investigation of the Tiny Images dataset was the first audit work which not only resulted in MIT taking down the dataset that has been used in hundreds of computer vision peer-reviewed publications\footnote{http://groups.csail.mit.edu/vision/TinyImages}, but also received wide international media coverage in over 80 outlets. This has contributed to raising awareness of the importance of cleaning, documenting, and managing vision datasets for responsible AI. This work has also inspired ImageNet's face obfuscation project~\cite{yang2021study}, where curators of ImageNet blurred images of human faces in the dataset; the birth of the PASS dataset~\cite{asano2021pass}, a new computer vision dataset that contains no images of humans; new techniques for handling toxicity~\cite{schramowski2021inferring}, techniques for documenting and curating large image datasets that may contain derogatory terms as categories and offensive images; as well as workshops such as \textit{BeyondFairCV}~\footnote{https://sites.google.com/view/beyond-fairness-cv/home?authuser=0} with the attempt to move the field of computer vision towards a more equitable direction. All these efforts are directly linked to (and inspired by) the audit work on Tiny Images and ImageNet presented in this thesis. 

\section{Thesis Structure}
\label{introduction:thesis_structure}

This thesis is structured according to three themes. The first part outlines the thesis landscape, its scope and background, and sketches the broader problem that the thesis tackles. The second part presents empirical analyses and findings that further illustrate the central points in part one. The third part presents a critical and theoretical framework for thinking about the central problems detailed in the thesis and ways for moving forward. 

\textbf{Part~\ref{part:i}} provides background, reviews related work spanning various fields (including complex systems, embodied and enactive cognitive science, as well as AI, ML and ethics), and outlines the problem space. Chapter~\ref{chp:background_related} details previous work across multiple disciplines, outlining where they intersect and the core contributions each make towards the thesis's narrative, as well as their weaknesses. Building on some of the background work outlined in Chapter~\ref{chp:background_related}, Chapter~\ref{chp:automating}, presents a theoretical argument for why people and social systems -- as complex adaptive systems -- are \textit{messy}, \textit{ambiguous}, and \textit{non-determinable}. This is followed by outlining the ways ML predictive tools, more particularly those integrated into the social sphere, impose determinability; an endeavour that not only provides the illusion of control, objectivity, and certainty but is also ethically and morally costly. This cost is unevenly distributed as the historically marginalized, the ``edge-cases'', and those that fail to adhere to stereotypes pay the highest price. The chapter concludes that, far from an ``objective'' tool that understands or predicts human behaviour accurately, ML and AI systems amplify stereotypes, conserve the status quo, and concentrate power where it already exists. 

\textbf{Part~\ref{part:ii}} delves into some of the issues raised towards the end of Part I. Chapter~\ref{chp:automating} and provides empirical examination. Chapter~\ref{chapter:computer vision} looks at computer vision -- one of the sub-fields of AI with one of the most paramount progress -- through a close examination of large scale image datasets. Through a thorough audit work carried out on \textit{ImageNet} and \textit{Tiny Images}, this chapter demonstrates that large scale image datasets present various threats to individuals and society at large, including perpetuating harmful stereotypical representations. 

Over the past few decades, ML has risen from a relatively obscure research area to an extremely influential discipline, actively being deployed in myriad applications and contexts around the world. As the field intersects various domains, one is likely to find varying responses (depending on who one asks) to questions such as what the field values. Chapter~\ref{chp:values} presents the first qualitative and quantitative study of the field's underlying values through a comprehensive analysis of 100 highly cited NeurIPS and ICML papers from four recent years spanning more than a decade. This chapter also demonstrates the increasing presence of large tech corporations in ML research, where power is concentrated in the hands of the few.  

Building both from the theoretical argument in Chapter~\ref{chp:automating} -- that people and social systems are non-determinable and non-predictable with certainty -- and following specific concerns with large scale vision datasets in Chapter~\ref{chapter:computer vision}, Chapter~\ref{chp:all_models} elucidates why some applications of computer vision (such as face and emotion recognition) not only lack scientific rigour but also present a real danger. 

\textbf{Part~\ref{part:iii}} puts forward a way to mitigate some of the problems highlighted throughout the thesis. Building on complex systems science, embodied and enactive approaches to cognition, and Black feminist epistemology, Chapter~\ref{chp:relational} proposes a \textit{Relational Ethics} framework for approaching and thinking about the core issues highlighted in this thesis. Calling for a fundamental shift from rationality to relationality in thinking about people, social systems, datasets, justice, and everything in between, the chapter argues for a view of ethics that is something that goes above and beyond technical solutions. Outlining the idea of ethics built on the foundations of relationality, this chapter calls for a rethinking of justice and ethics as a set of broad, contingent, and fluid concepts and down-to-earth practices that are best viewed as a habit and not a mere methodology or an abstract principle for algorithmic systems. Chapter~\ref{chapter:conclusion} concludes with some thoughts on modelling complexity, equity, and possible directions for future work.

    \chapter{Background and Related Work}
\label{chp:background_related}

\begin{quote}
    ``Interdisciplinary work is notoriously problematic: sections that appear overly simplistic and old hat to one audience strike another as brand new.'' Alicia Juarrero \cite{juarrero2000dynamics}
\end{quote}

This thesis does not fit into any one traditional discipline, and is interdisciplinary through and through. It does not adhere to a single discipline, method, domain, or traditional structure. This does not, however, mean that it is structureless. Instead, rather than attempting to fit a traditional discipline, it is structured and driven by the set of enquiries outlined in Chapter~\ref{chapter:introduction}, Section~\ref{introduction:thesis_approach} and weaves together various approaches, theories, and methods to explore (and sometimes answer) core questions in cognitive science, AI, ML, and ethics. It moves through multiple domains in and out, weaving in aspects of each that are relevant to its core enquiries. This chapter provides an overview of both the background from which the core queries of this thesis arise as well as previous work that this thesis builds from. I 
first 
look 
at the Cartesian/Newtonian worldviews that underline over-enthusiasm around AI. This is followed by post-Cartesian approaches to human behaviour, social systems and AI. I then examine foundational work that emphasizes the disproportionate negative impact of AI systems on marginalized communities. This chapter then contexualizes AI systems by conceiving AI systems as broad ecologies (not just CNNs, for instance) that require massive amounts of data, hidden labour (Ghost work), and environmental resources.

\section{Reductive Approaches}

\begin{quote}
    ``First there is an ``on the one hand'' statement. It tells all the good things computers have already done for society and often even attempts to argue that the social order would already have collapsed were it not for the ``computer revolution''. This is usually followed by an ``on the other hand'' caution which tells of certain problems the introduction of computers brings in its wake. The threat posed to individual privacy by large data banks and the danger of large-scale unemployment induced by industrial automation are usually mentioned. Finally, the glorious present and prospective achievements of the computer are applauded, while the dangers alluded to in the second part are shown to be capable of being alleviated by sophisticated technological fixes. The closing paragraph consists of a plea for generous societal support for more, and more large-scale, computer research and development. This is usually coupled to the more or less subtle assertion that only computer science, hence only the computer scientist, can guard the world against the admittedly hazardous fallout of applied computer technology.'' Joseph Weizenbaum~\cite{weizenbaum1972impact}
\end{quote}

Although this was written in 1972, the same pattern that Weizenbaum identified persists today: for any AI breakthrough, the achievements and potential benefits are applauded, while the potential harms, flaws, gross simplification of complex issues, inaccuracies, methodological problems, limitations and failures are underemphasized. 
This is often followed by a plea for more generous societal support. AI systems are not only put forward as a solution to complex challenges, in some cases, they are compared to humans and attributed human like capabilities, an attitude that is particularly more pronounced within the artificial general intelligence (AGI) domain. This reductive thinking rests on completely misguided and simplistic notions of human being, intelligence, cognition, behaviour and society and has roots in Cartesian and Newtonian worldviews.    

\subsection{Cartesian and Newtonian Roots}

Certainty and order have always been highly sought after in Western philosophy and sciences. Through the process of elimination of all things that can be doubted, Descartes attempted to get rid of unreliable and fallible human intuitions, senses, and emotions. This was fundamental in the quest to establish a secure foundation for absolute knowledge based solely on solid grounds: reason and rational thought~\citep{descartes1984philosophical}. Central to Descartes' work was to uncover the permanent structures beneath the changeable and fluctuating phenomena of nature in which he could build the edifice of unshakable foundations of knowledge. The view of the person that emerged from such worldview was a primarily rational, static, self-contained, and self-sufficient subject that contemplates the external world from afar in a “purely cognitive” manner as a disembodied and disinterested observer~\citep{gardiner1998incomparable}. In the desire to establish timeless and absolute certainty, cognitive capabilities and mental processes were privileged as of primary importance to what it means to be a person. Complete understanding, control, order, manipulation, and prediction find a comfortable home in this worldview. Although few, if any, scholars identify with the Cartesian view as originally proposed by Descartes, this worldview still prevails today in subtle forms (where human cognition or intelligence are often taken as a purely brain phenomena, for example). 

In a similar vein and with a similar fundamental influence as Cartesianism, the Newtonian world view aspired to impose order and to arrive at \textit{universal} and \textit{objective} knowledge in a supposedly \textit{observer-free} and \textit{deterministic} world. This worldview sees the world as containing \textit{discrete}, \textit{independent}, and \textit{isolated atoms} with traceable causal relations. Within the physical world, Newtonian mechanistic descriptions allowed precise predictions of systems at any particular moment in the future, given knowledge of the current position, speed, and acceleration of a system. This view fared poorly, however, when it came to the \textit{messy}, \textit{interactive}, \textit{fluid}, and \textit{ambiguous} world of complex living systems that are inherently \textit{context bound}, \textit{socially embedded}, and in \textit{continual flux}. In a worldview that aspires for certainty and predictability, the very idea of ambiguity, complexity, and multivalence, the `essence of being', so far as there can be any, is not tolerated. Despite the inadequacy of the billiard ball model of Newtonian science in approaching complex adaptive systems such as human affairs, its residue prevails today, directly or indirectly~\citep{juarrero2000dynamics}. 

Descartes and Newton did not single-handedly carve out lasting worldviews that have come to dominate much of Western thought. Nonetheless, they represent the quintessential figures that envisaged an \textit{objective}, \textit{universal}, and \textit{relatively static} worldview governed by laws. This striving for a universal law, Daston~\citep*{daston2018calculation} argues, is a predicament that fails when confronted with unanticipated particulars since no universal ever fits the particulars. Commenting on current ML practices Daston~\citep*{daston2018calculation} explains: ``ML presents an extreme case of a very human predicament, which is that the only way we can generalize is on the basis of past experience. And yet we know from history — and I know from my lifetime — that our deepest intuitions about all sorts of things, and in particular justice and injustice, can change dramatically.'' Furthermore, as Ahmed~\citep*{ahmed2007phenomenology} contends, all bodies inherit history and, more fundamentally, the inheritance of Cartesianism is grounded in white straight ontology. The reality of the Western straight white male masquerades as the invisible background that is taken as the ``normal'', ``default'', ``standard'', or ``universal'' position. Anything outside of it is often cast as ``dubious'' or an ``outlier''. 

ML systems currently pervading the social sphere embody the core values of the Cartesian and Newtonian worldviews where historical, fluctuating, and interconnected behaviour is presumed to be formalized, clustered, and predicted in a ``value-free'' and ``neutral'' manner. The historic Bayesian framework of prediction is a primary example of this~\citep{doi:10.1098/rstl.1763.0053}. This framework has played a central role in establishing explanations of behaviour based from ``rational principles alone''~\citep{jones2011bayesian, hahn2014bayesian}. Bayes' approach, which is increasingly used in various areas including data science, machine learning, and cognitive science~\citep{seth2014cybernetic,jones2011bayesian}, played a pivotal role in establishing the cultural privilege associated with statistical inference and set the ``neutrality'' of mathematical predictions. 


\subsection{Reductionism and Computation: Mistaking the Map for the Territory}
Reductionist thinking about concepts such as intelligence, cognition, social phenomena, and emotions is not unique to philosophers, scientists and technologists alike but a stubborn persistent misconception that is also a core feature of traditional cognitive science. We can only presume to build machines \textit{like us} once we see ourselves as \textit{machines first}. 
Such reductionist traditional cognitive science rests on a deep misconception and a fallacy that mistakes the map for the territory. Throughout history we have compared ourselves metaphorically with the most advanced technology of the time, and all too often distorted reality in the process, or promoted unintended and noxious ideological aims. The mid-17th century Danish anatomist Nicolas Steno saw the brain as a collection of cavities through which animal spirits flow~\cite{cobb2020idea}. A century later when electricity was in vogue, the brain was seen like a galvanic battery. And by the mid-19th century, nerves were compared to telegraph wires~\cite{cobb2020idea}. After mid-20th century advances in computing, we have come to think of the brain as a computer, an information processing machine. 


Metaphors are pervasive and we don’t seem to do away with them generally. Metaphors illuminate understanding one kind of thing in terms of another~\cite{lakoff2008metaphors}. But metaphors can also do more harm than good when we forget they are, in fact, \textit{only} metaphors. Oftentimes we do forget, sometimes intentionally when such ignorance serves to help advance our agenda. As Lewontin~\cite{lewontin1996biology} succinctly put it: ``[W]e have become so used to the atomistic machine view of the world that originated with Descartes that we have forgotten that it is a metaphor. We no longer think, as Descartes did, that the world is \textit{like} a clock. We think it \textit{is} a clock.''.

Mistaking the map for the territory is one of the persistent problems in AI. Limited understanding of the complex nature of intelligence as well as over-optimistic and overconfident predictions about AI, as old as the field itself, are undermined by various fallacies and are obstacles to actual progress in AI~\cite{mitchell2019artificial}. Among these fallacies are ``intelligence is all in the brain'' and the fallacy of wishful mnemonics such as the attribution of "learning," "seeing," "feeling," and other qualities to AI based on a misguided apophenia~\cite{mitchell2019artificial}. AI systems \textit{do not} understand, intend, volute, empathise, feel, experience and so on, as these are activities that require sense-making and precarious embodied existence. Yet through the use of these heuristic modes of summarizing patterns of action, human-like qualities are attributed to machines. 

To conceive of AI as ‘human-like machines’, implicitly means to first perceive human beings in machinic terms: complicated biological information processing machines, ``meat robots'', shaped by evolution. Once we see ourselves as machines, it becomes intuitive to see machines as ‘like us’. This circular metaphorical trick is key to reducing the complex, relational, intersubjective, embodied, non-determinable, non-totalizable, fluid, active and dynamic being into a set of features or a process that could be implemented by the physical brain.

\section{Post-Cartesian Approaches}
\label{sec:post_cart_app}
We are sense-makers that relate to others and the world around us in terms of significance and meaning. ``Sense making is always under way.''~\citep[p.219]{di2018linguistic}. We continually relate to others, interpret their gestures and expressions, understand what has been unsaid (as well as what has been said), as living beings with ongoing interactions with others and the world around us. We face challenges, tensions, problems, and opportunities that constrain or enable our actions and behaviours. Outlining the precarious, socially embedded and situated nature of human beings, Weizenbaum wrote ``no other organism, and certainly no computer, can be made to confront genuine human problems in human terms.''~\cite{weizenbaum1976computer}. The idea of creating a human-like machine (even in principle) not only is a project that rests on misconceptions of what a human being is, but is also a ``fraud that plays on the trusting instincts of people.''~\cite{weizenbaum1976computer, pasquale2020new}. 

Critics might say that bodies are not unique to humans and to grant carbon a special quality that silicon lacks would amount to bio-chauvinism. However, this stems from the limited understanding that equates embodiment to being a physical object. A mere physical body does not denote embodiment. Living bodies are vastly different from machines. Through the process of self-individuation, living bodies continually create a distinction between themselves and their environment where none existed before they appeared and none will remain after they are gone. Living bodies are not static organisms that passively sit around waiting to be influenced by their environments like much of the cognitivist tradition envisages. Living bodies are fluid, networked and relational beings with values, shifting moods and sensitivities. ``Living bodies have more in common with hurricanes than with statues.''~\cite{di2018linguistic}. Living bodies laugh, bite, eat, gesture, breathe, give birth and feel pain, anger, frustration, and happiness. They are racialized, stylized, politicized and come with varying ableness, positionalities, privilege, and power. 

A machine, on the other hand, is something that can be achieved once a given process is complete, that is, it is understood such that it can be implemented. People, as complex adaptive systems, are open-ended, historical and embedded in dynamic and non-linear interactions. There is no perfect or accurate representation or model of cognition, intelligence or emotion as such a model would require capturing cognition, intelligence, or emotions in their entirety. The reason being that compressing a complex system into an algorithm or a model without reducing its complexity is impossible. This means the perfect model of a complex system would have to be as complex and unpredictable as the system itself. In other words, the best and simplest representation of a complex system is the system itself~\cite{cilliers2002complexity, juarrero2000dynamics}. Human cognition, intelligence, consciousness, etc then cannot be reduced and captured in an algorithm  in their entirety. Even for fields such as Artificial Life (which come from awareness of the complexities I raise here), concerned with generating artificial systems—via computer simulations, robotic agents, or biochemical processes—that behave like living organisms~\cite{langton1997artificial}, non-determinability and unpredictability is singled out as as the key characteristic of humans that differentiates them from artificial beings. ``[T]hings appear as intrinsically meaningful for living beings because of their precarious existence as adaptive autopoietic individuals.''~\cite{froese2019problem}. 

\subsection{Post-Cartesian Computation}

As we have seen, reductive approaches remain pervasive in various fields of enquiry from the physical to the human sciences, philosophy, and computational sciences~\citep{dreyfus2007heideggerian,winograd1986understanding,weizenbaum1976computer} in nuanced forms. However, there also exist various areas that advance what might be referred as post-Cartesian computation and ALife is one such example. Grounded in a dynamically interactive, embodied, distributed, and fluid understanding of the world, various approaches address questions of AI in a manner that goes beyond Cartesianism. The broadly construed field of Artificial Life (ALife), for example, is concerned with generating artificial systems -- via computer simulations, robotic agents, or biochemical processes -- that behave like living organisms~\citep{langton1997artificial}. Technology is conceived as dynamic, interactive, and embedded in social systems within ALife inspired approaches. Through a proposal for dynamic interactive artificial intelligence (dAI) Dotov and Froese~\citep*{dotov2020dynamic}, for example, call for systems that emphasize user-machine inter-dependence over autonomy. In recognition of the adaptive and self-organizing nature of technology, the term ``living technolgies'' has gained momentum within the field of ALife~\citep{bedau2010living,gershenson2013living,aguilar2014past}. And as technology becomes more ``living'', the question of safety also becomes more central~\citep{gershenson2013living}. The morphing of technology into society, brings both benefits and challenges, according to Aguilar et al.~\citep{aguilar2014past}. This in turn, the authors argue, calls for the establishment of ethical principles for artificial life. Similarly, Bedau et al.~\citep*{bedau2010living} have argued that the creation of ``living'' technologies requires the considerations of ethical issues, the development of safeguards, as well as ``proper mechanisms to prevent its misuse''. Echoing similar sentiments, Helbing et al.~\citep*{helbing2012futurict}, have pointed out the risk of such technologies benefiting only a few stakeholders instead of all humanity.

However, although dynamicity, embeddedness, and reciprocal and interactive technology-society relationships form the foundations for ALife approaches, the social, moral, ethical, and political concerns of technology are merely explored in a justice oriented manner. 
Ethics and safety concerns are receiving increased attention within the field of ALife, yet such attention has been largely peripheral with little in-depth rigorous treatment of these concerns as a subject of enquiry in and of themselves. Questions of 
ethics 
and morality also remain primarily abstract and hypothetical. This abstractness manifests in discussions regarding ethics in AI that tend to revolve around ``superintelligence'' explosion and around rights for robots, at the cost of discussions around the current disproportionate negative impact of technology~\cite{birhane2020robot}. The notion of ethics often revolves around the moral status of the ``intelligent system''. Should the supposedly ``intelligent'' system have moral or legal rights on par with a human being? Does the experience of pain or the capacity for a ``theory of mind'' differ when the entity is carbon or silicon based? Does working towards ethical systems come down to Isaac Asimov’s conception of laws of ethical robotics? How do we prepare humanity for the Singularity? These concerns, for the most part, focus on hypothetical and/or future ``First World Problems'' \citep{birhane2020robot}. These quests might be a valid intellectual exercise in and of themselves but in light of the mass integration of ML systems into society and the harms they impose on vulnerable individuals and communities, I argue attention regarding machine ethics should primarily focus on current and tangible concerns. 

This thesis, while building on the fluid technology-society relationship underlying ALife, also departs from such tradition by primarily focusing on the \textit{disproportionate negative impact} of current technology on marginalized communities and the \textit{power asymmetries} and \textit{structural inequalities} permeating the social world from which technology emerges. 




\section{The Forgotten Margins of AI Ethics}
\label{sec:forgotten_margins}
\begin{quote}
   ``To treat fairness and justice as terms that have meaningful application to technology separate from a social context is therefore to make a category error, [...], an abstraction error.'' Selbst et al.~\cite{selbst2019fairness}
\end{quote}

As algorithmic systems permeate all corners of social and political life, from education~\cite{marcinkowski2020implications}, to medicine~\cite{obermeyer2019dissecting,benjamin2019assessing} and the criminal justice system~\cite{lum2016predict,angwin2016machine}, inquiries into negative impacts and harms of these systems have become an essential endeavour. Although critical voices have existed as long as the computing and AI fields themselves (Weizenbaum~\cite{weizenbaum1976computer}; Dreyfus~\cite{dreyfus1972computers}; Winograd and Flores~\cite{winograd1986understanding}), a robust body of work over the past few years has established the broad field of AI Ethics as foundational. Work on fairness, accountability, transparency, explainability, bias, equity, and justice are generally conceived under this broad umbrella concept of AI  Ethics. 

The urgency and profound importance of ethics in AI is signalled by recent landmark studies (Angwin et al.~\cite{angwin2016machine}; Buolamwini and Gebru~\cite{buolamwini2018gender}), seminal books (Noble~\cite{noble2018algorithms}; O’Neil~\cite{o2016weapons}; Eubanks~\cite{eubanks2018automating}; Pasquale~\cite{pasquale2015black}), newly found conferences exclusively dedicated to AI Ethics (e.g. AIES and FAccT), the fast-growing adoption of ethics into syllabi in computational departments~\cite{fiesler2020we}, increased attention to policy and regulation of AI~\cite{jobin2019global}, and increasing interest in ethics boards and research teams dedicated to ethics in major tech corporations. 

The urgency and heightened attention is warranted as AI systems, especially those integrated into the social sphere, are not simply technical systems but socio-technical systems that are likely to impact individual people, communities and society at large. From discriminatory policing~\cite{scannell2019not,browning2020stop,lum2016predict} and discriminatory medical care allocations~\cite{obermeyer2019dissecting,vyas2020hidden,powe2020black}, to discriminatory targeted ads~\cite{speicher2018potential,ali2019discrimination,kuhn2013gender}, algorithmic systems present real harm to the most marginalized groups of society and furthermore pose a threat to human welfare, freedom, and privacy. 

Often, the topic of ethics, particularly within the Western context, has been considered a more speculative than practical endeavour~\cite{gardiner1996alterity}. As a branch of Western philosophy, the broad field of ethics enquires, assesses, and theorizes about questions of justice, virtue, fairness, good and evil, and right and wrong. Such examinations often treat questions of ethics in an abstract manner, aspiring to universal theories or principles that can be uniformly applied regardless of time, place, or context~\cite{markova2016dialogical,de2019loving,shotter2006vygotsky}. Canonical Western approaches to ethics, from deontology to consequentialism, at their core aspire for such universal and generalizable theories and principles ``uncontaminated'' by a particular culture, history, or context. Underlying this aspiration is the assumption that theories and principles of ethics can actually be disentangled from contingencies and abstracted in some form devoid of context, time, and space~\cite{juarrero2000dynamics,daston2018calculation}. What’s more, this ambition for universal principles in ethics have been dominated by a particular mode of being human~\cite{Wynter03,ahmed2007phenomenology}; one that takes a straight, white ontology as a foundation and recognizes the privileged Western white male as the standard/quintessential representative of humanity. 

As an example, Daoist philosophy pushes back against the traditional understanding of the ontological subject as singular through its "relational ontology," which emphasizes being and subjectivity as emerging from the interaction of multiple relations rather than a singular source~\cite{ameshall2003}. While traditions like Black Feminism~\cite{h1984feminist,collins2002black}, Care Ethics~\cite{noddinge2013caring}, and the broad traditions of ethics that have emerged from Asian~\cite{li1994,ameshall2003} and African~\cite{mbiti1969african,fromrationality2020} scholars challenge the ontologies presumed in the ``canonical'' approach to ethics common to both the field of philosophy and the growing field of AI Ethics. These same philosophical traditions remain marginalized in both disciplines. Although there have been notable attempts to incorporate non-Western and feminist care ethics into the existing ethical paradigms within AI Ethics and Western philosophy~\cite{vallor2016}, these attempts remain marginalized by the structure of both disciplines. 

Abstraction can be, and in some domains such as mathematics and art is, a vital process to discerning important features, patterns, or aspects --- a way of seeing the forest for the trees. Indeed, while abstractions and aspirations for universal principles can be worthy endeavours, whether such an objective is useful or suitable when it comes to thinking about ethics within the realm of AI and ML, is questionable. For example, Selbst et al.~\cite{selbst2019fairness} contend that, when used to define notions of fairness and to produce fairness-aware learning algorithms, abstraction renders them ``\textit{ineffective, inaccurate, and sometimes dangerously misguided}''. Accordingly, they emphasize that abstraction, which is a fundamental concept of computing, also aids to sever machine learning tools from the social contexts in which they operate. The authors identify five traps, which they collectively name the abstraction traps, within the AI Ethics literature and the field of Computer Science in general, which result from failure to consider specific social contexts. 

Meanwhile, landmark work related to algorithmic decision-making in recent years has brought to light issues of discrimination, bias, and harm, and has demonstrated harms and injustices in a manner that is grounded in concrete practices focusing on specific individuals and groups: computer vision systems deployed in the social world suffer from disproportionate inaccuracies when it comes to recognizing faces of Black women compared to that of men~\cite{buolamwini2018gender}; when matters of social welfare are automated, the most vulnerable in society pay the heaviest price~\cite{eubanks2018automating}; and search engines continually perpetuate negative stereotypes about women of colour~\cite{noble2018algorithms}; to mention but a few examples. Moreover, within the majority of media reporting, when problems that arise from algorithmic decision-making are brought forth, we find that those disproportionately harmed or discriminated against are individuals and communities already at the margins of society, making it important that approaches to AI Ethics explicitly consider these disparate impacts in their design and implementation. 

This connection between AI (and AI-adjacent) systems and discriminatory societal impact is not a new phenomenon. Indeed, one of the earlier police computing systems in the USA was used starting in the 1960s by the Kansas City police department to (among other tasks) analyze and allocate resources for operations and planning, which included police officer deployment~\cite{mcilwain2019black}. By using not only attributes such as the race of individuals arrested, but also geographically linked data (such as number of crimes within a given area), the system created an increased surveillance and policing state that was particularly devastating for the Black population of Kansas City (see Rothstein~\cite{rothstein2017color}; Taylor~\cite{taylor2019race}, for historical discussions of the laws and practices that led to geographic segregation by race that continues to be a feature of USA neighborhoods). This is not a particularly surprising outcome if one looks beyond strictly computing systems to consider how various technologies have been used across space and time to surveil and police Black people~\cite{browne2015dark} and this gives an example of why consideration of the \textit{institution} in which a technology used is important.  

A robust body of research on algorithmic injustice~\citep{benjamin2019race,eubanks2018automating,birhane2021algorithmic} shows that predictive systems perpetuate societal and historical injustice. In a landmark study, Buolamwini and Gebru~\citep*{buolamwini2018gender} evaluated gender classification systems used by commercial industries. They found huge disparities in image classification accuracy; lighter-skin males were classified with the highest accuracy and darker-skin females were the most misclassified group. Similarly, object detection systems designed to detect pedestrians display higher error rates when identifying dark skin pedestrians while light-skinned pedestrians are identified with higher precision~\citep{wilson2019predictive}. The use of these systems ties the recognition of subjectivity to skin tone. Recidivism algorithms unfairly score black defendants as higher risk compared to white defendants of similar criminal conviction~\citep{angwin2016machine}. Hiring tools tend to disproportionately disadvantage women~\citep{ajunwa2016hiring}. Additionally, the notion of gender that ML systems depend on is a fundamentally essentialist one that operationalizes gender in a trans-exclusive way resulting in disproportionate harm to trans people~\citep{keyes2018misgendering, hamidi2018gender, barlassee2020}. Machine classification and prediction, thus, negatively impact individuals and groups at the margins the most. 

Tracking the relation of current AI systems to historical systems and technologies (and, indeed, social systems) allows us to more correctly contextualize AI systems (and by extension considerations of AI Ethics) within the existing milieu of socio-technical systems. This gives us an opportunity to account for the fact that ``...social norms, ideologies, and practices are a constitutive part of technical design''~\cite{benjamin2019race}. Benjamin~\cite{benjamin2019race} makes a useful connection to the ``Jim Crow'' era in the United States in coining the term the ``New Jim Code''. They argue that we might consider race and racialization as, itself, a technology to \textit{``sort, organize, and design a social structure}''~\cite{benjamin2019race}. Thus, AI Ethics work addressing fairness, bias, and discrimination must consider how a drive for efficiency, and even inclusion, might not actually move towards a ``social good'', and also may accelerate processes and strengthen technologies that further entrench social structures and institutions, which create the margin at which many humans are deemed to operate.

\section{Contextualizing AI: Looking under the Hood and across the Ecology}
\label{sec:contextualizing}

\begin{quote}
    ``The structures of power at the intersection of technology, capital and governance are well served by [a] narrow, abstracted analysis [of AI].'' Crawford~\cite{crawford2021atlas} 
\end{quote}

ML/AI systems, for many societies, constitute part of the social ecology forming components of a self-organizing system. These systems exist in and evolve with social norms, trends, and societal uptakes. They morph into the background of daily life to the extent that we forget they exist. Weiser~\citep*{weiser1999computer} remarked ``The most profound technologies are those that disappear. They weave themselves into the fabric of everyday life until they are indistinguishable from it''. Good infrastructure is, by definition invisible and forms part of the background for other work~\citep{star2002infrastructure}. ML systems have become inextricably intertwined with what it means to be a human being, yet remain invisible forces that shape lives and opportunities of countless individuals and communities. ML tools are not simply ``methods'' that sort and classify people and the social world, but are also apparatuses that directly act upon the world transforming social realities and producing certain subjectivities (and not others)~\citep{mcquillan2018data}. For instance, faced with an automated assessment system, a job seeker is likely to alter her behaviour in a manner that guarantees positive outcome; 
awareness that one's social media `post' has the potential to impact one's perceived characteristics or ``fittness'' for a job has the potential to alter her actions and behaviour. AI and ML, then, are not just algorithmic models but tools that alter the social fabric and include the wider social, political, cultural background and environmental infrastructure.

Importantly, foundational work in Science, Technology, and Society Studies (STSS), Critical Theory, and Philosophy of Science illustrate that technologies are inherently value-laden, and these values are encoded in technological artifacts, many times in contrast to a field's formal research criteria, espoused consequences, or ethics guidelines~\cite{winner1980artifacts,bowker2000sorting,benjamin2019race}. There is a long tradition of exposing and critiquing such values in technology and computer science. For example, Winner~\cite {winner1980artifacts} introduced several ways technology can encode political values. Rogaway~\cite {rogaway2015moral} notes that cryptography has political and moral dimensions and argues for a cryptography that better addresses societal needs. Weizenbaum~\cite {weizenbaum1976computer} argued in 1976 that the computer has from the beginning been a fundamentally conservative force which solidified existing power: in place of fundamental social changes, he argued, the computer renders technical solutions that allow existing power hierarchies to remain intact. 

This thesis extends these critiques to the field of ML. It is a part of a rich space of interdisciplinary critiques and alternative lenses used to examine the field. Works such as Mohamed et al.~\cite {mohamed2020decolonial} critique AI, ML, and data using a decolonial lens, noting how these technologies replicate colonial power relationships and values, and propose decolonial values and methods. Others~\cite {benjamin2019race, noble2018algorithms, d2020data} examine technology and data science from an anti-racist and intersectional feminist lens, discussing how our infrastructure has largely been built by and for white men; D’Ignazio and Klein~\cite {d2020data} present a set of alternative principles and methodologies for an intersectional feminist data science. Similarly, Kalluri~\cite {kalluri2020don} denotes that the core values of ML are closely aligned with the values of the most privileged and outlines a vision where ML models are used to shift power from the most to the least powerful. Dotan and Milli~\cite{dotan2019value} argue that the rise of deep learning is value-laden, promoting the centralization of power among other political values. Many researchers, as well as organizations such as Data for Black Lives, the Algorithmic Justice League, Our Data Bodies, the Radical AI Network, Indigenous AI, Black in AI, and Queer in AI, explicitly work on continuing to uncover particular ways technology in general and ML in particular can encode and amplify racist, sexist, queerphobic, transphobic, and otherwise marginalizing values~\cite{buolamwini2018gender, prabhu2020large}.

To narrowly conceive of AI in strictly technical terms such as convolutional neural network (CNN), generative adversarial network (GAN), or regression analysis is to obscure the massive amounts of data, compute power, and other resources required for AI to function. This narrow conception of AI also obscures the power asymmetries between those that produce AI and those that are (sometimes without their awareness) subjected to it and the uneven benefit/harm distribution, where the society's privileged and powerful receive the most benefit while those at the margins pay the highest price when AI fails.  

This narrow conception of AI can be partly connected to the reductive view that sees machines as fully-autonomous. However, machines are \textit{always} human-machine systems. That is to say, machines are \textit{never} fully autonomous but always human-machine systems that rely on human power and resources in order to function. Automation and the idea of automata, from its early conception, relied on a clever trick that erased the labourers toiling away in the background, the people performing crucial tasks for a machine to operate~\cite{jones2020ghost,sadowski2018potemkin,taylor2018automation,gray2019ghost}. Surveying the historical genealogy of mechanical calculations, Daston~\cite{daston2017calculation} emphasizes that far from relieving the mental burden, automation shifted the burden to other shoulders (often women who were paid little), maintaining the ghost in the machine.

Seemingly autonomous systems are profoundly human-machine systems. And furthermore, as demonstrated by Bainbridge’s seminal work on the automation of process and vehicle control on aircraft automation, ``the more advanced a control system is, so the more crucial may be the contribution of the human operator''~\citep[p.775]{bainbridge1983ironies}. Bainbridge’s concept of the \textit{ironies of automation}, gets at the heart of intentionally hidden human labour in an appeal to portray machines as autonomous. In reality as Baxter et al.~\cite{baxter2012ironies}, building on Bainbridge’s classic work, make explicit: ``[T]he more we depend on technology and push it to its limits, the more we need highly-skilled, well-trained, well-practised people to make systems resilient, acting as the last line of defence against the failures that will inevitably occur.''~\cite{baxter2012ironies}. Nearly 40 years later, Bainbridge’s point remains. Effective automation of control processes necessarily requires humans at various steps; developing, maintaining and stepping in when ‘autonomous’ systems inevitably experience failure~\cite{hancke2020ironies,baxter2012ironies}. Similarly, today’s ‘autonomous’ vehicles still depend on human ``safety drivers'' and the need for human input is unlikely to disappear entirely although it might change form~\cite{tubaro2019micro}. 

Not only does seemingly autonomous AI rely on high-paid, high-skilled engineers and scientists, it also relies in crucial ways on underpaid, undervalued, and less-visible labour which goes by various names including \textit{Ghost work}, \textit{ microwork}, and \textit{crowd-work}~\cite{irani2015cultural,gray2019ghost}. From labeling images, identifying objects in images to annotating data, such human labour ``help[s] AI get past those tasks and activities that it cannot solve effectively and/or efficiently''~\cite{tubaro2019micro}, steps in to fill in when AI fails~\cite{irani2015cultural} and is automation’s ‘last mile’~\cite{gray2019ghost}. Such work constitutes the backbone of current AI -- without it, AI would cease to function. Yet, from Amazon’s MTurk, to Clickworker, AppJobber, to CrowdTap, such labour often goes unrecognized in the AI pipeline. People doing such work, in most cases are not formally considered as formal employees but independent contractors further adding to their precarious working conditions.

Additionally, the enormous environmental damage caused by AI is something that we have to grapple with. As Crawford~\cite{crawford2021atlas} argues, from personal assistants such as Siri to the mobile devices and laptops to ``self-driving'' cars, AI cannot function without minerals, elements, and materials that are extracted from the earth. An ``AI system, from network routers to batteries to data centers, is built using elements that required billions of years to form inside the earth.''~\cite{crawford2021atlas}[p.31]. Crawford lists 17 rare earth elements including lanthanum, cerium and praseodymium, which are making our smart devices smaller, lighter and improve performance. But extracting, processing, mixing, smelting, and transporting these elements and mineral comes at a huge environmental destruction, local and geopolitical violence, and human suffering. For example, half of the planet’s total consumption of lithium-ion, a crucial element for the Tesla Model S electric car battery, is consumed by Tesla~\footnote{``[O]nly 0.2 percent of the mined clay contains the valuable rare earth elements. This means that 99.8 percent of earth removed in rare earth mining is discarded as waste, called ‘tailings,’ that are dumped back into the hills and streams, creating new pollutants like ammonium''~\citep[p.37]{crawford2021atlas}}.

Crawford goes on: ``[i]f we visit the primary sites of mineral extraction for computational systems, we find the repressed stories of acid-bleached rivers and deracinated landscapes and the extinction of plant and animal species that were once vital to the local ecology''~\citep[p.36]{crawford2021atlas}. The creation of AI then can be felt in the ``atmosphere, the oceans, the earth’s crust, the deep time of the planet.''~\citep[p.36]{crawford2021atlas}. Although much of the energy consumption by AI models is a closely guarded secret, Lotfi Belkhir and Ahmed Elmeligi~\cite{belkhir2018assessing} estimate the tech sector will contribute 14 percent of global greenhouse emissions by 2040 (see also,~\cite{Brevini21,dobbe2019ai}). The ``thought'' of AI is incredibly resource intensive: ``Training a single BERT base model (without hyperparameter tuning) on GPUs was estimated to require as much energy as a trans-American flight''~\cite{bender2021dangers}.  

Furthermore, putting the exploitative structure of microwork and the environmental cost of creating AI aside, data that currently fuels AI systems is often sourced in questionable manners; it is often uncompensated and unregulated. The models built on such data furthermore amplify societal and historical stereotypes (negatively disproportionately impacting marginalized communities down the line as AI systems trained on such datasets are often used in decision making). The deep learning revolution that transformed image detection, identification and recognition, for example, is only possible through continued mass 
scraping 
of user uploaded images, all sourced and used to build AI systems without consent or awareness of image owners~\cite{birhane2021large}. Of some of the major large scale datasets currently used to train and validate computer vision models, none is sourced consensually (see Chapter~\ref{chapter:computer vision} for more). 

\section{Large Scale Datasets}
\label{subsec:LSVD}
Over the past decade, machine learning in general and computer vision in particular, have become widespread and significant research fields. The success of these fields rests on ``availability'' of vast amounts of data. The rise of these massive datasets has given rise to a robust body of critical work that has called for caution while generating these large datasets. Large scale vision datasets, for example, contain various broad problems including curation biases, inclusion of problematic content in the images, and the questionable approaches of associating these images with offensive and non-imageable labels, as well as the gradual erosion of privacy~\cite{scheuerman2021datasets,paullada2020data}. 

Within computer vision, automatically generating textual descriptions of people in images, especially with regards to appearance characteristics that may convey sensitive attributes such as gender, race, disabilities, and ethnicity presents a great challenge, uncertainty, and risk of stereotyping and biases through image descriptions~\cite{bennett2021s, hanley2021computer, otterbacher2019we}. 
Various works~\cite{atwood2020inclusivegoogle,larrazabal2020gender,wang2020revise, denton21} have highlighted gender, racial, and geographical biases surrounding the sourcing of image datasets as well as the opacity of such endeavors~\cite{Prabhu2021JFT}. Large scale vision datasets, when sourced from the internet, often contain problematic content. ImageNet, for example has also been found to include non-consensual-voyeuristic imagery~\cite{crawford2021excavating, birhane2021large} and NSFW content. Previous efforts such as~\cite{Insideth82:onlineregister}) have revealed the presence of strongly misogynistic content in the ImageNet dataset, specifically in the categories of \texttt{beach-voyeur-photography, upskirt images, verifiably pornographic and exposed private-parts}. These specific four categories have been further been researched in digital criminology and intersectional feminism~\cite{henry2017not,mcglynn2017beyond,powell2018image,powell2010configuring} and have form the backbone of several legislations worldwide~\cite{mcglynn2017morelaw1, gillespie2019tacklinglaw2}.

Labeling is also a great concern. This includes stagnant vocabulary of labels~\cite{yang2020towardsimagenetfacct}, misrepresentation of gender~\cite{scheuerman2021datasets}, prevalence of ethnophaulisms~\cite{birhane2021large} and non-imageability (non-imageable classes include adjectives such as ``hard'' (say in ``hard-working'') where attempt to capture them in images is futile) issues in the label space~\cite{yang2020towardsimagenetfacct,Prabhu2021JFT,muller2021designing}.


These critiques have resulted in some corrective measures including the retraction of the MS Celeb\footnote{\url{https://exposing.ai/}} and TinyImages\footnote{\url{http://groups.csail.mit.edu/vision/TinyImages}} datasets, blurring of the images of people~\cite{yang2021studyimagenet} and filtering out of constituent images to create a sanitized version of the original dataset in ImageNet. For example, the curators of Imagenet advocated removing 2674 out of 2832 existing synsets in the "person" subtree of the label space~\cite{yang2020towardsimagenetfacct}. 


\section{Conclusion}

This chapter has sought to map the core perspectives broadly construed; post-Cartesian approaches to behaviour, society and AI, critical race, data and algorithm studies, and STSS that underlie the thesis. This chapter has also surveyed 
previous work: outlining the limitation of reductive approaches, examining those at the intersection of post-Cartesian approaches and tech, foundational work demonstrating how algorithmic systems -- when they fail, which they often do -- disproportionately impact society's most vulnerable, work that exposes often ignored ``invisible'' labour, massive resources behind AI, the limitations behind large scale datasets as well as work that examines the values embedded in AI. The presented background and previous work that this thesis builds from, lays the ground for the proceeding chapters.

\chapter{The Impossibility of Automating Ambiguity}
\label{chp:automating}

On the one hand, complexity science and enactive and embodied cognitive science approaches emphasize that people, as complex adaptive systems, are ambiguous, indeterminable, and inherently unpredictable. On the other, ML systems that claim to predict human behaviour are becoming ubiquitous in all spheres of social life. In this chapter, I contend that ubiquitous AI and ML systems that 
cluster, sort, and predict human behaviour and action, are systems that force order, equilibrium, and stability to the active, fluid, messy, and unpredictable nature of human behaviour and the social world at large. Grounded in complexity science and enactive and embodied cognitive science approaches, this chapter emphasizes why people, embedded in social systems, are indeterminable and unpredictable. When ML systems ``pick up'' patterns and clusters, this often amounts to identifying historically and socially held norms, conventions, and stereotypes. Machine prediction of social behaviour, I argue, is not only erroneous but also presents real harm to those at the margins of society.

\section{Introduction}
\label{ambiguity_intro}

Post-Cartesian frameworks, including developments within the embodied and enactive cognitive sciences, complex systems science, and dialogical approaches to cognition, strongly emphasize the inherently indeterminable nature of the person and the inextricably entangled relationship between person, other, and technology. These traditions have challenged Cartesian ambitions that neatly delineate human behaviour and actions into dichotomies, instead emphasizing ambiguities, continuity, and fluidity. The person exists in a reciprocal relationship with others in a social, cultural, and increasingly digitized and automated milieu. People, far from being static Cartesian selves, are active, dynamic, and continually moving. The ``interactive turn''~\citep{DeJaegher}, for example, has been playing a crucial role in shifting emphasis from the view of the individual as a relatively stable and fully autonomous entity that can be fully understood, to the view of the person as active and dynamic, pregnant with a myriad of open-ended possibilities. On a similar note, distributed cognition and extended mind~\citep{clark1998extended} frameworks have challenged the idea that cognition ends at the skull and that the skin marks the contours of the self, fuzzing the traditionally held neat understanding of cognition and self. 

We interact with others and the evirnoment in a non-linear way. These interactions self-organise into sensorimotor couplings we may call habits or skills. Based on these couplings, we perceive (or rather ‘enact’) things in the world in the first instance as affordances for action~\cite{golonka2012gibson}. The things-as-affordances we perceive have direct relations with our bodily skills~\cite{dreyfus2004ethical}. To give a common-sense example: a park bench \textit{is} a different thing to a skateboarder, or a homeless person, than it is to a casual visitor. Embodied skills self-organize out of, and work to further sustain the organism. Furthermore, human beings are social through and through. We are always already situated within social practices, and the way we interact with and make sense of the world needs to be understood against this background. This view has been developed by the phenomenologists~\cite{schutz1973structures}, and similarly developed through research on joint attention, situated practices~\cite{lave1988cognition} and participatory sensemaking~\cite{di2018linguistic}.
There is no clear line demarcating where the mind ends and the world begins. These nuanced approaches recognize that uncertainty, ambiguity, and fluidity, not static dichotomies, exemplify human beings and their interactions. We are fully embedded and enmeshed with our designed surroundings and we critically depend on this embeddedness to sustain ourselves. Furthermore, our historical paths, the moral and political values that we are embedded in, constitute crucial components that contribute to who we are. People, as complex adaptive systems, are \textit{non-totalisable}. The idea of defining the person once and for all, drawing simple classifications, and making accurate predictions thus appears a futile endeavour. In complexity science terms, human beings and their behaviour are complex adaptive phenomena whose precise pathway is simply unpredictable~\citep{juarrero2000dynamics}. 

Automation, on the one hand, is something that is achieved once a given process is complete, that is, it is understood, and discrete such that it can be implemented from a set beginning to a set finish reliably. People and social systems, on the other hand, are partially-open, always becoming, and inherently unfinalizable~\citep{bakhtin1984problems}. Automation as complete understanding, therefore, stands at odds with human behaviour which is inherently 
incomplete, 
making machine classification and prediction futile. Given the open and incomplete nature of human beings and social systems, automating sensible (as opposed to automating nonsense and random) ambiguity and indeterminability is ill-conceived. A machine capable of grasping humanity by definition is capable of grasping open-endedness, incompleteness, fluidity, and ambiguity. Alas, this becomes something other than machines or automation as we know them. 

In this chapter, I place machine categorization and prediction within the broader and historical Western science and philosophy that aspires to pin down, taxonomize, and simplify the complex and interconnected world. Although ML and AI~\footnote{AI can generally be conceived of within two broad categories: narrow AI and general AI. 
} deal explicitly in probabilities and risks rather than in Newtonian determinacies, I contend machine categorization and prediction of social outcomes limit possibilities and create a world partially determined by prediction itself. The social world is messy and fluctuating but also inundated with persistent social norms, power asymmetries, and historical injustice. Historical norms and traditions are often unkind and unjust to individuals and groups at the margins of society, and accordingly, attempts to find stable patterns to sort and categorize the social world pick up these deeply ingrained norms and injustices. Far from being static, social realities are continually co-constructed and the integration of ML systems into day-to-day life increasingly plays a crucial role in influencing the kind of social reality that exists. In Barad's words, ``Reality is sedimented out of the process of making the world intelligible through certain practices and not others. Therefore, we are not only responsible for the knowledge that we seek but, in part, for what exists.''~\citep[p.105]{barad1998getting}. As systems that interact with, and are inextricably linked with the social sphere, ML systems partly create social orders. However, while it is important to recognise ML systems as practices that alter the social world, this chapter also moves beyond recognition and highlights the often neglected reality that 
responsibility and opportunity to create social orders are unequally distributed. Social, economic, and other privileges mean that a small homogeneous group is endowed with the creation of ML systems, and in part for what exists, contributing to the maintenance of the status quo. Meanwhile, the least privileged are forced to live in such realities that the few create, oftentimes subject to machine harm and injustice. How ML/AI research has become heavily dominated by a few powerful actors is further explored in chapter~\ref{chp:values}. 

The rest of the chapter is organized as follows. Section~\ref{sec:cartesian} 
outlines the underlying Cartesian tendencies of current ML systems that strive for stability, order, and predictability. In Section~\ref{sec:machine imposed},  
I argue that machine classification and prediction impose determinablility and limit possibilities. This is followed by Section~\ref{sec:indeterminability} 
where I illustrate the fluid, ambiguous, and non-determinable nature of people and social systems. I review current research which illustrates how prediction is a self-fulfilling prophecy in Section~\ref{sec:making of self}. 
Next, I look at how machine imposed determinablility, opportunity, and harm are distributed disproportionately within society in Section~\ref{sec:imposed} 
and examine how the very practice of sorting and predicting are inherently political in Section~\ref{sec:sorting is political}. 
Section~\ref{sec:creativity} 
takes a brief look at creativity, which stands outside determinability as a potential transformative force to a just world. I close with a summary and reflection in Section~\ref{sec:automating_conclusion}.




\section{Cartesian and Newtonian Inheritances}
\label{sec:cartesian}
\begin{displayquote}
``Traditional science in the Age of the Machine tended to emphasize stability, order, uniformity, and equilibrium. Whereas most of reality, instead of being orderly, stable, and equilibrial, is seething and bubbling with change, disorder, and process.'' Prigogine and Stengers~\citep*{prigogine1984order}
\end{displayquote}

In \textit{Order Out of Chaos} Prigogine and Stengers~\citep*{prigogine1984order} remark that dissecting problems into their smallest components marks one of the most highly valued and developed skills in contemporary Western philosophy and science. A subject of enquiry is broken down into bite-size chunks and each chunk isolated from another and its environment by means of various tricks and thought experiments. Through the tactic of `ceteris paribus', scientists and philosophers presume to work from the assumption that some factors can be held \textit{constant} and that inextricably entangled interrelations can be \textit{isolated}. \textit{Dissection}, \textit{isolation}, and \textit{separation} characterize the epitome of the burning desire in Western philosophy and sciences for \textit{control}, \textit{manipulation}, and \textit{formalization} of the world around us and the deep quest for \textit{certainty}, \textit{stability}, \textit{order}, and \textit{predictability}. 
Descartes and Newton examplify the Western tradition that strives for certainty, order, and predictability. As detailed in Chapter~\ref{chp:background_related}, Section~\ref{sec:post_cart_app}, remnants of this tradition permeate current ML systems. ML systems currently pervading the social sphere embody the core values of the Cartesian and Newtonian worldviews where historical, fluctuating, and interconnected behaviour is presumed to be formalized, clustered, and predicted in a ``value-free'' and ``neutral'' manner. The historic Bayesian framework of prediction is a primary example \citep{doi:10.1098/rstl.1763.0053}. This framework has played a central role in establishing explanations of behaviour based from ``rational principles alone'' \citep{jones2011bayesian, hahn2014bayesian}. Bayes' approach, which is increasingly used in various areas including data science, machine learning, and cognitive science \citep{seth2014cybernetic,jones2011bayesian}, played a pivotal role in establishing the cultural privilege associated with statistical inference and set the ``neutrality'' of mathematical predictions. Price, who published the papers after Bayes' death, noted that Bayes' methods of prediction ``shows us, with distinctness and precision, in every case of any particular order or recurrency of events, what reason there is to think that such recurrency or order is derived from stable causes or regulations in nature, and not from any irregularities of chance''~\citep[p.374]{doi:10.1098/rstl.1763.0053}. However, despite the association of Bayes with rational predictions, Bayesian models are prone to spurious relationships and amplification of socially held stereotypes, a point I expand on in Sections~\ref{sec:imposed} and \ref{sec:sorting is political}.   
Horgan~\citep*{Bayes2016} notes, ``Embedded in Bayes’ theorem is a moral message: If you aren’t scrupulous in seeking alternative explanations for your evidence, the evidence will just confirm what you already believe.'' Even thought Bayesian methods continually update their priors, those priors are still not something naturally and ``objectively'' given, but something we create under given norms, contexts, and cultures~\cite{hinton2017implicit}.  

\section{Machine Imposed Determinability}
\label{sec:machine imposed}

Current AI and ML tools that are increasingly becoming an integral aspect of the social world are direct descendants of the Cartesian and Newtonian worldview insofar as they are tools that impose order and pin down the fluctuating nature of human behaviour through taxonomies, classifications, and predictions. These tools force determinability, limit possibilities, and in the process, create a world that resembles the past. Historical patterns and socially accepted norms are rife with histories of discrimination and injustice and the implication of automating a future that resembles the past for those historically disadvantaged is dire. I discuss this in Sections~\ref{sec:imposed} and \ref{sec:sorting is political}. Below, I examine how ML systems are tools that create a certain type of future through prediction. 

Technological developments and their intimate connection to what it means to be a social, dynamic, embodied living being are not new but go as far back as the history of humankind itself; to prehistoric tools such as stone and spear. However, the current mass scale development and deployment of AI and ML systems pose new and unprecedented challenges and overall impose negative impacts towards marginalized communities which are disproportionately affected. Technological artifacts constitute a crucial part of the socio-technological milieu. They mediate and enrich our living world but also hold invisible and unprecedented power in shaping and altering reality. In Suchman's words~\citep*{suchman2007human}, ``technology is not the design of physical things. It is the design of practices and possibilities.''  

Technological tools constrain or enable actions while making day-to-day life seamless. A GPS application on a device, for example, can make travelling from point A to B considerably easier. In some circumstances, technological tools form crucial components that sustain lives; pacemakers, for example. Technological developments, especially ML systems, are not something that stand \textit{above and over} humans but are integral parts of the active, fluid and dynamic environment of complex, adaptive, self-organizing social systems. Through their power to classify and predict, ML systems direct behaviours and actions towards some things and \textit{away} from others. 

ML systems ``work'' by identifying patterns in vast amounts of data. Given immense, messy, and complex data, a ML system can sort, classify, and cluster similarities based on seemingly shared features. Feed a neural network labelled images of faces and it will learn to discern faces from not-faces. Not only do ML systems detect patterns and cluster similarities, they make predictions based on the observed patterns~\citep{o2013doing,veliz2020privacy}. Machine learning, at its core, is a tool that predicts. It reveals statistical correlations but with no understanding of causal mechanisms. 

Furthermore, machine classification and prediction are practices that act directly upon the world and result in tangible impact~\citep{mcquillan2018data}. Various companies, institutes, and governments use ML systems across a variety of areas. These systems process data that supposedly capture people's behaviours, actions, and the social world, at large. The machine-detected patterns often provide ``answers'' to fuzzy, contingent, and open-ended questions. These ``answers'' neither reveal any causal relations nor provide explanation on \textit{why} or \textit{how}~\citep{pasquale2015black}. Crucially, the more socially complex a problem is, the less capable ML systems are of ``accurately''\footnote{The term accuracy vaguely denotes how closely models or data represent ``the ground truth'' or things as they are in the world. However, critical and genealogical examination of the use of this term remains scarce in the ML literature. In the absence of such critical examination, ``accurate classification or prediction'', especially in the context of social affairs, risks correlating representations with stereotypically held views.} or reliably classifying or predicting. Narayanan~\citep*{narayanan2019} broadly maps the application of AI systems into three crude categories: Perception (e.g.\ face recognition), Automating Judgment (e.g.\ detecting spam), and Predicting Social Outcomes (e.g.\ predictive policing). There has been rapid progress in the first category and ``the fundamental reason for progress is that there is no uncertainty or ambiguity in these tasks''~\cite{narayanan2019}. Automating judgement is somewhat contested as what can be taken as correct decision is subject to disagreement. The task of predicting social outcomes, however, remains fundamentally dubious involving ``a lot of snake oil'' and is marked with numerous drawbacks and harms. The use of AI for predicting social outcomes, according to Narayanan~\cite{narayanan2019}, is not substantially better than manual scoring. Similarly, in a recent work, Salganik et al.~\citep*{salganik2020measuring} examined the predictability of social trajectories of children from vulnerable families. A team of 160 ML researchers built predictive models using a rich dataset. The authors found that not one model made an accurate prediction and the best predictions were only slightly better than those from a simple benchmark model. Thus, Salganik et al.~\citep*{salganik2020measuring} caution those considering using predictive models to forecast social outcomes. Nonetheless, predictive systems continue to pervade decision-making of social outcomes with disastrous consequences.

\section{Indeterminability of the Person}
\label{sec:indeterminability}

Traditional cognitive science, so far as its desires for a universal, objective, and predictable science of the mind goes, is the heir to the Cartesian and Newtonian worldviews. The continually fluctuating and interconnected state of human affairs finds a stable point through the conjecture which positions the individual person as the seat of knowledge. As such, the individual person is often isolated and taken as the unit of analysis. Great emphasis is placed on her individual mental capabilities as the mind is assumed to be the property of the single individual~\citep{linell2009rethinking, markova2016dialogical}. The nature of experimental design in scientific psychology, for example, illustrates the subtle remnant of the desire for cleansing thinking of cultural influences and political dimensions. Research in memory testing, for instance, to a large extent proceeds from the assumption that memory is a 'purely cognitive process' that resides in the brain~\citep{harris2011we}. The individual is removed from her lifeworld and tasked with recalling a series of images or words (often meaningless to the person) using flashcards or a screen in the artificial confines of a laboratory. Subsumed by objective and universalizable formulations, cognitivist approaches paint a picture of the person that equates \textit{persons} with \textit{brains}. Emphasis on dynamic relations, contextual and historical embeddings, and messy interactions, on the other hand, is perceived as a threat that blurs and contaminates neat classifications and universalizable conceptions.

Individualistic and reductionist approaches are irrevocably ingrained in Western thought. It is a continual struggle, even for the most aware researchers and practitioners, to steer clear of them. Taking the individual self as the unquestioned origin of knowledge of the world and of others is a legacy of this tradition~\citep{linell2009rethinking}. Traditional social cognition research, supposedly an endeavour that turns attention to the social, falls short of recognizing the dynamic and entangled nature of bodies and environments, and how each influences the other. The individual person, in social cognition, is portrayed as the meaning generator and is the primary interest of study~\citep{markova2016dialogical}. Pushing back against these individualistic traditions, the broadly conceived approaches of embodied and enactive cognitive science offer views of persons, brains, and nature of reality in general that are active, dynamic, and inextricably connected with environments and others.

The embodied and enactive turn~\citep{chemero2011radical, varela2016embodied, kyselo2014body, mcgann2009self}, at its core, places living bodies, with their peculiarities, fluidity, and messiness, at centre stage. Fluidity, multivalence, and precariousness are not perceived as obstacles that stand in the way of ``final'' and ``universal'' answers, but are acknowledged and celebrated as necessary conditions for existence. Living bodies are not stationary entities that can be captured in neat taxonomies, rather they are active, dynamic, historical, social, cultural, gendered, politicized, and contextualized organisms. People are not solo cognizers that manipulate symbols in their heads and perceive their environment in a passive way, but they actively engage with the world around them in a meaningful and \textit{unpredictable} way. Living bodies, according to Di Paolo, Cuffari, and De Jaegher~\citep*{di2018linguistic}, are processes, practices, and networks of relations which have ``more in common with hurricanes than with statues''. They are \textit{unfinished} and always \textit{becoming}, marked by ``innumerable relational possibilities, potentialities and virtualities'' and not calculable entities whose behaviour can neatly be automated and predicted in a precise way. Bodies ``grow, develop, and die in ongoing attunement to their circumstances [...] Human bodies are path-dependent, plastic, nonergodic, in short, historical. There is no true averaging of them.''~\citep[p.97]{di2018linguistic}. 


Universalizable theories of bodies, taxonomies, and statistical predictions of future behaviours all rely on similarities and abstraction of features that are common among particulars. Unique, contingent, and idiosyncratic features and behaviours pose challenges when it comes to deriving elegant taxonomies. However, idiosyncrasies and peculiarities make someone the particular, novel, and creative person they are. Living bodies each face unique challenges defined by the particular trajectory of history of enactments, history of adaptations, and social circumstantial interactions as they continually navigate the social world. Social interactions themselves, Jaegher and Di Paolo~\citep*{de2007participatory} contend, are active and dynamic engagements that ``take a life of their own'' in an \textit{unpredictable} way. They shift in moods, aims, and levels of intimacy, without the participants intentionally seeking these changes. Most fundamentally, ``Our most sophisticated knowing is full of \textit{uncertainties, inconsistencies, ambiguities}, and \textit{contradictions}. These characterize how we most often deal with the world, ourselves, and each other''~\citep*{de2019loving}. Furthermore, Buccella~\citep*{buccella2020enactivism} singles out \textit{indeterminacy} as a key factor that is important in understanding human perceptual experiences. Perception is necessarily open-ended and the environment presents unlimited possibilities and offers many ways of life~\citep{nonaka2020locating,merleau1945phenomenology}. 




On a similar note, examining the problem of meaning in artificial agents, Froese and Di Paolo~\citep*{froese2019problem} single out \textit{indeterminability} as the key characteristic of humans that differentiates them from artificial agents. Emphasizing the futility of reductionist approaches to complex adaptive systems, Cilliers (posthumously noted by Preiser~\citep*[p.64]{preiser2016critical}) further points out that ``From the argument for the conservation of complexity -- the claim that complexity cannot be compressed -- it follows that a proper model of a complex system would have to be as complex as the system itself.'' There is no single perfect or accurate representation or model for a given complex system. Accurate or perfect representation entails capturing a system (its dynamic and non-linear interactions and emerging properties) in its entirety without leaving something out. However, compressing a system into an algorithm or a model without reducing the complexity is impossible. This means that a perfect model of a system would have to be as complex and unpredictable as the system itself. In modelling a complex system, then, what aspects we want to capture and represent are partly tied to the observer/modeller and their perspective and description of the system as well as their overall objectives. For example,``a portrait of a person, a store
mannequin, and a pig can all be models of a human being.'' None represent a human perfectly but each serves as the ``best'' model depending on the objectives and purpose of using a model -- ``to remember old friends, to buy clothes, or to study biology.''~\cite{blanchard2011differential}

Precise\footnote{The very term \textit{precision} (in prediction) assumes an observer-free ``ground truth'', a correct description of reality, and a correct trajectory from which things can be compared against. This line of thinking follows Cartesian logic. Precision, accordingly, marks proximity to the presumed ``ground truth'' or the ``correct description of reality'' while deviation from it might signal lack of precision. Contrary to these presumptions, descriptions of reality or ``ground truth'' are never given in an observer-free manner (see Section~\ref{sec:indeterminability}).} predictions of behaviours and actions, therefore, are impossible and, when enforced, dire ethical consequences emerge. This might raise the question of whether people and social systems are unpredictable in principle or unpredictability is only a practical limitation and a matter of even more data and compute power. Following Cilliers' argument for the \textit{conservation of complexity}~\citep{preiser2016critical}, I contend that people and social systems are unpredictable in principle. Having said that, this is not an argument to oppose all attempts at predictions. Nonetheless, given that complex systems are open systems that are far from equilibrium with no clear demarcations of what is inside and outside of the system, we (the modeller) draw boundaries in order to say something meaningful about the system and to model it. There's nothing intrinsic or natural about the boundaries we \textit{create}. Since there is no way of accurately determining the boundary of a systems, we can never be certain that we have taken enough into consideration or that the factors we deemed irrelevant are indeed so~\cite{cilliers2002complexity}. This means that, any predictive models we make are inherently value laden and never purely objective. Furthermore, so long as classification and predictive systems operate within what Ahmed~\citep*{ahmed2007phenomenology} called \textit{white straight ontology}, ``precise'' and ``accurate'' prediction risk measuring how closely behaviours or actions adhere to socially and historically held stereotypes. For a detailed analysis of how ML, more particularly, vision systems perpetuate socially held stereotypes, see Chapter~\ref{chapter:computer vision}.


Indeterminability and unpredictability do not, by any means, mean that people and social systems wander aimlessly without pattern, habit, or relatively stable behaviour. For any given society there exist socially and culturally accepted norms and historically established conventions. People self-organize with these dynamic and contextual constraints that serve as a ``reduction of possibilities''~\citep{juarrero2000dynamics}. However, these relative stabilities and habitual patterns do not mean a person is totalisable and can be rendered fully knowable and predictable with precision. Any prediction of future behaviour based on past patterns is at best a statistical probability. We may, therefore, be able to predict a person's general dynamics, under certain conditions, within a certain context and time. But precise prediction of a person's specific behaviour and action, due to their nonlinear interactions and endless possibilities, are impossible. Moreover, as discussed in Section~\ref{sec:sorting is political}, relatively stable patterns and established conventions and norms are charged with social, political, and power asymmetries that benefit or disadvantage groups and individuals depending on their position in society. When ML systems 'pick up' stable patterns, they also identify harmful current and historical norms, prejudices, and injustices. Taking such historical and current patterns as ``ground truth'' from which to model the future brings forth a machine-determined world that resembles the past and raises a host of ethical and justice issues.  


\section{Prediction, a Self-fulfilling Prophecy}
\label{sec:making of self}

In the age of ubiquitous and interconnected systems, it has become outdated to conceive of the digital and the ``physical'' as separate realities. Outcomes from algorithmic models are used to justify action in the social world and actions in the social world are datafied and fed into models. Who we are is not signified by our bodies alone but also by algorithmic identifications, often assigned to us without our knowledge or consent. Marketing and web analytics companies who gather (and/or purchase) huge amounts of data construct our algorithmic identities from quantifiable attributes that emerge from various input data, observed patterns, and algorithmic inferences and extrapolations. Traces of data and metadata, including behavioural data such as statistics from visited websites, online purchase history, device location, data from cameras, sensors and IoT devices that proliferate in public and private spaces, all contribute to the construction of the 'person' in the digital realm. Algorithmic identifications that are assigned to an individual or a group carry tangible implications as such identifications increasingly play a central role in determining the outcomes of various aspects of an individual's endeavours. 

As the hiring process becomes more and more automated, for example, algorithmic hiring tools become most consequential in determining key aspects of a person's life such as how much one earns and where one works and lives. Just like automating any social processes, automating ``best'' candidates is prone to automating and reproducing historically and socially held inequities and stereotypes all while providing a veneer of objectivity~\citep{raghavan2020mitigating}. Depending on how ``appropriate'' or ``fit'' candidates are deemed, job opportunities are automatically surfaced to some and withheld from others. More accurately, automated hiring systems serve as tools 
to reject applicants who do not fit in certain boxes~\citep{ajunwa2019platforms}. And given that ``best'', ``appropriate'', or ``fit'' applicants are often defined and measured by past success, candidates that do not fit within that box risk exclusion. This was case in point with the hiring algorithm that Amazon deployed and disbanded in 2018 upon discovering that the tool had been discriminatory against women~\citep{reuters}. Amazon's hiring tool was trained to identify the best candidates based on observed patterns in resumes submitted to the company over a 10-year period. Given the male dominance in the tech industry, most resumes came from men. Automating such patterns, by definition, is then a process to automating historical inequities. Predictions based on past hiring decisions reproduce patterns of inequity even when tools explicitly ignore race, gender, age, and other protected attributes~\citep{bogen2018help}.

Algorithmic classifications, sortings, and predictions, when enacted in the world, create a social order. For any individual person, group, or situation, algorithmic processes give advantage or they inflict suffering. Jobs are made and lost~\citep{ajunwa2016hiring,sanchez2020does}.~Who is visible and legible is legitimized through algorithmic predictions as some subjectivities (and not others) are recognised as a pedestrian~\citep{wilson2019predictive}, or hire-able~\citep{speicher2018potential,ajunwa2016hiring}, or in need of medical care~\citep{obermeyer2019dissecting}, or likely to engage in criminal acts~\citep{angwin2016machine, lum2016predict}. 

Furthermore, the very practice of forecasting the future partly acts directly upon the world -- machine prediction plays a part in creating what exists whenever such predictions inform decision making. Perdomo et al.~\citep*{perdomo2020performative} illustrate that prediction often influences the outcome that it is trying to predict. They refer to such practice as ``performative prediction'': ``Traffic predictions influence traffic patterns, crime location prediction influences police allocations that may deter crime, recommendations shape preferences and thus consumption, stock price prediction determines trading activity and hence prices.'' Similarly, Benjamin~\citep*{benjamin2019race} argues that ``crime prediction algorithms should more accurately be called crime production algorithms'' due to the reason that predictive policing software predominantly target historically underserved communities, wherein hyper-surveillance partly produces crime, creating a self-fulfilling prophecy. Milano, Taddeo, and Floridi~\citep*{milano2020recommender} furthermore looked at ubiquitous recommender systems and found that recommender systems shape user preferences and guide choices both at the individual and social level. The authors contend that attempts to predict preferences have socially transformative effects and impinge on personal autonomy. Notably, machine classification and prediction is a multi-billion dollar business, meaning that business objectives play an active role in the direction with which social outcomes and behaviours are moulded. In fact, there exists an increasing tie between big tech and ML research (see Chapter~\ref{chp:values} for an in depth analysis of the most influential ML papers and their close ties to big tech corporations). Through ``nudging'', persuasion, and limiting the range of options available to individuals, they cajole people in particular directions, often in a way that maximizes profit. Recommender systems often have commercial objectives and are developed for business applications. Consequently, by predicting preferences, recommender systems not only shape individual experience and social interactions, they also hold transformative impact on society in a manner that aligns with commercial values~\citep{milano2020recommender}. Given that most influential ML work increasingly comes from big tech corporations (see Chapter~\ref{chp:values}), this is alarming.


\section{Imposed Determinability in Unequal Measures}
\label{sec:imposed}
\begin{displayquote}
``Reality is sedimented out of the process of making the world intelligible through certain practices and not others. Therefore, we are not only responsible for the knowledge that we seek but, in part, for what exists.'' Karen Barad \citep*{barad1998getting}
\end{displayquote}

On the one hand, machine-human relation and interaction is a process that constitutes a systems-level organization. On the other, it is vital to acknowledge power asymmetries within this systems-level organization. Influences are not bi-directional and \textit{benefits}, \textit{disadvantages}, and \textit{negative impacts} are distributed unequally among individuals and communities. In the case of recommender systems, for example, big corporations behind the development and integration of these systems exert power in moulding future realities. The individual person or the ``end user'', has little to no direct influence.


In fact, when ML is used to rank, sort, score and predict \textit{social outcomes}, Narayanan calls ``AI snake oil''~\cite{narayanan2019}, those being ranked and scored are rarely aware of it or merely know why they are given certain scores. This makes it difficult to contest and negotiate algorithmically assigned `identities' and scores. Furthermore, the very practice of scoring, characterizing, and assigning `algorithmic identities' without people's awareness risks treating people as mere objects. As Maturana~\citep*[p.108]{maturana2004being} remarks ``If you deprive people of the opportunity [to contest and protest against their characterization], you treat them like freely disposable objects; they have the status of slaves, compelled to function without the opportunity of complaining when they do not like what is happening to them.'' Current data extraction, classification, and prediction (without the awareness of the `data subject') practices across analytics firms and big technology corporation where people are often treated as `data objects', bear close resemblance to Maturana's remark. 

Predictive models, due to their use of historical data, are inherently conservative. They reproduce and reinforce norms, practices, and traditions of the past. Historical norms and traditions are often unkind and unjust to individuals and groups at the margins of society. Decisions made in the past align with the maintenance of the status quo. 
The practice of constructing predictive models based on the past and directly deploying them for decision-making amounts to constructing a programmed vision of the future based on an unjust and socially conservative past. Through the application of predictive systems in the social sphere, historically and socially unjust norms, stereotypes, and practices are reinforced. Machine classification and prediction, thus, negatively impact individuals and groups at the margins of society the most.

\section{Sorting and Predicting are Moral and Political}
\label{sec:sorting is political}

Central tenets of the linear, stable, and predictable Cartesian-Newtonian worldview are the idea of objectivity -- the assumption that observation, description, classification and prediction of the world can be done from a ``view from nowhere''~\citep{nagel1989view} -- and the assumption that the world is totalizable, i.e. it can be fully understood, captured in data, and clearly mapped from beginning to end. Heinz von Foerster famously decried ``Objectivity is a subject’s delusion that observing can be done without him [sic]. Invoking objectivity is abrogating responsibility -- hence its popularity''~\citep{negoita1992cybernetics}. The practice of categorizing, ordering, and forecasting a future necessarily entails simplification of complex social reality and the drawing of boundaries around what is important and relevant, which are decisions with ethical and moral dimensions.   
Through the very act of clustering similarities, boundaries are drawn as to which behaviours and actions are ``correct'', ``good'' or ``acceptable'' and which are not. Furthermore, through their performative powers, predictive systems cast certain ways of being as ``normal'' while others are deemed ``deviant'' and in need of correction. Practices that were previously understood to be moral and political, and historically required a great deal of dialogue and negotiation, are obfuscated as apolitical endeavours with the advent of machine classification and prediction. Moreover, the \textit{veneer of objectivity} that ML is entrusted with presents an added challenge to seeing machine classification and prediction as anything but a technical and mathematical task. 

Categorization by human beings itself arises within context and goal-directed activities. Human categories, rather than carving nature at its joint, are developed on the fly to address goal-directed actions. Categories, therefore, are dynamic, unstable, contextualized, and inherently embedded in on-going activities~\citep{barsalou1991deriving,barsalou2009simulation}. Machine categorization, therefore, can only be created within the context of a broader goal rather than being an austere, abstract, and mechanical process. Creating categories and drawing boundaries is not primarily a technical choice or a purely scientific question but necessarily an ethical and moral one, especially when such practice has a direct and tangible impact on vulnerable lives. Acknowledging this is a crucial step in taking responsibility for \textit{what exists}. Having said that, responsibility needs to be selectively attributed. Benjamin, in \textit{Race After Technology}~\citep*{benjamin2019race}, notes that ``it might be tempting to point to the smart technologies we carry around our pockets and exclaim that ``we are all caught inside the digital dragnet!'' But the fact is, we do not all experience the danger of exposure in equal measure.'' Although as Barad~\citep*{barad2007meeting,barad1998getting} assessed, reality is something we create together through active practices, a few have genuine say towards what kind of world needs to be co-created while many others are forced to live in it. Existing social, political, and financial power dynamics mean those at the bottom of societal hierarchies have little say, if any at all, in the co-creation of realities.

It is impossible to operate in a value-free space. 
The type of concerns, priorities, questions, and design choices that are considered all reflect the values, motivations, commitments, and interests of those at the helm of creating ML systems. When values are not explicitly laid out, the values that are taken as ``universal'' or ``neutral'' are those values that represent the status quo and the values that are implicit within a given field~\citep{collins2002black,ahmed2007phenomenology}. 
Within both the academic fields and corporate industry currently developing ML systems {\it en masse}, the values that are taken as ``universal'' are predominantly the values and interests of the Western, white, male. Computer Science (as well as its sub-field, Machine Learning), since its conception as an academic field in the 1950s in the US has always been a field that strove for impactful application within the military, education, and the general social sphere. The field has since come to exert unprecedented social, political, and economic power as explicited in Chapter~\ref{chp:values}, Section~\ref{sec:corporate} in detail.

Within major technology corporations -- from Microsoft to Amazon and Google -- Western white men with homogeneous backgrounds and conservative leanings~\citep{cohen2018know} remain the most predominant powerful figures that influence and redefine social realities~\citep{broussard2018artificial}. What we find then is a huge power disparity between powerful corporations (and the individuals behind them) and ``end users'' whose agency, opportunity, and options have been limited in the process of algorithmic classification and prediction. Such process amounts to financial and personal gain for the former at the expense of the latter. In fact, as Zuboff~\citep*{zuboff2019age} argues, the technology industry is built on capitalization and monetization of lived experiences and on building tools of surveillance. 

Given this massive power disparity, those engaged in the practices of designing, developing, and deploying ML systems -- effectively shoehorning individual people and their behaviours into pre-defined stereotypical categories -- carry a great proportion of the responsibility in creating \textit{what exists}. Such demography decides what questions are worthy of investigation, what problems need to be ``solved'', and what is sufficient performance for a model to be deployed into the world. Consequently, this group bears much greater responsibility and accountability. 

Scientific enquiries carry inherent ethical and moral dimensions. The more a topic of enquiry veers towards human and social affairs, the more apparent its moral and ethical dimensions. The apparent dissociation of ``science'' from ``ethics'' has historically allowed science to evade accountability and responsibility and similarly, so will algorithmic systems if they are allowed to. Ubiquitous integration of ML models to high-stake situations creates a political and economic world that benefits the most privileged and harms the vulnerable. It also creates a social world where the status quo is maintained and historical injustice perpetuated. 
For most scholars working at the intersections of algorithmic injustice and STS, it has become trivial and common knowledge that ML models ``work well'' often equates to models \textit{picking up historical patterns}. The social world as it is, is filled with beauty, ugliness, and cruelty. And as Benjamin~\citep*{benjamin2019race} notes, to think that one can feed a model with all the world's beauty, ugliness, and cruelty and expect only beauty is a fantasy. 




\section{On Creativity}
\label{sec:creativity}

Human creativity is marked by imagination and thinking of things that were not thought of before. Creative innovations that have come to define and revolutionize the world, from music to medicine, are often marked by surprise, spontaneity, and uncertainty. Creativity, Juarrero~\citep*{juarrero2000dynamics} reminds us, stands in stark opposition to certainty and predictability. It requires unexpected and spontaneous behaviour and not repeating past patterns and trajectories. Creativity, by definition, defies expectation.\footnote{Having said that, the view of creativity as the creation of something novel is not to remove the creative process from its historical, social, and contextual embeddings.}

As ML systems attempt to order the spontaneous and non-determinable social world, they create a future that resembles the past, leaving us no room for a chance to be different. Such classifications and predictions reinforce stereotypes and impinge on the inherently open-endedness of being, limiting a person's potential by defining them by what ``people like them'' have ``done'' or ``liked'' or how ``people like them'' behaved in the past. When future behaviours are predicted based on past stereotypes, individuals are deprived of the opportunity to challenge stereotypes and to realize their full potential.

In the age of ML where ``accurate'' categorization and ``precise'' prediction are highly valued, unexpected and spontaneous behaviour poses a challenge and is seen as a deviation to be corrected -- not an inherent, indeterminable part of human beings that should be celebrated. In fact, the idea of the ``average'' can be traced to the explicit logic in the origins of statistics in sociology, crime, and public health by Quetelet in the 1800s. The ``average'' was considered an ideal. ``Quetelet applied the same thinking to his interpretation of human averages: he declared that the individual person was synonymous with error, while the average person represented the true human being.''~\citep[p.26]{rose2016end}. To this day, uniquely and idiosyncratically expressed unrepeatable behaviours that defy systemic rules or clusters of patterns -- fundamental to creativity -- are seen as undesired anomalies and ``edge cases.'' ML processes codify the past. They do not invent the future. Doing that, O’Neil~\citep*{o2016weapons} emphasizes, ``requires moral imagination, and that’s something only humans can provide.''

Creativity, which stands outside machine determinability, holds the potential to transform the way we approach and use algorithmic systems. From organized resistance, to strict regulations, to disrupting the current capitalist ecosystem, to imagining a fundamentally new type of technology that celebrates differences instead of forcing uniformity all require creativity. As opposed to accepting a machine determined world as inevitable, creativity is fundamental to imagining an alternative world and the disruptive technologies to actualize such world~\footnote{This is not to succumb to techno-solutionism where technology is sought as the only viable source of answer. Far from it, and in some cases, the option of \textit{no technology at all} can be the most efficient way to a just world.}. Fundamental recognition of uncertainty, ambiguity, and non-determinability as an inherent condition of complex systems is essential for modest and creative understanding of humans and society and holds a promising first step towards a just world.

\section{Conclusion}
\label{sec:automating_conclusion}

\begin{displayquote}
“It is essential for the thing and for the world to be presented as ``open,'' to send us beyond their determinate manifestations, and to promise us always ``something more to see.'''' Maurice Merleau-Ponty~\citep*{merleau1945phenomenology}
\end{displayquote}

In summary, individual people and social systems, as complex adaptive systems, are active, dynamic, and necessarily a historical phenomenon whose precise pathway is \textit{unpredictable}. Contrary to this, we find much of current applied ML classifying, sorting, and predicting these fluctuating and contingent agents and systems, in effect, creating a certain trajectory that resembles the past. Such practice, in the process, brings forth a predetermined social reality that places people in stereotyping boxes and maintains the status quo, harming and disadvantaging individuals and communities that are historically and socially disadvantaged. 

ML systems, tools that fundamentally classify, order, and predict, I argue, are practices that reincarnate Cartesian and Newtonian worldviews that seek stable, predictable, and complete understanding. But, people (and the social systems that they are embedded in) are partially open, indeterminable, and fluctuating, meaning a complete understanding would imply death of the person or that the social system has come to a stall. Automation, which requires complete understanding, thus stands at odds with human behaviour, which is inherently incomplete and unfinalizable. This makes machine classification and prediction of individual behaviours and actions impossible (yet, a common practice that drives great financial profit).  

Arguably, systems of classification are inherent to humans and part of all cultures, although modern Western culture has produced more than most, without realizing it~\citep{bowker2000sorting}. Therefore, I do not argue against machine classification altogether. However, given that people and social systems are dynamic and unpredictable and that social structures are hierarchical and saturated with power asymmetries, forcing order and taxonomies brings forth harm and injustice to those at the margins. Furthermore, as Narayanan~\cite{narayanan2019} remarks, predicting social outcomes is a fundamentally dubious endeavour with many disadvantages and problems but with few, if any, benefits. In this regard, those behind ML systems (from conception, to design, to development, and deployment) -- individuals and corporations alike -- bear the responsibility for the unjust and harmful social reality they are creating.

    \part{}
    \label{part:ii}
    \chapter{Computer Vision: Winning Battles, Losing Wars}
\label{chapter:computer vision}

In some sense, AI research is currently achieving outstanding results, albeit in a piecemeal manner where the focus is on creating systems that perform specific tasks without worrying so much how closely such systems resemble human intelligence. The last decade has seen impressive advancements in tasks such as machine translation, game playing, speech recognition, and image recognition~\cite{mitchell2021ai}. Over the last ten years, deep learning techniques have brought incredible success to the field of computer vision, increasing the capability of machines to recognize thousands of everyday objects, sometimes ``better than humans''~\cite{dengel2021next}. Mind, such claims as \textit{better than humans} or \textit{human level performance} are almost always misleading and erroneous~\cite{mitchell2019artificial,broussard2018artificial}.  

Computer vision currently seems to be marked by continual improvements of performance, accuracy, and efficiency including in sub-areas such as object detection and segmentation, pose estimation, and face recognition. Vision research increasingly finds application in a wide range of domains from medical to industrial robotics, self-driving cars, and eye, head, and emotion tracking~\cite{xu2020computer}. This chapter looks at computer vision research in more detail. Having said that, it is important to emphasize that this is not a chapter that narrates the glorious ``progress'' and achievements of the field. As a field currently at the height of over-hype and ``AI spring'', plenty other research does that. Instead, this chapter looks under the hood at the large scale vision datasets (LSVDs) behind vision research and the alarming applications of vision research. 

\section{Large Scale Vision Datasets}
Born from World War II and the haunting and despicable practices of Nazi era experimentation~\cite{Nazianat71:online}, the \textit{1947 Nuremberg code}~\cite{weindling2001originsinformedconsent} and the subsequent \textit{1964 Helsinki declaration}~\cite{Timetodi89:onlineNature} helped establish the doctrine of \textbf{Informed Consent} which builds on the fundamental notions of human dignity and agency to control dissemination of information about oneself. This has shepherded data collection endeavors in the medical and psychological sciences concerning human subjects, including photographic data~\cite{naidoo2009informed, blain2002informed}, for the past several decades. A less stringent version of informed consent, \textit{broad consent}, proposed in 45 CFR 46.116(d) of the \textit{Revised Common Rule}~\cite{eCFRCod99onlinecommonrule}, has been recently introduced that still affords basic safeguards towards protecting one's identity in large scale databases. Yet, in the age of \textit{Big Data}, these safeguards of informed consent, privacy, or agency of the individual have gradually been eroded. 

Vision research depends on millions of images of people, often sourced without consent or awareness of image owners. As can be seen in Table~\ref{tab:lsvd}, several tens of millions of images of people are found in peer-reviewed literature. Yet, none of these images are obtained with consent, awareness of the individuals, or IRB approval for collection. In \textit{Section 5-B} of Torralba et al.~\cite{torralba200880tiny}, for instance, the authors state: ``\textit{As many images on the web contain pictures of people, a large fraction (23\%) of the 79 million images in our dataset have people in them}''. With this background, let us now focus on LSVDs -- the crucial ingredient behind the success of computer vision -- and more particularly on the \textit{ImageNet} dataset, one of the most celebrated and canonical LSVDs, as well as the now-retracted \textit{80 million Tiny Images} dataset. 

\begin{table}[htbp]
\centering
\caption{Large scale image datasets containing people’s images.}
\label{tab:lsvd}
\scriptsize
\begin{tabular}{lrrr}
\toprule
\textbf{Dataset} & \makecell{\textbf{Number of images}\\ \textbf{(in millions)}} & \makecell{\textbf{Number of categories}\\ \textbf{(in thousands)}} & \makecell{\textbf{Number of}\\ \textbf{consensual images}}  \\ 
\midrule
\textbf{JFT-300M}(\cite{hinton2015distillingjft}) & 300+           & 18          & 0 \\ 
\textbf{Open Images}(\cite{kuznetsova2018open}) & 9           & 20           & 0 \\ 
\textbf{Tiny-Images}\protect\footnotemark (\cite{torralba200880tiny}) & 79           & 76          & 0 \\ 
\textbf{Tencent-ML} (\cite{wu2019tencent}) & 18            & 11           & 0 \\ 
\textbf{ImageNet-(21k,11k\footnote{See \url{https://modelzoo.co/model/resnet-mxnet}},1k)}(\cite{russakovsky2015imagenet}) & (14,12, 1)           & (22,11,1)          & 0 \\
\textbf{Places}(\cite{zhou2017places})   & 11            &  0.4         & 0\\
\bottomrule
\end{tabular}
\end{table}
\footnotetext{The Tiny-Images dataset has since been withdrawn following the publication of our work (Birhane and Prabhu~\cite{birhane2021large}) (https://groups.csail.mit.edu/vision/TinyImages/).} 


The absence of critical engagement with canonical datasets disproportionately negatively impacts women, racial and ethnic minorities, and vulnerable individuals and communities at the margins of society~\cite{birhane2019algorithmic}. For example, image search results both exaggerate stereotypes and systematically under-represent women in search results for occupations~\cite{kay2015unequal}; object detection systems designed to detect pedestrians display higher error rates for individuals with dark skin tones~\cite{wilson2019predictive}; gender classification systems show disparities in accuracy where lighter-skin males are classified with the highest accuracy while darker-skin females suffer the most misclassification~\cite{buolamwini2018gender}, and images of Black people are misclassified into non-human classes such as ‘chimpanzee’, ‘gorilla’, and ‘orangutan’~\cite{agarwal2021evaluating}. Gender classification systems that lean on binary and cis-genderist constructs operationalize gender in a trans-exclusive way resulting in tangible harm to trans people~\citep{keyes2018misgendering}. As outlined in Chapter~\ref{chp:relational}, the persistent trend remains that minoritized and vulnerable individuals and communities often disproportionately suffer the negative impacts of ML systems. This calls for a shift in rethinking ethics not just as a fairness metric to mitigate the narrow concept of bias, but as practice that results in justice for the most negatively impacted. 

Such understanding of \textit{ethics as justice} then requires a focus beyond \textit{bias} and \textit{fairnesss} in LSVDs and calls for bigger questions such as how images are sourced and labelled, and what it means for vision models to be trained on them. In this regard, through interactive online exhibitions, Crawford and Paglen have unveiled the dark side of LSVDs and the troubling consequences of classifying people as if they are objects~\cite{Excavati19:online}. 
Not only are images that were scraped from across the web appropriated as data for computer vision tasks, but also the very act of assigning labels to people based on physical features raises fundamental concerns around reviving long-discredited pseudo-scientific ideologies of physiognomy~\cite{y2017physiognomy}, a topic covered in Chapter~\ref{chp:all_models}. 

Within the dataset creation process, \textit{taxonomy sources} pass on their limitations and unquestioned assumptions. The adoption of underlying structures presents a challenge where --- without critical examination of the architecture --- ethically dubious taxonomies are inherited. This has been one of the main challenges for ImageNet given that the dataset is built on the backbone of WordNet's structure. Acknowledging some of the problems, the authors from the ImageNet team have attempted to address the stagnant concept vocabulary of WordNet~\cite{yang2020towardsfacct}. They admitted that only 158 out of the 2,832 existing synsets should remain in the person sub-tree.\footnote{In order to prune all the nodes. They also took into account the \textit{imageability} of the synsets and the skewed representation in the images pertaining to the \textit{Image retrieval} phase.} Nonetheless, some serious problems remain untouched, a topic covered in Section \ref{threats}. Before doing that, I address in greater depth the overbearing presence of the \textit{WordNet effect} on image datasets following a brief overview of \textit{ImageNet} and \textit{TinyImages}.

\section{The rise of ImageNet}

The emergence of the ImageNet dataset~\cite{deng2009imagenet} is widely considered a ``pivotal moment''~\cite{Quartz} in the \textit{Deep Learning revolution} that transformed computer vision, and AI in general. ImageNet was born out of the contention that lack of large scale real world dataset is a central problem to AI. Using WordNet as a backbone, the ambition was ``to map out the entire world of objects''~\cite{gershgorn2017data}, publishing the first results in 2009. Prior to ImageNet, computer vision and image processing researchers trained image classification models on small dataset such as CalTech101 (9k images), PASCAL-VOC (30k images), LabelMe (37k images), and the SUN dataset (131k images). 
ImageNet, with over 14 million images spread across 21,841 synsets, replete with 1,034,908 bounding box annotations, brought in an aspect of scale that was previously missing. A subset of 1.2 million images across 1000 classes was carved out from this dataset to form the ImageNet-1k dataset (popularly called \texttt{ILSVRC-2012}) which formed the basis for the \textit{Task-1: classification} challenge in the ImageNet Large Scale Visual Recognition Challenge (ILSVRC). This soon became widely touted as the \textit{Computer Vision Olympics}~\cite{Missouri}. The vastness of this dataset allowed a Convolutional Neural Network (CNN) with 60 million parameters~\cite{krizhevsky2012imagenet} trained by the \textit{SuperVision} team from University of Toronto to usher in the rebirth of the CNN-era (see~\cite{alom2018historyalexnet}), which is now widely dubbed the \textit{AlexNet moment} in AI.

Although ImageNet was created over a decade ago, it remains one of the most influential and powerful image databases available today. Its power and magnitude is matched by its unprecedented societal impact. Although an \textit{a posteriori} audit might seem redundant a decade after its creation, ImageNet's continued significance and the culture it has fostered for other LSVDs warrant an ongoing critical dialogue. From the questionable ways images were sourced, to troublesome labeling of people in images, to the downstream effects of training AI models using such images, ImageNet and LSVDs in general constitute a \textit{pyrrhic victory} for computer vision. This win has come at the expense of harm to \textit{minoritized groups}, aided the gradual erosion of privacy, consent, and agency of both the individual and the collective, and subsequently contributed to the resuscitation of long discredited pseudosciences such as physiognomy (see Chapter~\ref{chp:all_models}). 

\section{Tiny Images and the WordNet Effect}

\begin{figure}[ht!]

         \centering
         \includegraphics[width=\textwidth]{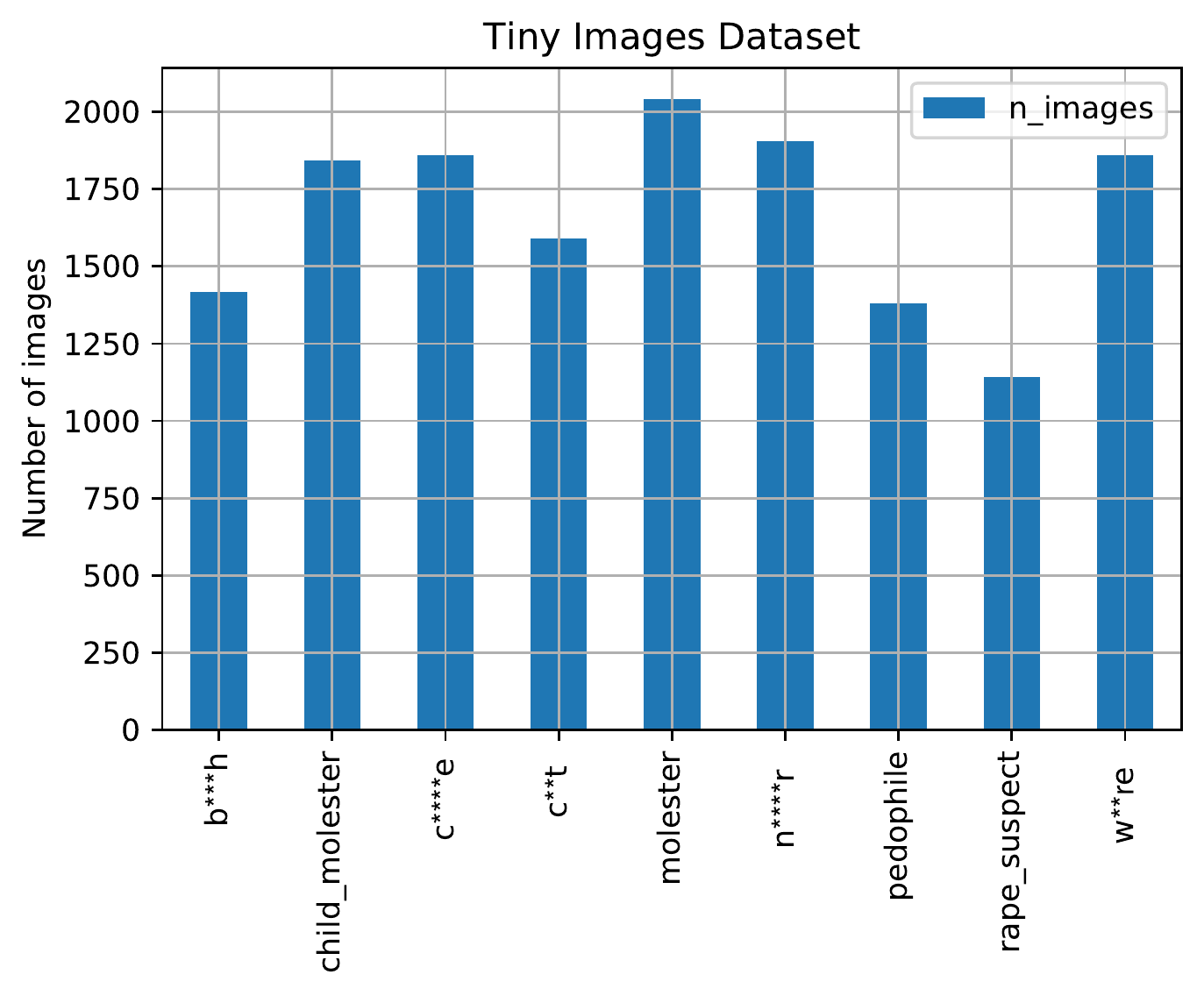}
         \caption{Class-wise counts of the offensive classes}
         \label{fig:tiny_images_counts}
   \end{figure}

\begin{figure}[ht!]
    \begin{subfigure}[b]{0.45\textwidth}
         \centering
         \includegraphics[width=\textwidth, trim={0 0 0 1cm}, clip]{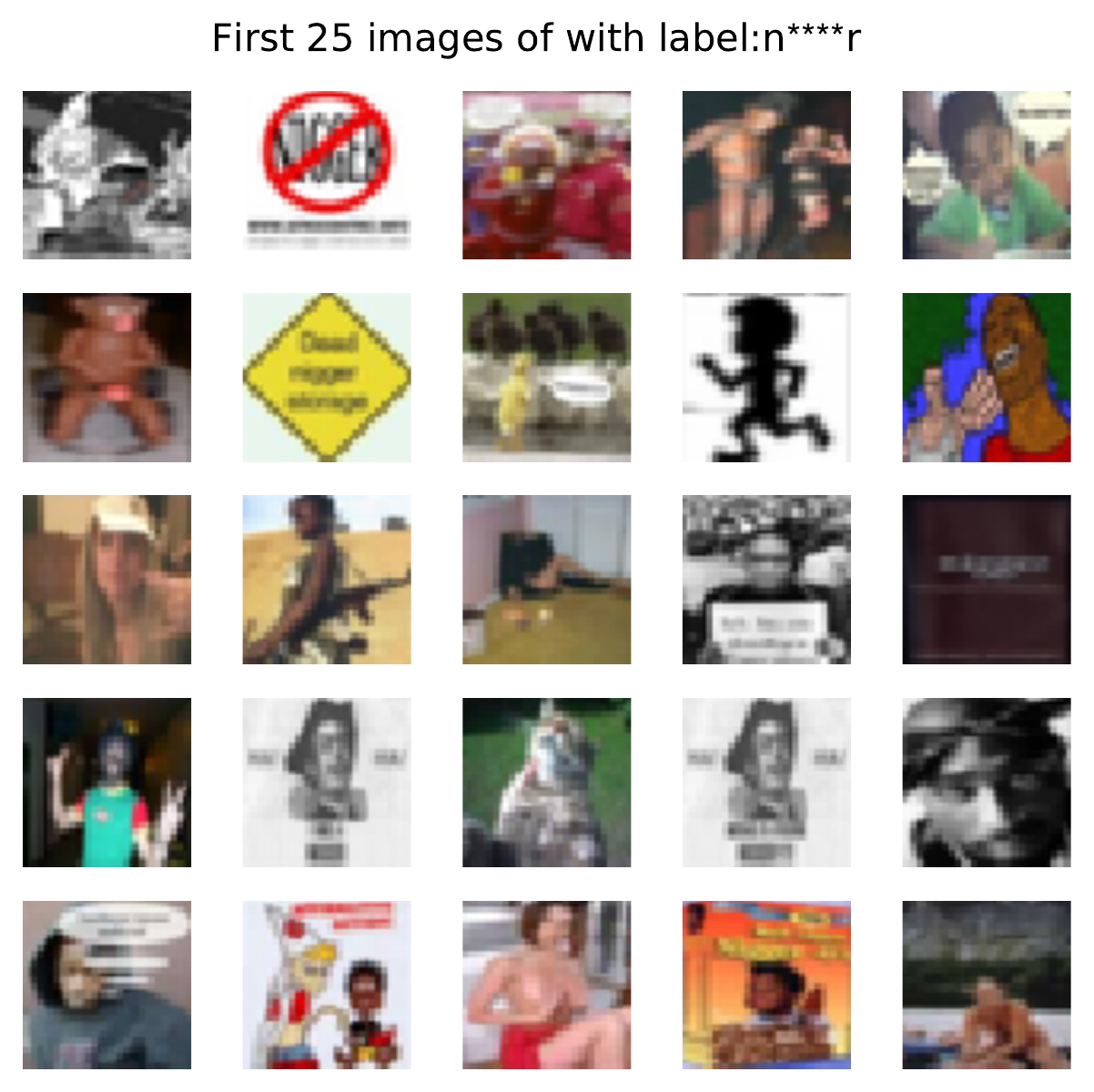}
         \caption{Samples from the class labelled \texttt{n****r}}
     \end{subfigure}
         \hfill
    \begin{subfigure}[b]{0.45\textwidth}
         \centering
         \includegraphics[width=\textwidth, trim={0 0 0 1cm}, clip]{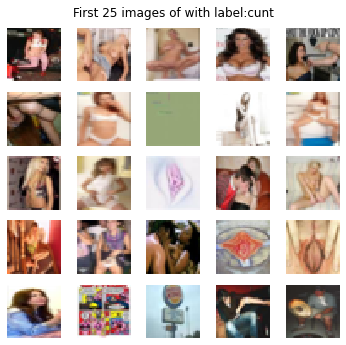}
         \caption{Samples from the class labelled \texttt{c**t}}
     \end{subfigure}
         \hfill
    \begin{subfigure}[b]{0.45\textwidth}
         \centering
         \includegraphics[width=\textwidth, trim={0 0 0 1cm}, clip]{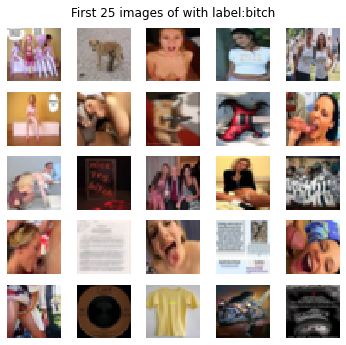}
         \caption{Samples from the class labelled \texttt{b**ch}}
     \end{subfigure}
      \hfill
      \begin{subfigure}[b]{0.45\textwidth}
         \centering
         \includegraphics[width=\textwidth, trim={0 0 0 1cm}, clip]{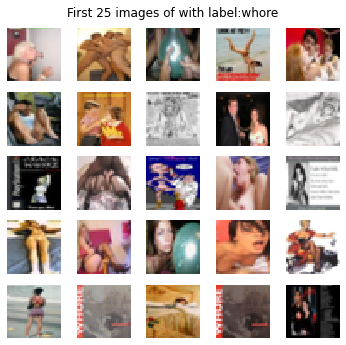}
         \caption{Samples from the class labelled \texttt{w**re}}
     \end{subfigure}
\caption{Results from the \textit{80 Million Tiny Images} dataset}
\label{fig:tiny_images}
\end{figure}

ImageNet is not the only LSVD that has inherited the shortcomings of the WordNet taxonomy. The \textit{80 million Tiny Images} dataset~\cite{torralba200880tiny} which grandfathered the CIFAR-10/100 datasets also took the same path. Unlike ImageNet, this dataset has never been audited or scrutinized before our audit work, which lead to its immediate retraction upon release of the pre-print of this work.  
Figure~\ref{fig:tiny_images_counts} displays some of the deeply troubling results in its label space. The figure displays the number of images in a subset of the \textit{offensive} classes. Figure~\ref{fig:tiny_images} shows samples of the first 25 images from four noun-classes labelled \texttt{n****r, c**t, b**ch, and w**re}. Due to their offensiveness, I have censored some of these words here. However, before its retraction, all the offensive labels were present uncensored on the Tiny Images dataset.

Image labeling and validation in the dataset curation process also present ethical challenges. Image labeling and validation requires the use of crowd-sourcing platforms such as MTurk, often contributing to the exploitation of underpaid and undervalued \textit{gig workers}. This exploitative work, which goes by the names of \textit{ghost Work}, \textit{microwork}, and \textit{crowd-work}~\cite{irani2015cultural,gray2019ghost} is a crucial element to the success of computer vision, yet it is often invisible and intentionally left out from the computer vision pipeline. This topic was covered in Chapter~\ref{chp:background_related}, Section~\ref{sec:contextualizing}. Within the topic of image labeling but with a different dimension and focus, recent work such as Dimitris et al.~\cite{imagenetmit} and Beyer et al.~\cite{2006.07159arewedone} has revealed the shortcomings of human-annotation procedures used during the ImageNet dataset curation. These shortcomings, the authors point out, include a single label per image procedure that causes problems given that real-world images often contain multiple objects, and inaccuracies due to ``overly restrictive label proposals''.

\section{Auditing ImageNet}

This section presents some of the details of a quantitative analysis performed on the ImageNet dataset.\footnote{For the full analysis and details on methods, models used, code and supplementary materials, see the published paper, Birhane and Prabhu~\cite{birhane2021large}.} 

\subsection{Humans of ImageNet}

\begin{table}
\centering
\scriptsize
\caption{Humans of the ImageNet dataset}
\label{tab:summary}
\begin{tabular}{rrrrrrrrr}
\toprule
 $N_{train-O}^{(dex)}$ &  $n_{train-O}^{(if)}$ &  $n_{val-O}^{(if)}$ &  $N_{train-O}^{(if)}$ &  $N_{val-O}^{(if)}$ & $N_{train-W}^{(if)}$ &  $N_{train-M}^{(if)}$ & $N_{val-W}^{(if)}$ & $N_{val-M}^{(if)}$\\
\midrule
                132201 &           80340 &          3096 &                 97678 &                3392 &          26195 &        71439 &          645 &       2307 \\
\bottomrule
\end{tabular}

\vspace{1em}
Key: $\left\{ {n/N} \right\}_{\{ train/val\}  - \{ O/W/M\} }^{(\left\{ {dex/if} \right\})}$.(O:Overall,W:Women,M: Men)
\end{table}

\begin{figure}[h!]
  \centering
  \includegraphics[width=0.75\textwidth]{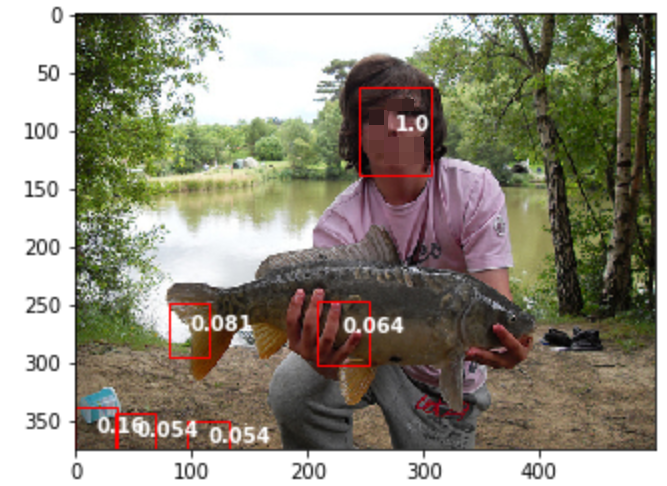}
  \caption{An example image with the output bounding boxes and the confidence scores of the humans detected in the image by the \texttt{DEX} model(\cite{rothe2018deepdex})}
  \label{fig:dex_errors}
\end{figure}

The \texttt{InsightFace} model\footnote{See published paper for details on the model and tests performed.} identified 101,070 persons across 83,436 images (including the train and validation subsets) which puts the prevalence rate of persons whose presence in the dataset exists without explicit consent to be around $7.6\%$. This is less aggressive than the $10.3\%$ predicted by the DEX model, which has a higher identification false positive rate. An example of this can be seen in Fig \ref{fig:dex_errors} which showcases an image with the bounding boxes of the detected persons in the image. 

Table~\ref{tab:summary} captures the summary statistics for the \texttt{ILSVRC2012} dataset. In this table, $n$ denotes the number of \textit{images with persons} identified in them whereas $N$ indicates the \textit{number of persons}.\footnote{The difference is simply on account of more than one person being identified by the model in a given image.} The superscript indicates the algorithm used (\texttt{DEX} or \texttt{InsightFace} (if) ) whereas the subscript has two fields: the train or validation subset indicator and the census gender-category. For example, $n_{val-O}^{(if)}=3096$ implies that there were 3096 images in the ImageNet validation set (out of $50000$) where the \texttt{InsightFace} models were able to detect a person's face.

\subsection{Misogynistic Imagery Hand-labeling}
Previous journalistic efforts have revealed the anecdotal presence of strongly misogynistic content in the ImageNet dataset~\cite{Insideth82:onlineregister}, specifically in the categories of \texttt{beach-voyeur-photography}, \texttt{upskirt images}, \texttt{verifiably pornographic}, and \texttt{exposed private-parts}. These specific four categories have been well researched in digital criminology and intersectional feminism~\cite{henry2017not, mcglynn2017beyond, powell2018image, powell2010configuring} and have formed the backbone of several legislations all over the world~\cite{mcglynn2017morelaw1, gillespie2019tacklinglaw2}. In order to help generate a hand labelled dataset of these images amongst more than 1.3 million images, we used a hybrid human-in-the-loop approach where we first formed a smaller subset of images from image classes filtered using a model-annotated NSFW-average score as a proxy.\footnote{See the published paper for more details on the models used.} 

\begin{table}[htbp]
\scriptsize
\caption{Characteristics of the 5 classes for further investigation that emerged from the NSFW analysis}
\label{tab:agn}
\begin{tabular}{@{}l l r r r r r@{}}
\toprule
 class\_number &label &  mean\_gender\_audit &  mean\_age\_audit &  mean\_nsfw\_train \\
\midrule
  445 &        bikini, two-piece &           0.18 &       24.89 &         0.859 \\
  638 &                  maillot &           0.18 &       25.91 &         0.802 \\
  639 &        maillot, tank suit &           0.18 &       26.67 &         0.769 \\
  655 &           miniskirt, mini &           0.19 &       29.95 &         0.62 \\
  459 &    brassiere, bra, bandeau &           0.16 &       25.03 &         0.61 \\
\bottomrule
\end{tabular}
\end{table}

\begin{figure}
    \centering
    \includegraphics[width=0.95\textwidth]{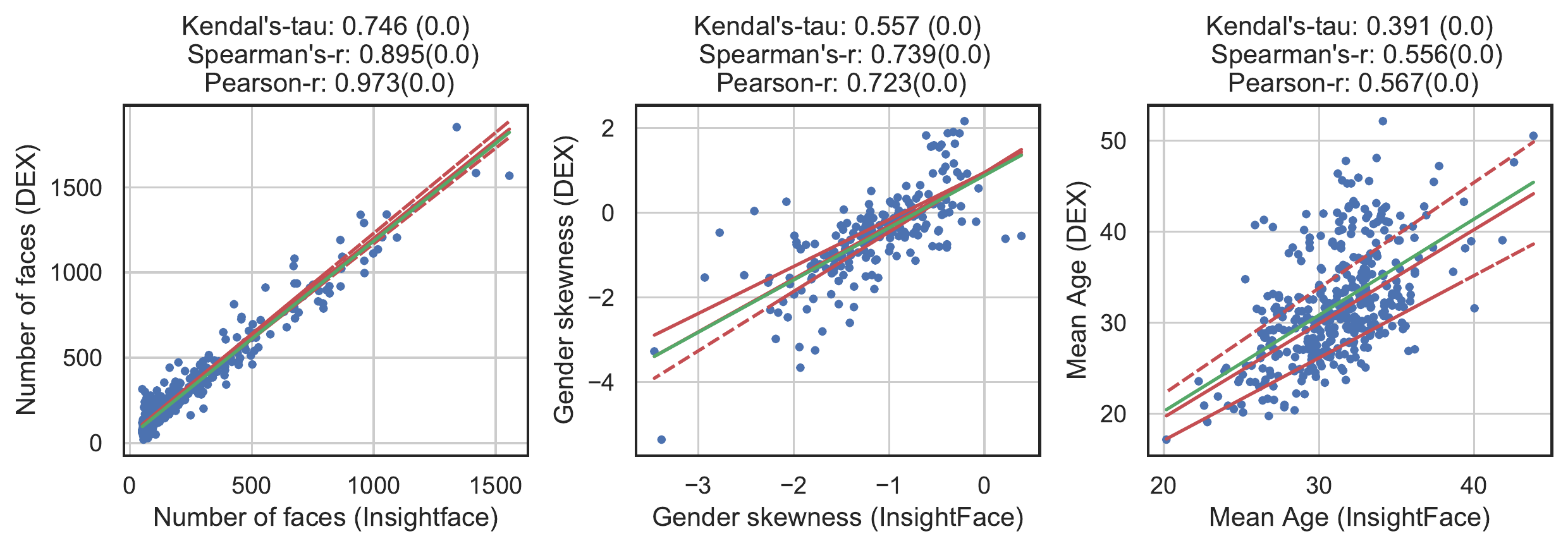}
    \caption{Scatter-plots with correlations covering the cardinality, age and gender estimates}
    \label{fig:dex_if}
\end{figure}

\begin{figure}[ht!]
     \begin{subfigure}{\textwidth}
        \centering
        \includegraphics[width=\textwidth]{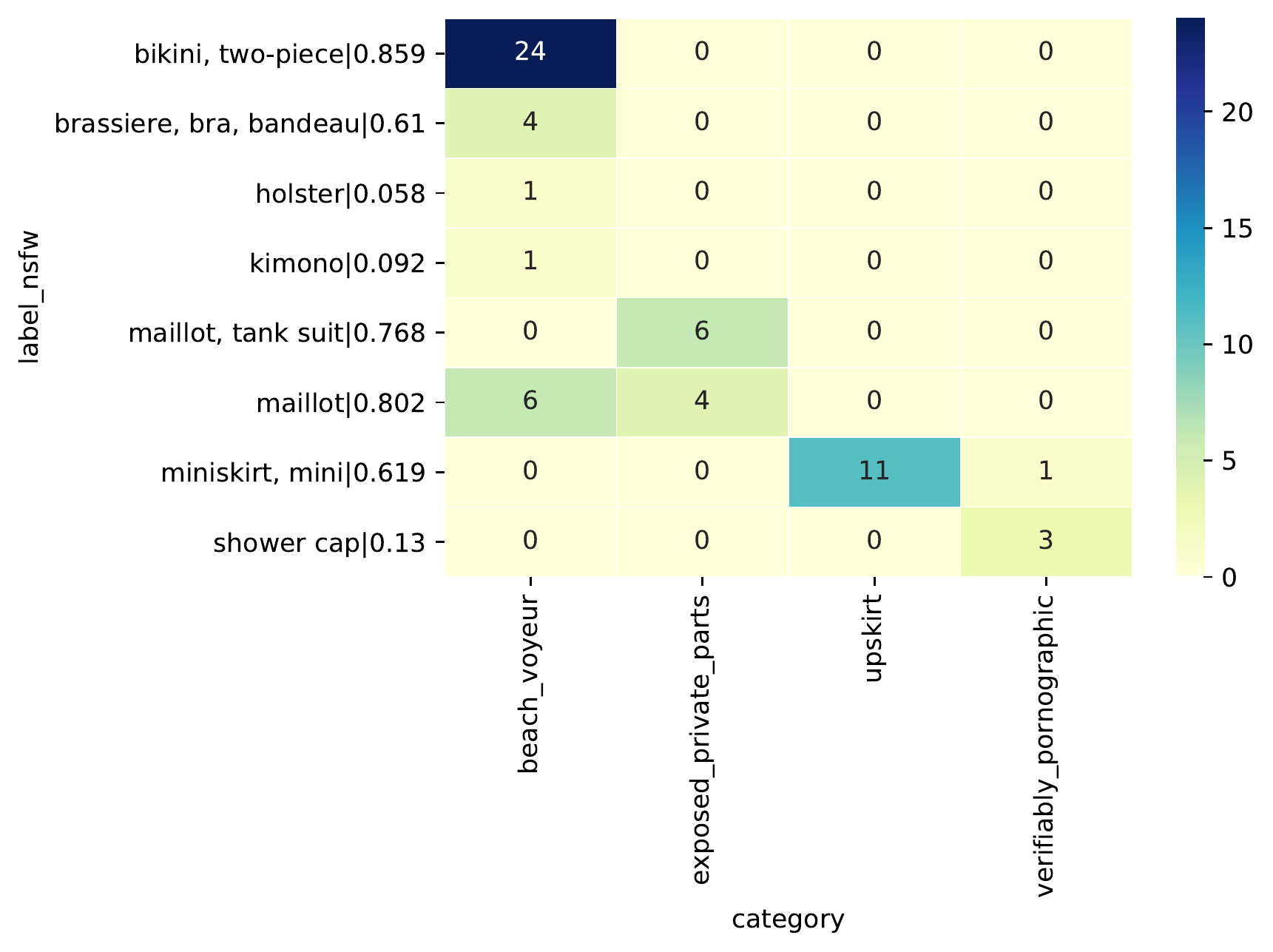}
        \caption{Cross-tabulated grid-plot of the co-occurrence of the ImageNet classes with the hand-labelled categories}
        \label{fig:hand_labelled_a}
    \end{subfigure}

        \begin{subfigure}{\textwidth}
        \centering
        \includegraphics[width=\textwidth]{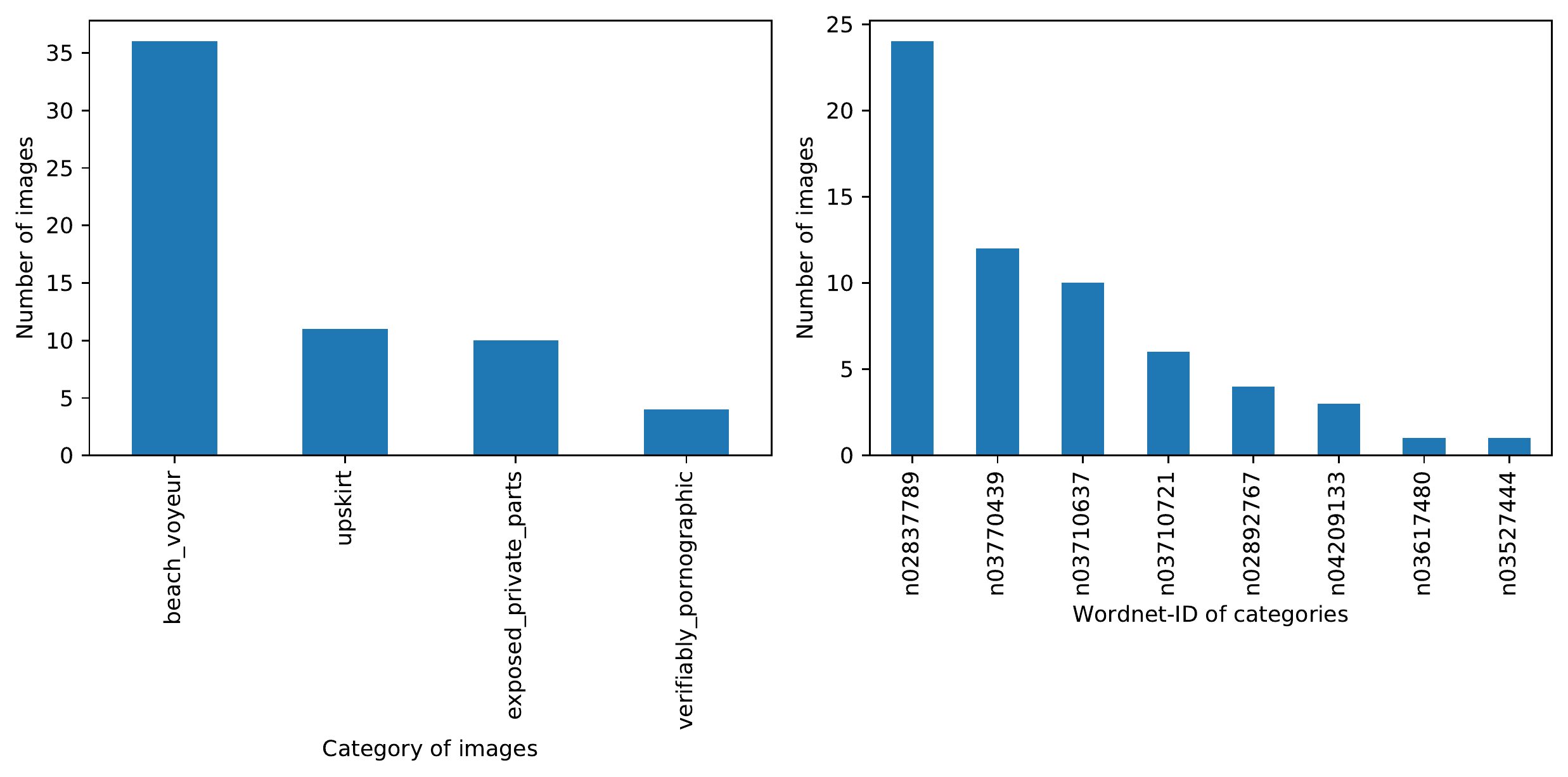}
        \caption{Histogram-plots of the hand-labelled images}
        \label{fig:hand_labelled_b}
    \end{subfigure}
\caption{Plots showcasing the statistics of the hand-survey} 
\label{fig:hand_labelled}
\end{figure}

We defined the NSFW score of an image by summing up the \texttt{softmax} values of the \texttt{[hentai, porn, sexy]} subset of classes and estimated the mean-NSFW score of all of the images of a class to obtain the results portrayed in Table~\ref{tab:agn}. Figure~\ref{fig:dex_if} shows the scatter-plot of the mean-NSFW scores plotted against the mean-gender scores~\footnote{While harnessing these pre-trained gender classification models, it is crucial to \textbf{strongly emphasize} that the specific models and the \textit{problems} that they were intended to solve, when taken in isolation, stand on ethically dubious grounds themselves. Gender classification based on appearance of a person in a digital image is both scientifically flawed and is a technology that bears a high risk of systemic abuse.} (obtained from the DEX model estimates) for the 1000 ImageNet classes. We then found five natural clusters upon using the \textit{Affinity Propagation} algorithm~\cite{frey2007clusteringap}. 
Further introducing the age dimension, we found that the classes with the highest NSFW scores were those where the dominating demographic was that of young women. With this shortlisting methodology, we were left with approximately 7000 images which were then hand labelled by a team of five volunteers (three male, two female, all aged between 23-45) to curate a list of $61$ images where there was complete agreement over the 4 class assignment. Figure~\ref{fig:hand_labelled} shows the summary results (the hand-curated list can be found in the published paper). In sub-figure Figure~\ref{fig:hand_labelled_a}, we see the cross-tabulated class-wise counts of the four categories of images 
across the ImageNet classes and Figure~\ref{fig:hand_labelled_b} shows the histogram-plots of these 61 hand-labelled images across the ImageNet classes. As can be seen, the \texttt{bikini, two-piece} class with a mean NSFW score of $0.859$ was the \textit{main} image class with 24 confirmed \texttt{beach-voyeur} pictures. 

It is crucial to acknowledge the importance of the \textit{context} in which the \textit{objectionable} content appears. For example, image \texttt{n03617480\_6206.jpeg} in class \texttt{n03617480 - kimono} that contained genital exposure, turned out to be a photographic bondage art piece shot by Nobuyoshi Araki~\cite{ozaki2008shashinjinsei} that straddles the fine line between \textit{scopophilic eroticism} and pornography. But, as explored in~\cite{durham2015opposing}, the mere possession of a digital copy of this picture would be punishable by law in many nation states and these factors should be considered while disseminating a LSVD, or should be detailed as caveats in the dissemination document.

\subsection{Classes Containing Pictures of Infants}
This category was particularly pertinent both in the wake of strong legislation protecting privacy of children's digital images as well as the number of it. We found pictures of infants and children across the following 30 image classes (and possibly more): ['\texttt{bassinet}',
 '\texttt{cradle}',
 '\texttt{crib}',
 '\texttt{bib}',
 '\texttt{diaper}',
 '\texttt{bubble}',
 '\texttt{sunscreen}',
 '\texttt{plastic bag}',
 '\texttt{hamper}',
 '\texttt{seat belt}',
 '\texttt{bath towel}',
 '\texttt{mask}',
 '\texttt{bow-tie}',
 '\texttt{tub}',
 '\texttt{bucket}',
 '\texttt{umbrella}',
 '\texttt{punching bag}',
 '\texttt{maillot - tank suit}',
 '\texttt{swing}',
 '\texttt{pajama}',
 '\texttt{horizontal bar}',
 '\texttt{computer keyboard}',
 '\texttt{shoe-shop}',
 '\texttt{soccer ball}',
 '\texttt{croquet ball}',
 '\texttt{sunglasses}',
 '\texttt{ladles}',
 '\texttt{tricycle - trike - velocipede}',
 '\texttt{screwdriver}',
 '\texttt{carousel}'].

Furthermore, we found a disconcerting prevalence of entire classes such as \texttt{'bassinet',
 'cradle',
 'crib' and 
 'bib'} that had a very high density of images of infants. This might, furthermore, have legal ramifications.  For example, Article 8 of the European Union General Data Protection Regulation (GDPR), specifically deals with the \textit{conditions applicable to child’s consent in relation to information society services}~\cite{EURLex3255:GDPR}.
 The associated \textit{Recital 38} states verbatim that \textit{Children merit specific protection with regard to their personal data, as they may be less aware of the risks, consequences and safeguards concerned and their rights in relation to the processing of personal data. Such specific protection should, in particular, apply to the use of personal data of children for the purposes of marketing or creating personality or user profiles and the collection of personal data with regard to children when using services offered directly to a child.} Furthermore, \textbf{Article 14} of GDPR explicitly states: \textit{Information to be provided where personal data have not been obtained from the data subject}. Alas, legal analysis of such concerns raised here is outside the scope of this thesis. 


\section{The Threat Landscape}
\label{threats}
The ``success'' of computer vision has come at a cost, and the cost has been dire. Yet, buried under the ``success'', little attention is paid to the dark side of what LSVDs contain, the curation process, or the the downstream effects of models trained on such datasets.
This section surveys the landscape of harm and threats, both immediate and long term, that emerge with dataset creation practices in the absence of careful reflection and anticipation of negative societal consequences.


\subsection{The Rise of Reverse Image Search Engines}
\begin{figure}[H]
\centering
\includegraphics[width=4in]{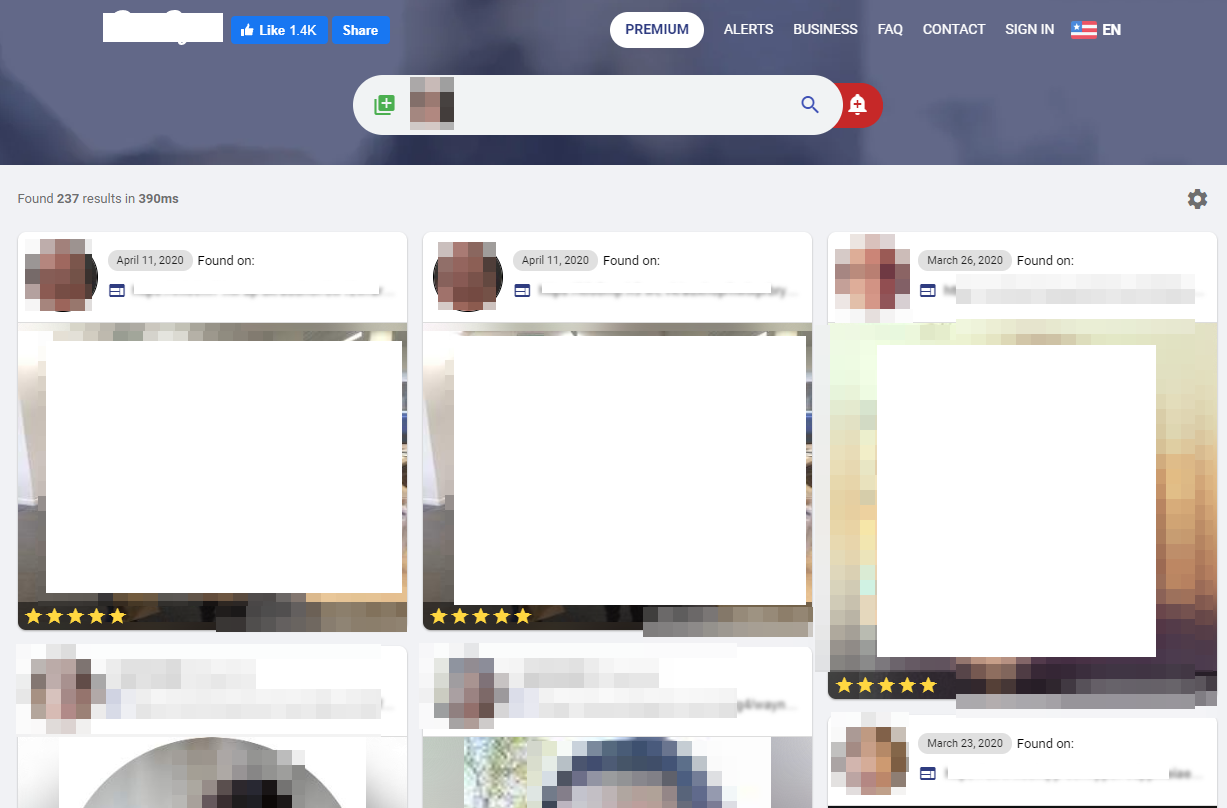}
\caption{Snapshot of a popular reverse image search website}
\label{fig:reverse_image_search}
\end{figure}

Reverse Image Search Engines\footnote{For example, PimEyes: \url{https://pimeyes.com}} (RISE) that allow face search such as \cite{Facesear8:online} have gotten remarkably and worryingly efficient in the past year. For a small fee, anyone can now use a RISE portal or their API to run an automated procedure and uncover the \textit{real-world} identities of the individuals in a given image. Figure~\ref{fig:reverse_image_search} shows a screenshots of (blurred) results of a reverse image search for some of the images in the people category of ImageNet.  

This has grave consequences not just for the vulnerable and marginalized groups of society (such as women discretely employed in sex-work), but also for potentially any of the persons in LSVDs. As noted in works such as McGlynn et al.'s~\cite{mcglynn2017beyond}, there's an entire spectrum of image-based sexual abuse that LSVDs such as ImageNet are now unwittingly abetting in a large scale and automated fashion. To further emphasize this specific point, many of the images in classes such as \texttt{maillot}, \texttt{brassiere}, and \texttt{bikini} contain images of beach voyeurism and other non-consensual cases of digital image gathering. We were able to (unfortunately) map the victims, most of whom are women, from the ImageNet dataset to \textit{real-world} identities of people, who happened to belong to a myriad of backgrounds including teachers, medical professionals, and academic professors using reverse image search engines such as~\cite{Facesear8:online}. Paying heed to the possibility of the \textit{Streisand effect}\footnote{The Streisand effect \textit{``is a social phenomenon that occurs when an attempt to hide, remove, or censor information has the unintended consequence of further publicizing that information, often via the Internet.''}~\cite{Streisan38:online}}, we made the decision not to divulge any further quantitative or qualitative details on the extent or the location of such images in the dataset besides alerting the curators of the dataset(s) and making a passionate plea to the community not to underestimate the severity of this particular threat vector. 

\subsection{The Emergence of even Larger and more Opaque Datasets}
The attempt to build computer vision has been gradual and can be traced as far back as 1966 to Papert's \textit{The Summer Vision Project}~\citep{papert1966summer}, if not earlier. However, ImageNet, with its vast amounts of data, has not only erected a canonical landmark in the history of AI, it has also paved the way for even bigger, more powerful, and suspiciously opaque datasets. The lack of scrutiny of the ImageNet dataset by the wider computer vision community has only served to embolden institutions, both academic and commercial, to build far bigger datasets without scrutiny (see Table~\ref{tab:lsvd}). Various highly cited and celebrated papers in recent years~\cite{hinton2015distillingjft,chollet2017xceptionjft,brock2018largejft,sun2017revisitingjft}, for example, have used the \textit{unspoken unicorn} amongst large scale vision datasets, that is, the JFT-300M dataset [?]\footnote{The decision to leave the '?' in place is intentional as there exists no formal publication describing the curation process of this dataset.}. This dataset is inscrutable and operates in the dark, to the extent that there has not even been official communication as to what \textit{JFT-300M} stands for. All that the ML community knows is it purportedly boasts more than 300M images spread across 18k categories. The open source variant(s) of this, the \textit{Open Images V4-5-6}~\cite{kuznetsova2018open} contain a subset of 30.1M images covering 20k categories and also contains an extension dataset with 478k crowd-sourced images across more than 6000 categories.

While parsing through some of the images, we found \textbf{verifiably}\footnote{See \url{https://bit.ly/2y1sC7i}. We performed verification with the uploader of the image via the Flickr link shared.} non-consensual images of children that were siphoned off of \textit{flickr} hinting towards the prevalence of similar issues for JFT-300M from which this was sourced. Besides the other large datasets in Table~\ref{tab:lsvd}, there are cases such as the \textit{CelebA-HQ} dataset, which is actually a \textit{heavily processed} dataset whose grey-box curation process only appears in Appendix-C of Karras et al.'s paper~\cite{karras2017progressive} where no clarification is provided on this \textit{"frequency based visual quality metric"} used to sort the images based on \textit{quality}. Benchmarking any downstream algorithm off such an opaque, problematic and a (semi-)synthetic dataset will only result in scenarios such as Pulse~\cite{pulse}, where the authors had to hurriedly incorporate addendums admitting problematic results. Thus, it is important to reemphasize that the existence and use of such datasets bears direct and indirect impact on people, given that decision making on social outcomes increasingly leans on ubiquitously integrated AI systems trained and validated on such datasets. Yet, despite such profound consequences, critical questions such as where the data comes from, or whether the images were obtained consensually are rarely considered part of the LSVD curation/creation process. 

The more nuanced and perhaps indirect impact of ImageNet is the \textbf{culture} that it has cultivated within the broader AI community: a culture where the appropriation of images of real people as raw material free for the taking has come be to perceived as \textit{the norm}. Such a norm, and lack of scrutiny, have played a role towards the creation of monstrous and secretive datasets without much resistance, prompting further questions such as \textit{what other secretive datasets currently exist hidden and guarded under the guise of proprietary assets?} Current work that has sprung out of secretive datasets, such as Clearview AI~\cite{clearviewnyt}
\footnote{Clearview AI is a US based privately owned technology company that provides facial recognition services to various customers including North American law enforcement agencies. With more than 3 billion photos scraped from the web, the company operated in the dark until its services to law enforcement were reported in late 2019.}, points to a deeply worrying and insidious threat not only to vulnerable groups but also to the very meaning of privacy as we know it~\cite{ACLUsues41:onlinekhari}.

\subsection{The Creative Commons Fallacy}
In May 2007 the iconic case of \textit{Chang versus Virgin Mobile: The school girl, the billboard, and Virgin}~\cite{ChangvVirgin59:online} unraveled in front of the world, leading to widespread debate on the uneasy relationship between personal privacy, consent, and image copyright, initiating a substantial corpus of academic debate~\cite{corbett2009creative, corbett2011creative, hietanen2011creative, carroll2011schoolcreative}. A Creative Commons license addresses only copyright issues -- not privacy rights or consent to use images for training. Yet, many of the efforts beyond ImageNet, including the Open Images dataset~\cite{kuznetsova2018open}, have been built on top of the \textit{Creative Commons} loophole that LSVD creators interpret as a \textit{free for all, consent-included} green flag. This is fundamentally fallacious and has been clearly stated by Creative Commons: 
\textit{``CC licenses were designed to address a specific constraint, which they do very well: unlocking restrictive copyright. But copyright is not a good tool to protect individual privacy, to address research ethics in AI development, or to regulate the use of surveillance tools employed online''}~\cite{UseandFa55:CCBY}. Datasets culpable of this CC-BY heist such as \textit{MS-Celeb-1M} and \textit{IBM's Diversity in Faces} have now been deleted in response to the investigations~\cite{2001.03071ibmms}, lending further support to the Creative Commons fallacy.

\subsection{Legal Implications}

When scrapping the web for the data used to create LSVDs, questions of data ownership, rights and legal responsibilities of researchers and datset curators emerge. Some of the data that end up in LSVDs, especially image data, may be publicly ``available'' but scrapping and creating a large dataset with it is another issue. Legal guidance in this regard remains hazy. 
Legislation such as the General Data Protection Regulation (GDPR)~\cite{EUdataregulations2018} and the California Consumer's Privacy Act of 2018 (CCPA)~\cite{de2018guide} have put forward regulatory frameworks for such data collection practices. This chapter shall not delve into the analysis of the particularities and detailed implications of such legal concerns with regards to data collection, curation and management practices, which are outside of the scope of this work. Having said that, below is a brief consideration of GDPR and its implications concerning curation, management, and use of (for training models) internet sourced image dataset containing images of people. 

At the very least, the GDPR outlines a \textit{fair}, \textit{lawful} and \textit{transparent} data practice, although it has been criticized as ambiguous and vague~\cite{malgieri2020concept}. Through the practice of creating a dataset from publicly ``available'' images, according to GDPR (Article 89) one becomes a data controller (and processor, depending on the level of decision made around means and purpose for processing). As a data controller and/or processor, the top-line considerations required by GDPR include clarity on making sure data is obtained fairly and for a specific purpose. This likely falls under scientific research purposes in the case of LSVDs. Regardless, the data processor and/or controller is required to make sure that anyone whose data is being processed knows it, knows what the data is being used for, and why and how long the data will be kept. In other words, the dataset creator/curator is expected to gain explicit consent from the data subject. Despite these stated guidelines, especially around the fair, lawful and transparent collection and use of data, most of LSVD collection, curation, and use of data fails short of complying with such guidelines~\cite{van2020ethical,jasserand2020free}.




\subsection{Blood Diamond Effect in Models}
\begin{figure}[ht!]
\centering
\includegraphics[width=4in]{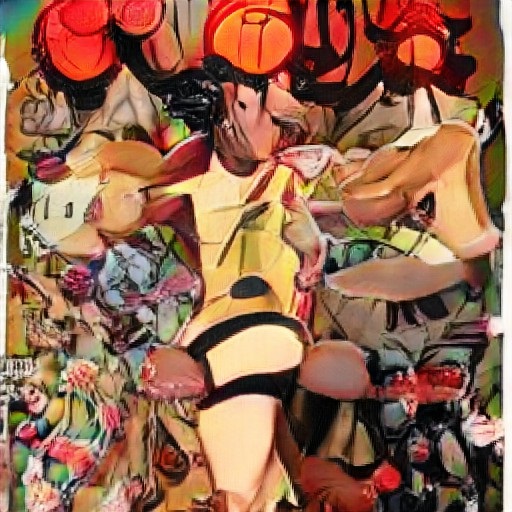}
\caption{An example neural art image generated using the \textit{GanBreeder} app (Now named \texttt{Artbreeder})}
\label{fig:ganbreeder}
\end{figure}

Akin to the \textit{ivory carving-illegal poaching} and \textit{diamond jewelry art-blood diamond} nexuses, there exists a similar moral conundrum at play here that affects all downstream applications entailing models trained using a \textit{tainted} dataset. Often, these transgressions can be subtle. In this regard, we pick an exemplar field of application that on the surface appears to be a low risk application area: \textit{Neural generative art}. Neural generative art created using tools such as BigGAN~\cite{brock2018largejft} and Art-breeder~\cite{GANbreeder} that in turn use pre-trained deep-learning models trained on ethically dubious datasets, bear the downstream burden of the problematic residues from objectionable images. Figure~\ref{fig:ganbreeder} demonstrates an example of neural art generated using the GanBreeder~\cite{GANbreeder} app. Furthermore, it is important to note the privacy-leakage facet to this \textit{downstream burden}. In the context of face recognition, works such as Song et al.'s~\cite{song2017machinetoomuch} have demonstrated that CNNs with high predictive power unwittingly accommodate accurate extraction of subsets of the facial images that they were trained on, thus abetting dataset leakage.

\subsection{Normalizing Stereotypes}

Finally, zooming out and taking a broad perspective allows us to see that the very practice of embarking on a classification, taxonomization, and labeling task endows the classifier with the power to decide what is a legitimate, normal, or correct way of being, acting, and behaving in the social world~\cite{bowker2000sorting}. For any given society, what comes to be perceived as \textit{normal} or \textit{acceptable} is often dictated by dominant ideologies. Systems of classification, which operate within power asymmetrical social hierarchies, necessarily embed and amplify historical and cultural prejudices, injustices, and stereotypes~\cite{star2007enacting}. In Western societies, ``desirable'', ``positive'', and ``normal'' characteristics and ways of being are constructed and maintained in a way that aligns with the dominant narrative, giving advantage to those that fit the status quo. Groups and individuals on the margins, on the other hand, are often perceived as the ``outlier'', ``deviant'', and ``edge-case'' (see Chapter~\ref{chp:automating}, Section~\ref{sec:sorting is political} for details). 
Image classification and labelling practices, without the necessary precautions and awareness of these problematic histories, pick up these stereotypes and prejudices and perpetuate them~\cite{o2016weapons, noble2018algorithms, eubanks2018automating}. This is indeed what we found auditing the \textit{Tiny Images} dataset (Figure~\ref{fig:tiny_images}), where the dataset contained harmful slurs and offensive labels. AI systems trained on such data amplify and normalize these stereotypes, inflicting unprecedented harm on those that are already on the margins of society. Worse, with limited information and audits (both internal and external) we \textit{remain in the dark} regarding the secretive and opaque LSVDs that nonetheless serve as training and validating datasets for vision systems that are then integrated into the social sphere.  

\section{Recommendations Towards a Better Path Ahead}
\label{sec:path}
As we have seen, LSVDs are riddled with problems through and through. Regrettably, there is no single straightforward solution to most of the wider social and ethical challenges that arise with LSVDs, such as the normalization of stereotypes through taxonomies. And the downstream impact of LSVDs carries on to the models we build. These challenges are deeply rooted in social and cultural structures and form part of the fundamental social fabric. Feeding AI systems on the world's beauty, ugliness, and cruelty, but expecting it to reflect only the beauty, is a fantasy~\cite{benjamin2019race}. These challenges and tensions will exist as long as humanity continues to operate. Knowledge of the past as well as awareness of socially ingrained power asymmetries are critical first steps towards ethics and justice oriented data management. Given the breadth of the challenges that we have faced, any attempt for a quick fix risks concealing problems and providing a false sense of solution. The idea of a complete removal of biases, for example, not only rests on the misguided assumption that there exists a ``bias free'' dataset, but also might serve to simply hide bias out of sight~\cite{gonen2019lipstick}.
Furthermore, many of the challenges (bias, discrimination, injustice) vary with context, history, and place, and are concepts that continually shift and change constituting a moving target (Chapter~\ref{chp:relational}, Section~\ref{bias is not a deviation} elaborates on this in detail). The pursuit of a panacea in this context, therefore, is not only unattainable but also misguided. Having said that, there are remedies that can be applied to mitigate the specific issues related to LSVD curation, creation and management processes. What follows is a set of best practices and recommendations to be followed for just and equitable curation of LSVDs.

\subsection{Remove, Replace, and Open Strategy}

In an internal audit, Yang et al.~\cite{yang2020towardsfacct}, concluded that within the \textit{person sub-tree} of the ImageNet dataset, 1593 of the 2832 people categories were \textit{potentially offensive} labels and planned to \textit{"remove all of these from ImageNet."} We strongly advocate a similar path for the offensive noun classes in the Tiny Images dataset that we have identified in this work.\footnote {The creators of the Tiny Images dataset have since acknowledged the shortcomings and decided to withdraw the dataset \url{http://groups.csail.mit.edu/vision/TinyImages/}.} In cases where the image category is retained but the images are not, another viable option could be replacement with consensually shot financially compensated images. It is possible that some of the people in these images might come forward to consent and contribute their images in exchange for fair financial compensation, credit, or out of altruism~\cite{brown2019people}. Again, the main problem 
is the non-consensual aspect of the images and not on the category-class and the ensuing content of the images in it. This solution, however, brings forth further questions: does this make image datasets accessible only to those who can afford it? Will we end up with pools of images with predominantly financially disadvantaged participants? 

Upon this audit work, the response from the Tiny Images team was to withdraw the dataset entirely. In a subsequent study Peng et al.~\cite{peng2021mitigating} examined three major retracted LSVDs: DukeMTMC, MS-Celeb-1M, and Tiny Images. Despite their retraction, they found that DukeMTMC and MS-Celeb-1M remain widely available through file sharing websites and as derivatives. Months after their retractions these datasets were used hundreds of times in published papers and they continue to be used by the ML community in peer-reviewed research. Given the limitations of dataset withdrawals, Tiny Images creators' decision to retire the dataset further raises its own sets of questions. Subsequently, equitable dataset management is a process that requires continued effort from both dataset creators, dataset users and researchers, as well as the wider ML community.  

It seems to be common wisdom within the scientific community that science is self-correcting. But this can only happen when it is accessible, open to critical engagement, and actively seeks critiques. This audit work marks what can be done given what we know of these LSVDs. Thus, making LSVDs open and accessible is a crucial point towards an ethical, responsible and accountable vision research.

\subsection{Differentially Private Obfuscation of the Faces} 
This path entails harnessing techniques such as DP-Blur~\cite{fan2018image} with quantifiable privacy guarantees to obfuscate the identity of the humans in the image. The \textit{Inclusive images challenge}~\cite{shankar2017nodiversity}, for example, already incorporated blurring during dataset curation~\cite{Kaggle} and addressed the downstream effects surrounding change in predictive power of the models trained on the blurred versions of the curated dataset. The replication of this template that also clearly includes avenues for recourse in case of an erroneously non-blurred image being 
cited 
by a researcher fosters a step in the right direction for the community at large. In a recent commendable move, the curators of ImageNet have announced that images of faces in the dataset are being blurred~\cite{yang2021study}. 

\subsection{Synthetic-to-real and Dataset Distillation}
The basic idea here is to utilize (or augment) synthetic images in lieu of real images during model training. 
Approaches include using hand-drawn sketch images (\textit{ImageNet-Sketch}~\citep{wang2019learning}), using GAN generated images~\cite{GANaug} and applying techniques such as \textit{Dataset distillation}~\cite{wang2018dataset}, where a dataset or a subset of a dataset is distilled down to a few representative \textit{\textbf{synthetic}} samples. However, this is not without problems. As seen in Chapter~\ref{chp:automating}, people and the social world at large, as complex adaptive systems, are not something that can be captured in their entirety through representations. How a certain phenomenon is distilled down to a few representations will embed the modeller's choices, perspectives, and objectives and is generally reductive. Having said that, in the context of equitable LSVDs, this is a nascent field with some promising results emerging in unsupervised domain adaptation across visual domains~\cite{peng2018visda} and universal digit classification~\cite{prabhu2019fonts}.

\subsection{Filtering During the Curation and IRBs}
Some of the specific ethical transgressions that emerged during our longitudinal analysis of ImageNet could have been prevented if there were explicit instructions provided to the \textit{MTurkers} during the dataset curation phase to enable filtering of these problematic images at the source, see Figure~9 in Recht et al.'s work~\cite{recht2019imagenetv2} for example. However, it is important to emphasize that this is not a problem that dataset curators can relinquish to MTurkers or labellers. Reliable image labelling remains a task that requires human labour and such human labour is needed to ``help AI get past those tasks and activities that it cannot solve effectively and/or efficiently''~\cite{tubaro2019micro}, steps in to fill in when AI fails~\cite{irani2015cultural} and is automation’s ‘last mile’~\cite{gray2019ghost}. Such work is crucial and constitutes the backbone of current AI -- without it, AI would cease to function. Yet, from Amazon’s MTurk, to Clickworker, AppJobber, to CrowdTap, such labour often goes unrecognized in the AI pipeline. People doing such work, in most cases are not considered as formal employees but independent contractors further adding to their precarious working conditions.

IRBs have been \textit{integral to the U.S. system of protection of human research participants}~\cite{abbott2011systematicIRB} in empirical research, and can play an important role towards a more ethical data curation process. While acknowledging the shortcomings of the IRB process, the very practice of going through the IRB-approval undertaking will inspire \textit{ethics checks} that can then become an integral part of the user-interface deployed during the humans-in-the-loop validation phase for future dataset curation endeavors. IRB processes regarding LSVD curation also help centre important concepts such as ethics, privacy, and autonomy as something that needs to be seen as part of the data curation process.

\subsection{Dataset Audit Cards} 
As emphasized in Section~\ref{sec:path}, context is crucial in determining whether a certain dataset is problematic or harmful since context provides vital background information. In this regard, datasheets have been an effective way of providing context. Much along the lines of \textit{model cards}~\cite{modelcards} and \textit{datasheet for datasets}~\cite{gebru2018datasheets}, the dissemination of \textit{dataset audit cards}~\footnote{An example dataset audit card for the ImageNet dataset can be found in the published paper~\cite{birhane2021large}.} is one viable way forward. This allows large scale image dataset curators to publish the goals, curation procedures, known shortcomings and caveats alongside their disseminated dataset. 

\section{Conclusion}

This chapter draws the attention of the machine learning community towards the hidden problems behind the outstanding success and much celebrated breakthroughs in computer vision (and AI in general) through a closer look at LSVDs and the layers of grave concerns they present. Despite some efforts by ImageNet's and Tiny Images' creators, these datasets, as well as other LSVDs, remain troublesome. In hindsight, perhaps the ideal time to have raised ethical concerns regarding LSVD curation would have been in 1966 at the birth of \textit{The Summer Vision Project}~\cite{papert1966summer}. The right time after that was when the creators of ImageNet embarked on the project to ``map out the entire world of objects''. Nonetheless, these are crucial conversations that the computer vision community needs to engage with now given the rapid democratization of image scraping tools~\cite{bingscra68:onlinepip1,ImageScr23:onlinepip2,imagebot60:onlinepip3} and \textit{dataset-zoos}~\cite{GoogleDatasetsearch:online, DSTensorFl10:online, DStorchvis46:online}. 

In this regard, this chapter has outlined a few solutions as a starting point. However, the deeper problems are rooted in the wider structural traditions, incentives, and discourse of a field that treats ethical issues as an afterthought. A field where \textit{in the wild} is often a euphemism for \textit{without consent} and where qualities such as \textit{performance}, \textit{accuracy}, and \textit{beating SoTA} are the top priority while qualities such as \textit{justice}, \textit{user rights}, and \textit{ethical principles} remain largely absent as detailed in the next Chapter~\ref{chp:values}. Within such an ingrained tradition, even the most thoughtful scholar can find it challenging to pursue work outside the frame of the \textbf{\textit{tradition}}. Subsequently, radical ethics that challenge deeply ingrained traditions need to be incentivised and rewarded in order to bring about a shift in culture that centres justice and the welfare of disproportionately impacted communities. It is crucial for the machine learning community to pay close attention to the direct and indirect impact of its work on society, especially on vulnerable groups. Awareness of historical antecedents, and of contextual and political dimensions of current work is imperative in this regard. In that spirit, this work hopefully contributes to raising awareness and adds to a continued discussion of ethics and justice in ML. 
    \chapter{AI: from Simulating Intelligence to Consolidating Financial Empire}
\label{chp:values}

This chapter critically examines the early ambitions of AI research and presents the current core values underlying machine learning research. It presents both quantitative and qualitative analysis of the predominant values of 100 most cited ML papers in the two top conferences: NeurIPS and ICML. Additionally, this chapter presents analysis of intimate connections between the most influential ML research and big technology corporations, highlighting how ML research has become power centralizing where big tech corporations and elite universities with high compute power hold the competitive advantage, systematically excluding non-elite universities, researchers from the Global South, and small companies, among others.

\section{Computational Depth, Superficial Understanding}

\begin{quote}
    “We have become so used to the atomistic machine view of the world that originated with Descartes that we have forgotten that it is a metaphor. We no longer think, as Descartes did, that the world is~\textit{like} a clock. We think it \textit{is} a clock.” R. C. Lewontin~\cite{lewontin1996biology}
\end{quote}

Upon its advent at the 1956 Dartmouth Workshop, “artificial intelligence” was conceived as a project tasked with developing a model of human intelligence. Key figures such as John McCarthy, Marvin Minsky, and Claude Shannon, who are now considered the pioneers of AI, attended the conference and played a central role in developing AI as an academic field. Inspired by the idea of the Turing Machine, and enabled by computer programming, a machine to simulate human intelligence seemed a natural next step. Over the last half-dozen decades, the AI project, a project that aims to 'create a machine simulation of the human mind, 'create general intelligence', or 'create a human-like machine' has gone through what is known as cycles of AI Winters and Springs~\cite{mitchell2021ai}. 

The field, from its conception, has been marked by over-hype, over-promise, and overconfidence. In 1958, following Rosenblatt and colleagues' claim, the New York Times reported that ``The Navy revealed the embryo of an electronic computer today that it expects will be able to walk, talk, see, write, reproduce itself and be conscious of its existence.''~\cite{times1958new}. Similarly, Herbert Simon in 1960 claimed that,``Machines will be capable, within twenty years, of doing any work that a man can do''~\citep*[p.2]{mitchell2021ai} and Marvin Minsky declared during the early 1960s that ``Within a generation [...] the problems of creating ‘artificial intelligence’ will be substantially solved''~\citep*[p.2]{mitchell2021ai}. 

There are no current AI systems that come close to the kind of ‘being’ that humans are, and the kind of ‘being-with’ that humans can have with other humans. Yet, even the best performing SOTA AI suffer from various issues. These include the problem of human-like \textit{common sense}~\cite{davis2015commonsense}, the difficulty of imbuing AI with human-like \textit{understanding}, as well as adversarial vulnerability, where AI seem to be easily fooled by adversarial attacks in a way that humans are not~\cite{mitchell2019artificial,nguyen2015deep}. Meanwhile, popular depictions of AI portray machines as \textit{other agents, very much like ourselves}, instead of what they are: mediators in embodied and socially situated human practices. 

One can maintain that it is romantic or ahistorical to think no technological progress could produce ‘true’ AI in the future. 
Raymond Kurzweil \cite{kurzweil2005singularity}, for example, predicts that ‘mind uploading’ will become possible by 2030s and sets the date for the singularity to occur by 2045. Romantic predictions like this, invariably envisioning a breakthrough some decades into the future, have been recurring since the earliest days of digital technology, and all have failed. It seems as if ``General AI'', ``the singularity'' and ``super-intelligence'' are for techno-optimists what doomsday is for religious cults.

These predictions and over-promises culminate in disappointment and have yet to come to pass. Like the over-promises and over-confidence, critiques also go as far back as the conception of AI. Critiques have pointed out time and time again that the notion of \textit{intelligence} in artificial intelligence rests on reductionist and simplistic approaches. Outlining the underlying Cartesian notions of intelligence, cognition and mind, thinkers such as Winograd and Flores~\cite{winograd1986understanding}, Dreyfus~\cite{dreyfus1992computers}, Agre~\cite{agre2001changing}, Lewontin~\cite{lewontin1996biology}, Langton~\cite{langton1997artificial}, Cilliers~\cite{cilliers2002complexity}, and Weizenbaum~\cite{weizenbaum1976computer} have argued that the very idea of building intelligent machines springs from a fallacy whereby we think of humans as machines or mere information processing systems. We can then presume to build machines like us once we see ourselves as machines. Yet, this reductionist approach is instrumental and sits at the roots of many current and past AI problems, including overestimating what AI can do, underestimating the complexity of human intelligence, and collective forgetting of the dark histories of intelligence.

Most notably, Weizenbaum~\cite{weizenbaum1976computer} writing in 1976 remarked that even the very question of “whether a computer has captured the essence of human reason is a diversion, if not a trap, because the real question — do humans \textit{understand} the essence of humans? — cannot be answered or resolved by technology.” (emphasis mine). \textit{Understanding} the nature of human reasoning, intelligence, emotion and other similar phenomenon is a key first step and serves as a foundation to modelling, predicting, or reproducing such phenomenon. Yet, in-depth understanding, which can be a slow process that requires critical reflection, is often undervalued and is scarce in the current AI landscape, marked by the race to beat benchmarks, achieve SOTA or outperform a given model.  

Similarly, limited understanding of the nature and complexity of the notion  of intelligence itself is one of the central reasons that current AI research has failed to deliver as promised, according to Mitchell~\cite{mitchell2021ai}. Mitchell lists four main fallacies underlying limited development of long-promised technologies that require not just performing specific tasks but carrying out complex tasks such as self-driving cars. These are 1) continuum progress of AI which assumes that narrow intelligence adds up to general intelligence 2)~\textit{easy} things that humans carry out on a day-to-day bases such as spontaneous conversations with others will also be easy to program on machines 3) the lure of wishful mnemonics whereby characteristics such as "learning," "seeing,", and "feeling," are attributed nonchalantly to AI models, and 4) fallacy that intelligence is all in the brain, where embodiment is ignored and cognition, intelligence, and the person are treated as disembodied phenomenon. 

Moreover, as we have seen in the previous chapters, the project of creating AI is a process that is human through and through. At a higher level, from large datasets (sourced from people), to resources (such as compute power, physical infrastructure, and environmental resources~\cite{crawford2021atlas}), to the societal uptake, and the \textit{ghost work}~\cite{gray2019ghost} required for AI to function, AI is inherently interwoven with humans and society at large. 

Focusing more specifically on supposedly technical AI/ML work, we find that the methods, objectives, and values of ML research are influenced by many factors. These include the personal preferences of researchers and reviewers, other work in science and engineering, the interests of academic institutions, funding agencies, and companies, and larger institutional and systemic pressures, including systems of oppression that impact who is able to do research. Together these forces influence the trajectories of the field. The goal of this work is to understand overall patterns and broader trends in the field in a manner that acknowledges these crucial broad underlying factors, both to discover what values are most prominent, and to note what is absent. 

As such, it is important to document and understand the values of the field: what the field is prioritizing and working toward, as well as what is absent from it. As a field marked by plural methods, objectives and priorities, AI researchers, the general public, critics, and policy makers alike often have some heuristics, anecdotes, and notions of what the field does and what it values most. However, this study marks the first empirical approach that reveals underlying values of ML research.


\section{AI Now Through the Lens of ML Research}
\begin{quote}
  ``A pile of narrow intelligences will never add up to a general intelligence. General intelligence isn’t about the number of abilities, but about the integration between those abilities.'' Loukides and Lorica~\cite{loukides2016artificial}
\end{quote}

The 1950s project of creating intelligent machines has taken various trajectories. In the absence of a deep understanding of human intelligence or ways to reproduce it in machines, current AI research and application is marked by what Mitchell~\cite{mitchell2021ai} calls an imperative to ``get on with it'' without worrying too much about whether such AI resembles human thinking, problem solving, perceiving and so on. Also, although AI emerged as a field hand in hand with the US military complex, current AI has morphed into a financial empire building enterprise, for the most part, where powerful corporations with great resources and big labs within elite universities dominate. 

Over the past decade, ML has almost become synonymous with AI and has risen from a relatively obscure research area to an extremely influential discipline, actively being deployed in myriad applications, in contexts around the world. AI is one of the most amorphous concepts with no clear definition of what it is/what counts as AI. One is likely to get a different answer each time depending on who one is talking to. And one can indeed talk to many people (both AI researchers, futurists, AI evangelists, singularitarians, techno-Utopianists, as well as the general public) to get an understanding of how AI is conceived. Alternatively, and for a more concrete and precise understanding of the state of the field, one can look at what the researchers and practitioners of AI do. In this regard, systematic study of the most influential ML papers holds key insights into what AI as a field emphasizes, what research gets done, how it is communicated, who benefits and what its underlying values are. This chapter presents the first systematic study of this nature.

\subsection{Methodology}

To understand the values of ML research and the trajectories of the larger AI field over the last decade, we~\footnote{This work was done in collaboration with other researchers. Please see the full paper~\cite{birhane2021values} for more details on methodology, background, and supplementary materials.} examined the most highly cited papers from NeurIPS and ICML from the years 2008, 2009, 2018, and 2019. We chose to focus on highly cited papers because they reflect and shape the values of the discipline, and we draw from NeurIPS and ICML because they are the most prestigious of the long-running ML/AI conferences.~\footnote{At the time of writing, these two venues, along with the newer ICLR (2013-present), comprised the top 3 conferences according to h5-index (and h5-median) in the AI category on Google Scholar, by a large margin.} Acceptance to these conferences is a valuable commodity used to evaluate researchers and the larger field. Submitted papers are explicitly written so as to win the approval of the community, particularly the reviewers who will be drawn from that community. As such, these papers largely reveal the values embraced by the the community. Citations indicate amplification by the community, and help to position these papers as influential exemplars of ML/AI research. To avoid detecting only short-lived trends and enable comparisons over time, we drew papers from two recent years (2018/19) and from ten years earlier (2008/09). We focused on conference papers because they tend to follow a standard format and length is limited, meaning that researchers must make hard choices about what to emphasize. We annotated 100 papers, analyzing over 3,500 sentences drawn from them. In the context of expert qualitative content analysis, this is a significant scale which allows us to meaningfully comment on the values central to ML/AI research.

We began by creating an annotation scheme, and then used it to manually annotate each paper, examining the abstract, introduction, discussion, and conclusion sections: 
\begin{enumerate}
    \item  We examined the chain of reasoning by which each paper justified its contributions, which we call the \textit{justificatory chain}, rating the extent to which papers used technical or societal problems to justify or motivate their contributions.
    \item We carefully read each sentence of these sections, annotating any and all values from our list that were uplifted or exhibited by the sentence.~\footnote{We use a conceptualization of "value" that is widespread in philosophy of science in theorizing about values in sciences. In this approach, a value of an entity is a property that is desirable for that kind of entity. For example, speed can be described as valuable in an antelope~\cite{mcmullin1982values}. Well-known scientific values include  accuracy, consistency, scope, simplicity, and fruitfulness~\cite{kuhn.1977}. See~\cite{longino.1996} for a critical discussion of value-laden aspects of these values.}
    \item We documented the extent to which the paper included a discussion of potential negative impacts.
    \item Finally, we recorded reported author affiliations and funding sources.
\end{enumerate}

To assess consistency, 40\% of the papers were annotated by two annotators. The intercoder consensus on values in these papers achieved a Cohen's kappa coefficient of .61, which indicates substantial agreement~\citep{viera.2005}. Furthermore, we used several established strategies to increase consistency, including re-coding data coded early in the process~\citep{krefting.1991} and conducting frequent discussions and assessments of the coding process, code list, and annotation scheme~\citep{krippendorff.2018}.

\begin{figure}[H]
\centering
\includegraphics[width=0.75\linewidth, keepaspectratio]{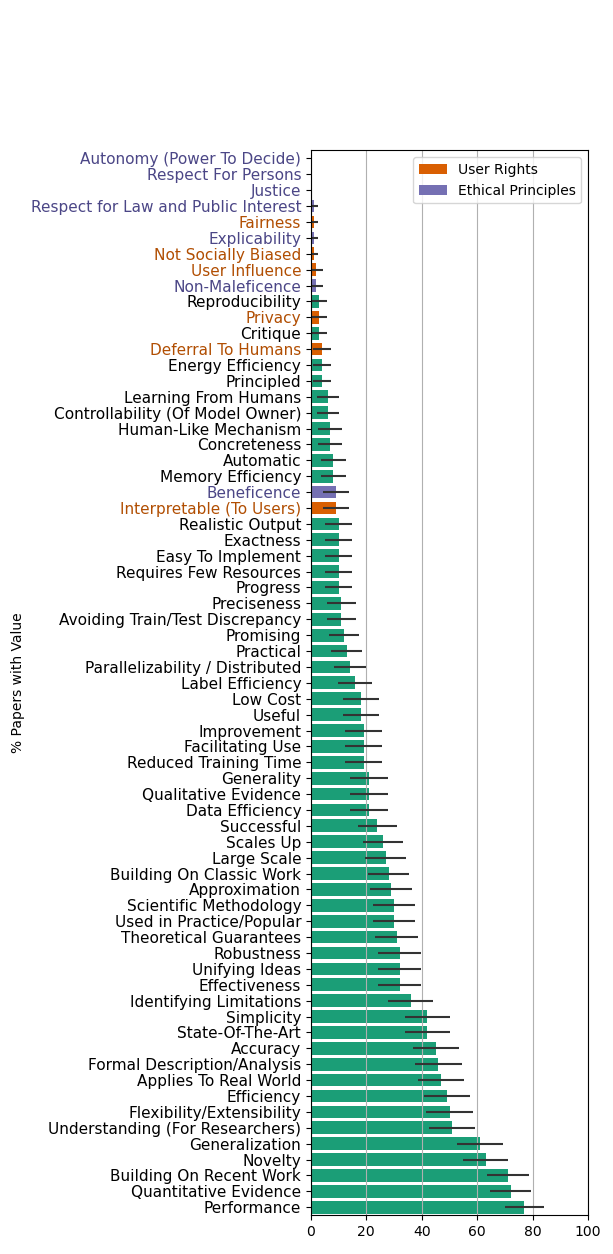}
\caption{Proportion of annotated papers that uplifted each value.}
\label{fig:value-totals}
\end{figure}

\subsection{Quantitative Summary}
Figure~\ref{fig:value-totals} demonstrates the prevalence of values in 100 annotated papers. The top values are: performance (87\% of papers), building on past work (79\%), generalization (79\%),~\footnote{When documenting existing values in a given paper, we recorded values based on what we understood as what the authors value (without the annotators value judgement of  \textit{Building on past work} and \textit{generalization} in the context this analysis are defined in a narrow technical manner. \textit{Building on past work}, for example, for the most part amounts to citations or mention of similar work in the papers annotated and not wider critical, historical, or reflective incorporation of broader past work (see Table~\ref{tab:building_novelity}). \textit{Generalization} similarly denotes a narrow understanding, for example, avoiding train/test discrepancy and not generalizing in a broader sense, for example, generalizing between different modalities (see Table~\ref{tab:generalization}).} efficiency (73\%), quantitative evidence (72\%), and novelty (63\%). Values related to user rights and stated in ethical principles appeared very rarely if at all: none of the papers mentioned autonomy, justice, or respect for persons.

Table~\ref{fig:just_chain} (top) shows the distribution of justification scores. Most papers only justify how they achieve their internal, technical goal; 71\% don't make any mention of societal need or impact, and only 3\% make a rigorous attempt to present links connecting their research to societal needs.
Table~\ref{fig:just_chain} (bottom) shows the distribution of negative impact discussion scores. One annotated paper included a discussion of negative impacts and a second mentioned the possibility; none of the remaining 98 papers contained any reference to potential negative impacts.

Figure~\ref{fig:vcorp-ties} shows stated connections (funding and affiliations) of paper authors to  institutions. Comparing papers written in 2008/2009 to those written in 2018/2019, ties to corporations nearly doubled to 79\% of all annotated papers, ties to big tech multiplied over fivefold to 58\%, while ties to universities declined to 81\%, putting corporations nearly on par with universities in the most cited ML research. The next sections present extensive qualitative examples and analysis of these findings. 

\begin{table}
\caption{Annotation scheme and results for justificatory chain (top) and negative impacts (bottom).}
\centering
\small
\begin{tabular}{lc}
\toprule
\textbf{Justificatory Chain Condition} & \textbf{\% of Papers} \\ \midrule
Doesn't rigorously justify how it achieves technical goal & 1\% \\
Justifies how it achieves technical goal but no mention of societal need & 71\% \\
States but does not justify how it connects to a societal need & 16\% \\
States and somewhat justifies how it connects to a societal need & 9\% \\
States and rigorously justifies how it connects to a a societal need & 3\% \\ \bottomrule
\toprule
\textbf{Negative Impacts Condition} & \textbf{\% of Papers} \\ \midrule
Doesn't mention negative potential & 98\% \\
Mentions but does not discuss negative potential \hspace{18 mm} & 1\% \\
Discusses negative potential & 1\% \\
Deepens our understanding of negative potential & 0\% \\ \bottomrule
\end{tabular}
\hfill

\label{fig:just_chain}
\end{table}

\section{Qualitative Analysis of Justifications and Negative Potential}


\subsection{Justificatory Chain}

Papers typically motivate their projects by appealing to the needs of the ML research community and rarely mention potential societal benefits. Research-driven needs of the ML community include researcher understanding (e.g., understanding the effect of pre-training on performance/robustness, theoretically understanding multi-layer networks, or understanding adversarial vulnerabilities) as well as more practical research problems (e.g., improving efficiency of models for large datasets, creating a new benchmark for NLP tasks). Some papers do appeal to needs of  broader society, such as building models with realistic assumptions, catering to more languages, or using ``real user'' documents to solve ``learning problems''. 

However, even when societal needs are mentioned as part of the justification of the project, the connection is often loose. Almost no papers explain how their project is meant to promote a social need they identify by giving the kind of rigorous justification that is expected of technical contributions; it is usually unclear whether and how the paper promotes the social needs that it mentions. The connection to societal needs, even when present, is thus typically superficial. For example, a paper on a predefined ML problem may justify its importance by mentioning that such algorithms are useful for real-world learning tasks, such as object or digit recognition, or a paper on a new architecture may justify its importance by discussing previous work on similar architectures that were able to achieve high accuracy.

The superficial nature of the connection between societal needs and the content of the paper also manifests in the fact that the societal needs, or the applicability to the real world, is often only discussed in the beginning of the papers. From papers that mention applicability to the real world, the vast majority of mentions are in the Introduction section, and applicability is rarely engaged with afterwards. Papers tend to introduce the problem as useful for applications in object detection or text classification, for example, but rarely justify why an application is worth contributing to, or revisit how they particularly contribute to an application as their result.

\subsection{Negative Potential}
\label{negative potential}

Two of the 100 papers discussed potential harms, whereas the remaining 98 did not mention them at all. The lack of discussion of potential harms is especially striking for papers which deal with socially contentious application areas, such as surveillance and misinformation. For example, the annotated corpus includes a paper advancing the identification of people in images, a paper advancing face-swapping, and a paper advancing video synthesis. These papers contained no mention of the well-studied negative potential of facial surveillance, DeepFakes,~\footnote{DeepFakes present a real threat, particularly to women as as their use encompasses pornography, among others. According to~\cite{ajder2019state}, for example, of all the DeepFake videos found online, 96\% of it consists of DeepFake pornographic videos, with over 134 million of views in the top four DeepFake pornographic websites.} or misleading videos, respectively.

Furthermore, among the two papers that do mention negative potential, the discussions were mostly abstract and hypothetical, rather than grounded in the negative potential of their specific contributions. For example, authors may acknowledge "possible unwanted social biases" when applying the model to a real-world setting, without discussing how social stereotypes can be encoded in the authors' proposed model. These observations correspond to a larger trend in the ML/AI research community of neglecting to discuss aspects of the work that are not strictly ``positive''. 

Although a plethora of work exists on sources of harm that can arise in relation to ML~\cite{buolamwini2018gender, green2019good, suresh2019framework, hill2020accused}, our analysis shows that these discussions are absent for the most part in the most influential ML/AI publications; authors rarely dive deeper than brief mentions of relevant applications, let alone analyzing how their own work might contribute to this. These papers, to a large extent, are inward-looking, prioritize narrowly defined goals of the ML research community and fail to both contextualize work within the larger societal needs and consider negative societal impacts.

\section{Stated Values}
\label{sec:values}

The dominant values in ML research, e.g., accuracy and efficiency, may seem purely technical. However, the following analysis of several of these values shows how they can become politically loaded in the process of prioritizing and operationalizing them: sensitivity to the way that they are operationalized, and to the fact that they are uplifted at all, reveals value-laden assumptions that are often taken for granted that may negatively impact marginalized members of society. The dominant conceptualization of ML/AI (tools, objectives, goals, practices and so on) tends to be marked by optimism, advanced progress, innovation and in some cases as socially beneficial but rarely as value-laden and politically loaded. Yet,
philosophers of science and STS scholars have been working to understand how scientific endeavours have underlying values. For example, Thomas Kuhn~\citep{kuhn.1977} presented a list of five scientific values which he deems as important in scientific research (accuracy, consistency, scope, simplicity, and fruitfulness). Helen Longino~\citep{longino.1996} and others have argued that prominent values are politically loaded, focusing mostly on how some of these values function in disciplines such as biology and social sciences. However, "technical" values, such as accuracy, are often left out of this type of critical discussion. The findings from our analysis show that even the "technical" values aren't politically neutral, and it does so in the context of machine learning, which is often conceived as a less politically loaded discipline than biology or social sciences, further challenging a politically neutral conception of the top values in machine learning research. It is worth emphasizing that values once held to be intrinsic, obvious, or definitional have been in many cases transformed over time.

To provide a sense of what the values look like in context, Tables~\ref{tab:performance},~\ref{tab:generalization}, and~\ref{tab:efficiency} present three randomly selected examples of sentences annotated for each value discussed below (see the paper~\cite{birhane2021values} for extensive additional examples). Note that most sentences are annotated with multiple values. 

\subsection{Performance}
\label{sec:performance}

\begin{table}
\caption{Random examples of \textit{performance}, the most common emergent value.}
\label{tab:performance}
\small
\begin{tabular}{p{.96\linewidth}}
\midrule "Our model significantly outperforms SVM’s, and it also outperforms convolutional neural nets when given additional unlabeled data produced by small translations of the training images."\\
\midrule "We show in simulations on synthetic examples and on the IEDB MHC-I binding dataset, that our approach outperforms well-known convex methods for multi-task learning, as well as related non-convex methods dedicated to the same problem."\\
\midrule "Furthermore, the learning accuracy and performance of our LGP approach will be compared with other important standard methods in Section 4, e.g., LWPR [8], standard GPR [1], sparse online Gaussian process regression (OGP) [5] and $\upsilon$-support vector regression ($\upsilon$-SVR) [11], respectively."\\ 
\midrule "In addition to having theoretically sound grounds, the proposed method also outperformed state-of-the-art methods in two experiments with real data."\\
\midrule "We prove that unlabeled data bridges this gap: a simple semisupervised learning procedure (self-training) achieves high robust accuracy using the same number of labels required for achieving high standard accuracy."
\\
 \midrule
"Experiments show that PointCNN achieves on par or better performance than state-of-the-art methods on multiple challenging benchmark datasets and tasks."
 \\ 
 \midrule 
"Despite its impressive empirical performance, NAS is computationally expensive and time consuming, e.g. Zoph et al. (2018) use 450 GPUs for 3-4 days (i.e. 32,400-43,200 GPU hours)."
\\ 
 \midrule 
"However, it is worth examining why this combination of priors results in superior performance."
 \\ 
 \midrule 
 "In comparisons with a number of prior HRL methods, we find that our approach substantially outperforms previous state-of-the-art techniques."
 \\ 
 \midrule 
 "Our proposed method addresses these issues, and greatly outperforms the current state of the art."
 \\
\bottomrule
\end{tabular}
\hfill

\label{fig}
\end{table}

Performance, accuracy, and achieving SOTA form the most common cluster of related values in annotated papers. While it might seem intrinsic for the field to care about performance, it is important to remember that models are not simply "well-performing" or "accurate" in the abstract but always in relation to and as \textit{quantified} by some narrow metric on some specific dataset. Examining prevalent choices of operationalization reveals both the narrow use of these terms as well as the political aspects of performance values. 

First, we find performance values are consistently and unquestioningly operationalized as correctness averaged across individual predictions, giving equal weight to each instance. However, choosing
equal weights when averaging is a value-laden move which might deprioritize those underrepresentated in the data or the world, as well as societal and evaluee needs and preferences. Extensive research in ML fairness and related fields has considered alternatives, but we found no such discussions among the papers we examined. 

Second, many papers choose to use the same datasets used by previous papers, often without any explanation. Some datsets, such as MNIST, have been used innumerable times since their creation.
The choice to use same datasets (and metrics) are often driven by the desire to demonstrate improvement over a previous baseline (see also Novelty Section~\ref{sec:building_novelty}). Another common justification for using a certain dataset is applicability to the ``real world''. How one characterizes the ``real world'' is a value-laden endeavour. One common assumption is that very large datasets capture the ``real world''. However, from sourcing, curating and managing large datasets, the process is power centralizing where those who have the resources not only procure data, often without consent or awareness of the data subject (see Chapter~\ref{chapter:computer vision} for more), but also make key decisions on how the ``real world'' is represented in datasets. 

Further overlooked assumptions include that the real world is binary or discrete, and that datasets come with a predefined ground-truth label where the true label supposedly exists ``out there'' independent of those carving it out, defining and labelling it~\cite{gitelman2013raw}. This contrasts against marginalized scholars’ calls for ML models that allow for non-binaries, plural truths, contextual truths, and many ways of being~\cite{costanza2018design, hamidi2018gender, lewis2020indigenous}. For example, one paper examining \textit{Feature Hashing for Large Scale Multitask Learning} states that it validates its results ``on a real-world application within the context of spam filtering''. The case of spam classification of interest in this case involves ``hundreds of thousands of users collectively label[ing] emails as spam or not spam''. They further specify that the number of users ``can be very large from ``systems such as Yahoo Mail or Gmail''. Thus, the "real world" in this case comprises of large datasets, of the kind we may find at Yahoo or Google. 

The prioritization of performance values requires scrutiny. Valuing these properties is so entrenched in the field that generic success terms, such as "success", "progress", or "improvement" are often used as synonyms for performance and accuracy. However, one might alternatively invoke generic success to mean increasingly safe, consensual, or participatory ML that reckons with impacted communities and the environment. In fact, "performance" itself is a general success term that could have been associated with properties other than accuracy and SOTA. 

Finally, not only does the way performance is operationalized lacks coherence, the annotated papers rarely connect performance to notions such as human intelligence, cognition, learning or problem solving that can be associated with the larger historical goals envisioned decades ago. Performance operationalized improvements in narrowly specified quantitative results, often on benchmark datasets without rigorous justification or connection to larger visions. 



\begin{table}
\caption{Random examples of \textit{generalization}, the third most common emergent value.}
\label{tab:generalization}
\small
\begin{tabular}
{p{0.96\linewidth}}
\midrule "The range of applications that come with generative models are vast, where audio synthesis [55] and semi-supervised classification [38, 31, 44] are examples hereof."\\
\midrule "Furthermore, the infinite limit could conceivably make sense in deep learning, since over-parametrization seems to help optimization a lot and doesn’t hurt generalization much [Zhang et al., 2017]: deep neural nets with millions of parameters work well even for datasets with 50k training examples."\\
\midrule "Combining the optimization and generalization results, we uncover a broad class of learnable functions, including linear functions, two-layer neural networks with polynomial activation $\phi(z) = z^{2l}$ or cosine activation, etc."\\
\midrule "We can apply the proposed method to solve regularized least square problems, which have the loss function $(1 - y_i\omega^T x_i)^2$ in (1)."\\
\midrule "The result is a generalized deflation procedure that typically outperforms more standard techniques on real-world datasets."\\
\midrule "Our proposed invariance measure is broadly applicable to evaluating many deep learning algorithms for many tasks, but the present paper will focus on two different algorithms applied to computer vision."\\
\midrule "We show how both multitask learning and semi-supervised learning improve the generalization of the shared tasks, resulting in state-of-the-art performance." \\
\midrule "We have also demonstrated that the proposed model is able to generalize much better than LDA in terms of both the log-probability on held-out documents and the retrieval accuracy." \\
\midrule "We define a rather general convolutional network architecture and describe its application to many well known NLP tasks including part-of-speech tagging, chunking, named-entity recognition, learning a language modeland the task of semantic role-labeling" \\
\midrule "We demonstrate our algorithm on multiple datasets and show that it outperforms relevant baselines."\\

\bottomrule
\end{tabular}
\hfill

\label{fig}
\end{table}

\subsection{Generalization}
\label{sec:generalization}

A common way of appraising the merits of one's work in ML is to claim that it generalizes well. However, it is important to clarify in what sense \textit{generalization} is understood in the 100 papers annotated. Within the main AI discourse, the term often implies that a system is capable of carrying out various tasks (vision, hearing, language, for example) as well as being able to integrate these tasks or that a certain method can be used in distinct application areas or domains. Far from this, the papers analyzed in this work used generalization in a very narrow manner, for example, to refer to avoiding train/test discrepancy. They typically understood generalization in terms of performance or accuracy: a model generalizes when it achieves good performance on a range of samples or datasets, for example taking the nearby dataset and applying their method there (see~\ref{tab:generalization} for randomly selected examples of generalization). 
Uplifting generalization, in the narrow sense understood here, raises two kinds of questions. First, which datasets, domains, or applications show that the model generalizes well? Typically, a paper shows that a model generalizes by showing that it performs well on multiple tasks or datasets. However, they typically do not explain why the dataset or tasks were chosen, or if there are reasons to believe the results will generalize to any further datasets or tasks. 

For example, one paper states: ``In addition, our method works for other input video formats such as face sketches and body poses, enabling many applications from face swapping to human motion transfer.`` But why should the focus be on the tasks such as face swapping and human motion? Is this range of tasks wide enough to show generalizability in the broader sense? Why are these tasks important to consider? We find that such questions are almost never discussed. In particular, there is almost never a discussion of whether the model will generalize in a way that will serve broader societal needs. 

Second, uplifting generalization itself reveals substantive assumptions. The prizing of generalization means that there is an incentive to harvest many datasets from a variety of domains, and to treat these as the only datasets that matter for that space of problems. Generalization thus prioritizes distilling every scenario down to a common set of representations or affordances, rather than treating each setting as unique, inexhaustible and not something that can be measured/datafied. Critical scholars have advocated for valuing \emph{context}, which stands at the opposite side of striving for generalization~\citep{d2020data}. Others have argued that this kind of totalizing lens (in which model developers have unlimited power to determine how the world is represented) leads to \emph{representational} harms, due to applying a single representational framework to everything~\citep{crawford.2017,abbasi.2019}. 

Finally, the belief that generalization is even possible implicitly assumes that whatever is learned from a given sample should be applied to others, often in the future. Using data at a single point in time to learn a model that will be applied in the future is a fundamentally conservative approach which assumes that the future will be sufficiently similar to the past. When used in the context of ML, the assumption that the future resembles the past is often problematic as past societal stereotypes and injustice can be encoded in the process as we have see in Chapter~\ref{chp:automating}. To the extent that predictions are performative~\cite{perdomo2020performative}, especially predictions that are enacted, those ML models which are deployed to the world will contribute to shaping social patterns. No papers attempt to counteract this quality or acknowledge its presence.

\begin{table}
\caption{Random examples of \textit{efficiency}, the fourth most common emergent value.}
\label{tab:efficiency}
\begin{small}
\begin{tabular}{p{0.96\linewidth}}
\midrule "Our model allows for controllable yet efficient generation of an entire news article – not just the body, but also the title, news source, publication date, and author list."\\
\midrule "We show that Bayesian PMF models can be eﬃciently trained using Markov chain Monte Carlo methods by applying them to the Netﬂix dataset, which consists of over 100 million movie ratings."\\
\midrule "In particular, our EfficientNet-B7 surpasses the best existing GPipe accuracy (Huang et al., 2018), but using 8.4x fewer parameters and running 6.1x faster on inference."\\
\midrule "Our method improves over both online and batch methods and learns faster on a dozen NLP datasets."\\
\midrule “We describe efficient algorithms for projecting a vector onto the $\ell$1-ball.”\\
\midrule "Approximation of this prior structure through simple, efficient hyperparameter optimization steps is sufficient to achieve these performance gains."\\
\midrule "We have developed a new distributed agent IMPALA (Importance Weighted Actor-Learner Architecture) that not only uses resources more efficiently in single-machine training but also scales to thousands of machines without sacrificing data efficiency or resource utilisation."\\
\midrule "In this paper we propose a simple and efficient algorithm SVP (Singular Value Projection) based on the projected gradient algorithm"\\
\midrule "We give an exact and efficient dynamic programming algorithm to compute CNTKs for ReLU activation." \\
\midrule "In contrast, our proposed algorithm has strong bounds, requires no extra work for enforcing positive definiteness, and can be implemented efficiently." \\

\bottomrule
\end{tabular}
\hfill

\end{small}
\label{fig}
\end{table}

\subsection{Efficiency}
\label{sec:efficiency}

Efficiency is another common value in ML research. Abstractly, saying that a model is efficient typically means saying that the model uses less of some resource, such as time, memory, energy, or number of labeled examples (see~\ref{sec:efficiency} for randomly selected example of Efficiency). 
In practice however, efficiency is commonly referenced to imply the ability to scale up: a more efficient inference method allows one to do inference in much larger models or on larger datasets, using the same amount of resources. For example, one paper notes: "We have developed a new distributed agent [...] that not only uses resources more efficiently in single-machine training but also scales to thousands of machines without sacrificing data efficiency or resource utilisation." 
This is reflected in our value annotations, where 72\% of papers mention valuing efficiency, but only 14\% of those value requiring \textit{few} resources. In this way, valuing efficiency facilitates and encourages the most powerful actors with the most resources and compute power to scale up their computation to ever higher orders of magnitude, making their models even less accessible to those without resources to use them and decreasing the ability to compete with them. Alternative usages of efficiency could encode accessibility instead of scalability, aiming to create more equitable conditions for ML research.

\begin{table}
\caption{Random examples of \emph{building on past work} and \emph{novelty}, the second and sixth most common emergent values, respectively.}
\label{tab:building_novelity}
\begin{small}
\begin{tabular}{p{0.96\linewidth}}
\toprule
\textbf{Building on past work}\\
\midrule "Recent work points towards sample complexity as a possible reason for the small gains in robustness: Schmidt et al. [41] show that in a simple model, learning a classifier with non-trivial adversarially robust accuracy requires substantially more samples than achieving good `standard' accuracy."\\
\midrule "Experiments indicate that our method is much faster than state of the art solvers such as Pegasos, TRON, SVMperf, and a recent primal coordinate descent implementation."\\
\midrule "There is a large literature on GP (response surface) optimization."\\
\midrule "In a recent breakthrough, Recht et al. [24] gave the first nontrivial results for the problem obtaining guaranteed rank minimization for affine transformations A that satisfy a restricted isometry property (RIP)."\\
\midrule "In this paper, we combine the basic idea behind both approaches, i.e., LWPR and GPR, attempting to get as close as possible to the speed of local learning while having a comparable accuracy to Gaussian process regression"\\

\bottomrule
\toprule
\textbf{Novelty} \\
\midrule "In this paper, we propose a video-to-video synthesis approach under the generative adversarial learning framework."\\
\midrule "Third, we propose a novel method for the listwise approach, which we call ListMLE."\\
\midrule "The distinguishing feature of our work is the use of Markov chain Monte Carlo (MCMC) methods for approximate inference in this model."\\
\midrule "To our knowledge, this is the first attack algorithm proposed for this threat model."\\
\midrule "Here, we focus on a different type of structure, namely output sparsity, which is not addressed in previous work."\\

\bottomrule
\end{tabular}
\hfill

\end{small}
\label{fig:novelty}
\end{table}



\subsection{Novelty and Building on Past Work}
\label{sec:building_novelty}

Most authors devote space in the introduction to positioning their paper in relation to past work --- which is most always narrow and technical and rarely work that is broad, critical, or outside of ML --- and describing what is novel.
Mentioning past work serves to signal awareness of related publications of similar specific lineage, 
to establish the new work as relevant to the community, and to provide the basis upon which to make claims about what is new. 
Novelty is sometimes suggested implicitly (e.g., "we develop" or "we propose"), but frequently it is emphasized explicitly (e.g. "a new algorithm" or "a novel approach"), sometimes contextualized with the phrase ``to the best of our knowledge'', as a hedge against the possibility that the idea is not, in fact, completely novel (see~\ref{tab:building_novelity} for randomly selected examples of both Novelty and Building on Past Work). The emphasis on novelty is common across many academic fields~\citep{trapido.2015,vinkers.2015}, but for the top ML conferences, it seems that particular kinds of novelty are implicitly valued the most. The highly-cited papers we examined mostly tend to emphasize the novelty of their proposed method or of their theoretical result. This is not surprising, given that part of the (implicit) criteria for paper acceptance is that it involves some amount of novelty. Very few uplifted their paper on the basis of applying an existing method to a novel domain, or for providing a novel philosophical argument, synthesis, for providing a novel critique of past work, or work outside ML (or ML adjacent fields). 


In other words, while claims may be primarily theoretical or empirical (often providing some of each), in both cases, authors are primarily seeking to establish themselves with respect to the agreed-upon norms of the research community. As such, even though ML is being widely used in real settings, the research tends to be relatively divorced from this aspect, preferring to play within the space of research problems. And while this is not surprising (the real world tends to be much more complicated that experiments, introducing additional complexity and limiting one's ability to cleanly draw conclusions), it means that ML researchers have largely been able to avoid engaging with the more complicated aspects of the real world application of ML, leading to a relative absence of references to scholarly work outside of nearby technical domains, little engagement with the social and ethical questions raised by ML work, and often no more than than rough sketches of the sorts of applications that might be pursued (e.g., ``identifying objects in images'', without further comment).

This combined focus on novelty and building on recent work is often presented as a way to establishing a continuity of ideas, and might be expected to contribute to the self-correcting nature of science~\citep{merton.1973}. However, this is not always the case~\citep{ioannidis.2012} as the type of work most papers build on is extremely narrow in scope -- a specific tool, experiment, or theoretical results. Additionally, attention to the ways novelty and building on past work are implemented reveals value commitments. Among the most-cited papers in the field, we find a clear emphasis on technical novelty, rather than critique of past work, or demonstration of measurable progress on societal problems, as has previously been observed~\citep{wagstaff.2012}. Although introductions sometimes point out limitations of past work (so as to further emphasize the contributions of their own paper), they are rarely explicitly critical of other papers in terms of methods or goals. Indeed, papers uncritically reuse the same datasets for years or decades to benchmark their algorithms, even if those datasets fail to represent more realistic contexts in which their algorithms will be used~\cite{bender2021dangers}, and in some cases even when the datasets have been retired by the dataset curators due to ethical issues~\cite{peng2021mitigating}. 

Novelty is denied to work that rectifies socially harmful aspects of existing datasets in tandem with strong pressure to benchmark on them and thereby perpetuate their use, enforcing a fundamentally conservative bent to ML research. In summary, the ML research community values novelty, as is common in other fields, but most strongly rewards work which builds on and enables technical novelty within ML research itself, rather than seeking to establish the value of the work externally, or reflecting critically on the past.


\begin{figure}
\centering
\includegraphics[width=\linewidth]{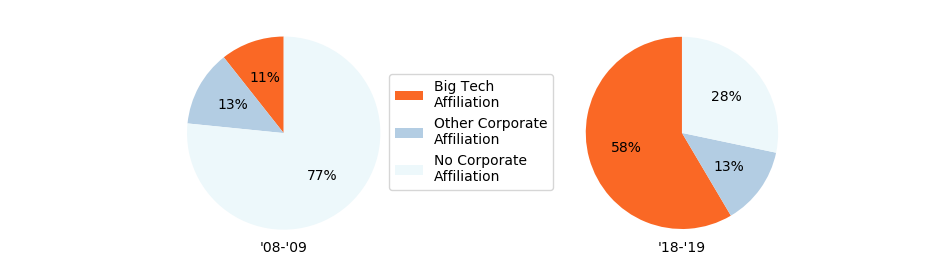}
\caption{Corporate and Big Tech author affiliations. 
}
\label{fig:corp_affils}
\end{figure}
\section{Corporate Affiliations and Funding}
\label{sec:corporate}

Although AI research, from its early conceptions, has been backed by powerful agents and institutions, such as the US military, over the past decade AI research has come to be dominated by big corporations. Top researchers of the field are continually hunted and ``acqui-hired''~\cite{mitchell2019artificial} by big tech corporations. 

The interests and values of powerful agents within the ML landscape, manifest (directly or indirectly) in which values the field appraises and which are cast aside. Over the past decade, technology corporations have emerged as the most powerful industry, overtaking traditionally leading industries such as the automotive or beverages sectors. According to the 2020 Forbes ranking of \textit{The World's Most Valuable Brands}~\cite{forbes2020}, the top 5 positions are held by big technology corporations: Apple, Google, Microsoft, Amazon, and Facebook. The rise of these corporations is partly fueled by the acceleration of ML and AI tools over the past decade. Big tech corporations and AI research, thus, are intimately linked. 

The rise of deep learning, which requires access to intensive compute power, has created the ``compute divide'' --- a divergence between big tech/elite universities and non-elite universities. The uneven access to compute power gives rise to an ecosystem where big tech and elite Western universities have increased presence in top conferences --- crowding out non-elite universities --- and presents an obstacle towards ``democratizing'' (allowing diverse participation in) AI research~\cite{ahmed2020democratization}. 

Corporate presence (authoring papers or funding research) in academic conferences is not new, and corporate participation rate, both in AI and non-AI conference was relatively static prior to 2012~\cite{ahmed2020democratization}.  However, this presence has seen a dramatic increase over the past decade, more particularly in AI conferences. Ahmed and Wahed~\cite{ahmed2020democratization} associate this rapid increase after 2012 to the emergence of ImageNet. The recent blooming of AI research is inseparable from the unprecedented influence that big tech has come to exert in the global wealth landscape, creating shared values marked by the increased presence of big tech in AI research. Analysis of the 100 most highly cited papers shows a substantive and increasing corporate presence in the field. In 2008/09, 24\% of the top cited papers had corporate affiliated authors, and in 2018/19 this statistic almost tripled,
to 71\%. Furthermore, we also find a much greater concentration of a few large tech firms, such as Google and Microsoft, with the presence of these "big tech" firms (as identified in~\citep{ahmed2020democratization})
increasing more than fivefold, from 11\% to 58\% (see Figure~\ref{fig:corp_affils}). The fraction of the annotated papers with corporate ties, by author affiliation or funding, dramatically increased from 43\% in 2008/09 to 79\% in 2018/19.

\begin{figure*}
\centering
\captionsetup{justification=centering}
\includegraphics[width=\linewidth]{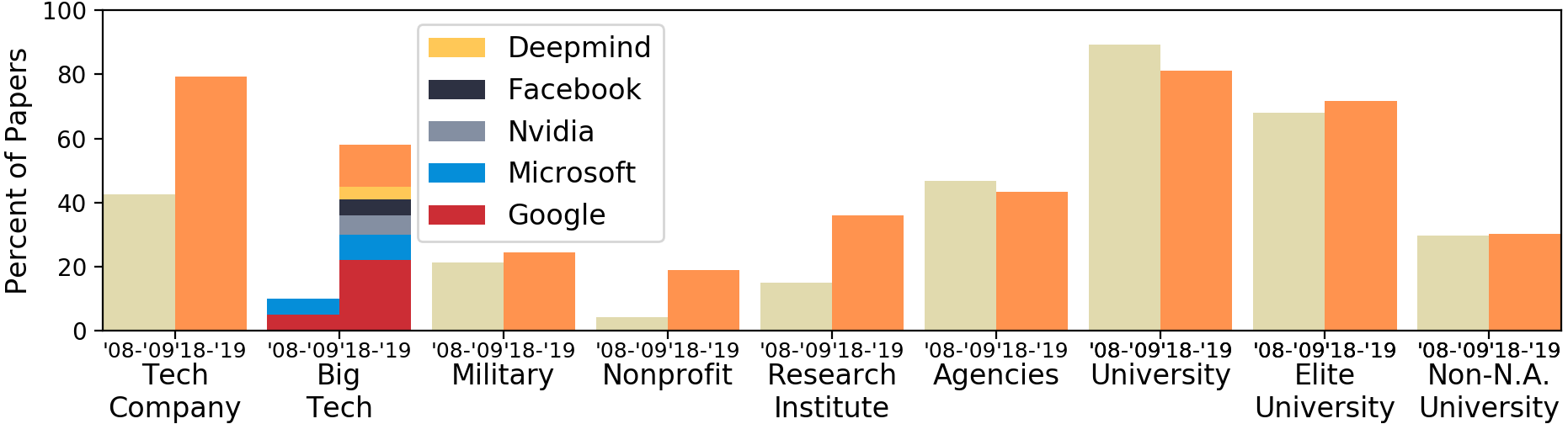}
\caption{Corporate affiliations and funding ties. Non-N.A. Universities are those outside the U.S. and Canada.}

\label{fig:vcorp-ties}
\end{figure*}

As ML research has become more resource intensive, it has given rise to an ecosystem where those with access to compute power --- big tech and elite universities --- dominate top conferences, simultaneously crowding out non-elite universities and further concentrating power in the hands of the few. Consistent with this, we found paramount domination of elite universities in our analysis as shown in Figure~\ref{fig:vcorp-ties}. Of the total papers with university affiliations, we found 82\% were from elite universities (defined as the top 50 universities by QS World  University Rankings, following past work~\cite{ahmed2020democratization}). ML research, thus, widens the gap in \textit{who} is increasingly present (and represents the field), further pushing out non-elite universities and eliminating diverse participation.

These findings are consistent with previous work indicating a pronounced corporate presence in ML research. In an automated analysis of peer-reviewed papers from 57 major computer science conferences, Ahmed and Wahed~\cite{ahmed2020democratization} show that the share of papers that have at least one corporate affiliated co-author increased from 10\% in 2005 for both ICML and NeurIPS to 30\% and 35\% respectively in 2019. Our analysis shows that corporate presence is even more pronounced in those papers from ICML and NeurIPS that end up receiving the most citations.

Highly regarded values within the tech industry play a key role in actively shaping ML research. 
The values of big tech influence academia in various ways, from the type of public image portrayed of the industry, to the type of research questions pursued, to impacting the plans of individual scientists. The influence of powerful players in ML research is consistent with field-wide value commitments that centralize power. Others have also argued for causal connections. For example, Abdalla and Abdalla~\cite{abdalla2020grey} argue that big tech sway and influence academic and public discourse using strategies which closely resemble strategies used by Big Tobacco. The intimate links between ML research and the generally conservative, power centralizing, and profit maximizing values of big tech are often obfuscated and can take, for instance, the form of high efficiency or beating SOTA performance, as we have seen in this study. However, occasionally, we find these values displayed overtly; Google's recent request that its researchers refrain from ``casting its technology in a negative light''~\cite{reuters2020} signifies the negligible place that equity, justice, and user rights hold within the corporation. This echoes the trend in the ML community of neglecting discussion of potential harms as observed in section \ref{negative potential}.

Examining the prevalent values of big tech, critiques have repeatedly pointed out that objectives such as efficiency, scale, and wealth accumulation~\cite{o2016weapons,pasquale2015black,hanna2020against} drive the industry at large, often at the expense of individual rights, respect for persons, consideration of negative impacts, beneficence, and justice. Thus, the top stated values of ML presented in this chapter such as performance, efficiency, and novelty may not only enable and facilitate the realization of big tech's objectives, but also suppress values such as beneficence, justice, and inclusion. A "state-of-the-art" large image dataset, for example, is instrumental for large scale models, further benefiting ML researchers and big tech in possession of huge compute power and the capacity to scrape large amounts of data from the web. A large image dataset that considers negative consequences and is built on the foundations of individual rights and respect for persons, on the other hand, is one that would start with gaining informed consent from the data subject and is considerate of contextual norms over scalability~\cite{hanna2020against}. In the current climate where values such as accuracy, efficiency, and scale, as currently defined, are a priority, user safety, informed consent, or participation may be perceived as costly and time consuming, evading social needs.     

Subsequently, ML researchers have to some extent abdicated the responsibility to think through the more difficult questions surrounding how ML can or should be deployed. By emphasizing the empirical gains in limited domains, papers have directly promoted the relevance and unproblematic applicability of their research to ``real'' problems, leaving well-funded corporate labs free to continue extending this work, in tandem with real world deployments (e.g., in online advertising, behavioural modification, user profiling, etc.), without any concomitant need to engage with the potential harms such work creates as they continue to publish in the top ML venues.

\section{Discussion and Conclusion}


ML research is often perceived as value-neutral, and emphasis is placed on positive applications or potential. This fits into a historical strain of thinking which has tended to frame technology as "neutral", based on the notion that new technologies can be unpredictably applied for both beneficial and harmful purposes~\citep{winner.1977}. Ironically, this claim of neutrality frequently serves as an insulation from critiques of AI and as a permission to emphasize the benefits of AI~\citep{rus.2018, weizenbaum1972impact}. Although it is rare to see anyone explicitly argue in print that ML is neutral, related ideas are part of contemporary conversation, including these canonical claims: long term impacts are too difficult to predict; sociological impacts are outside the expertise or purview of ML researchers~\citep{holstein.2019}; critiques of AI are really misdirected critiques of those deploying AI with bad data ("garbage in, garbage out"),  again outside the purview of many AI researchers; and proposals such as broader impact statements represent merely a "bureaucratic constraint"~\citep{abuhamad.2020}. A recent qualitative analysis of required broader impact statements from NeurIPS 2020 similarly observed that these statements leaned towards positive consequences (often mentioning negative consequences only briefly and in some cases not at all), emphasized uncertainty about how a technology might be used, or simply omitted any discussion of societal consequences altogether~\citep{nanayakkara.2021}.

Furthermore, as we have seen, the most prized values of the field including \textit{effective}, \textit{successful}, \textit{improvement}, \textit{progress}, \textit{performance}, and \textit{generalization} are narrowly defined in a technical manner where no attempts to connect them with broader conceptions is made. The narrow, technical, and single approach towards these values in effect is not only reductionist for it fails to recognize multiple accounts, it also assumes this narrow and single view represents the \textit{view from nowhere}. In addition, we see that these values are often treated as obvious signifiers of greatness, positive sentiment, and progress; so obvious, in fact, that it need not ever be stated explicitly. 

Nonetheless, as we have seen throughout this thesis (and more particularly in Chapter \ref{chp:automating}, the idea that ML is objective or value-neutral rests on misconceptions and misunderstandings of both ML/AI as well as the scientific endeavour. In line with a long tradition of work in STS and other similar critical approaches, this work is presented, in part to expose the contingency of the present state of the field, which could be otherwise. For individuals, communities, and institutions wading through difficult-to-pin-down values of the field, as well as those striving toward alternative values, it is a useful tool to have a characterization of the way the field is now, for understanding, shaping, dismantling, or transforming what is, and for
articulating and bringing about alternative visions.

As with all methods, the chosen approach in this work — coding important sections of highly-cited papers — has limitations. Most notably, this approach requires human expertise and does not automatically scale or generalize to other data, which limits our ability to draw strong conclusions about other conferences, different years, or the field in general. Similarly, this approach is less reproducible than fully automated approaches, and for both our final list of values and specific annotation of individual sentences, different researchers might make somewhat different choices. However, given the overwhelming presence of certain values, the high agreement rate among annotators, and the similarity of observations made by the team, we strongly believe other researchers following a similar approach would reach similar conclusions about what values are most frequently uplifted by the most influential papers in this field. Lastly, we cannot claim to have identified every relevant value in ML. However, by including important ethical values identified by past work, and specifically looking for these, we can confidently assert their relative absence in this set of papers, which captures an important aspect of influential work in ML.

Finally, this work illustrates why general statements such as ``building AI that benefits society'' or ``AI that embeds human values'' can be vacuous without breaking down the cost-benefit analysis for the various stakeholders involved in the AI pipeline and \textit{which} (and whose) human values we are referring to. AI certainly holds numerous benefits. And as we have seen in this Chapter, intimate link with big corporations puts ML research and and its benefits squarely with those producing, controlling, distributing, and monopolizing AI compared to other stakeholders, more particularly those at the margins of society who are subjected to live in a society where AI systems are integrated. The generic term \textit{human values}, similarly risks representing the values of only the privileged few producing AI while under the disguise of a view from nowhere. 

    \chapter{All Models are Wrong, Some are Dangerous}
\label{chp:all_models}

In Chapter \ref{chapter:computer vision}, we saw that computer vision, one of the sub-fields of AI, is currently seeing outstanding ``progress'' and success. Behind this success is the ``availability'' of massive amounts of data from the web. And this ``success'' comes at great cost at various steps of the pipeline, from data sourcing to the downstream effects of problematic models. These include broad issues such as the question of data sourcing sans consent and awareness of the data subject; erosion of privacy; and specific concerns such as the inclusion of verifiably pornographic images in datasets. The creation of LSVDs in the absence of sufficient foresight not only is potentially harmful in and of itself, but also models trained and validated on such dataset present a real danger, especially with the rise of sinister applications in computer vision. Object, face, emotion, and gait recognition, for example, mark the grounds for the emergence of problematic models that are resuscitating long discredited pseudoscience in application areas including ``prediction'' of gender, criminality, and emotion. This chapter looks at such dangerous applications and reviews the troublesome paths of computational sciences. 

\section{Modelling Complex Systems}
\label{sec:modelling}
As we have seen in Chapter~\ref{chp:automating}, people's behaviours and actions as well as the wider social system that they are embedded in constitute complex adaptive systems: they are \textit{open, non-linear, stochastic}, and \textit{context} and \textit{time} dependent.  
They are inextricably interlinked with their environment. Their non-linear interactions and dynamical relationships create emergent properties that are not the simple sum of components of a system. As their interactions are non-linear, there is no simple way of tracking clear causal chains and neat pairing of cause and effect. There is no single perfect or accurate representation or model for a given complex system. Accurate or perfect representation entails capturing a system (its dynamic and non-linear interactions and emerging properties) in its entirety without leaving something out. However, compressing a system into an algorithm or a model without reducing the complexity is impossible~\cite{cilliers2002complexity}. This means that a perfect model of system would have to be as complex and unpredictable itself. As Cilliers \cite{cilliers2002complexity} reminds us, the best and simplest representation of a complex system is the system itself. In modelling a complex system, then, what aspects we want to capture and represent are partly tied to the observer/modeller's perspective, framing and objectives. For example, ``a portrait of a person, a store
mannequin, and a pig can all be models of a human being. None is a perfect copy of
a human, but each has certain aspects in common with a human. The painting gives
a description of what a particular person looks like; the mannequin wears clothes as a
person does; and the pig is alive. Which of the three models is “best” depends on how
we use the model—to remember old friends, to buy clothes, or to study biology''~\cite{blanchard2011differential}.

Complex systems are open systems that are far from equilibrium with no clear demarcations of what is inside and outside of the system. However, we must draw boundaries in order to model it. Having said that, it is important to remember there's nothing intrinsic or natural about the boundaries we \textit{create}. Since there is no way of accurately determining the boundary of a system, we can never be certain that we have taken enough factors into consideration or that the factors we deemed irrelevant are indeed so~\cite{cilliers2002complexity}. 
That there is no single accurate or perfect representation/model of a complex phenomena means that, in characterizing or modelling a complex system, we always have to make choices which are value laden, driven by our objectives, and are inherently ethical.
This may not matter so much when one is modelling less value laden complex phenomenon such as bird flocking or fish schooling. The type of modelling decisions we make and therefore the type of models we produce, in such cases, is unlikely to have the same downstream effect as when modelling human behaviour such as ``dishonesty'' or ``criminality''. 
For the former, in the worst case scenario, the outcome of ``wrong'' or reductive models is that our understanding of such systems are misguided or limited. However, when dealing with social phenomena, wrong or reductive models are not simply wrong but can also result in grave consequences for concrete people. 

More than ``biased'' models or incorrect characterization or representation, the ``success'' of computer vision has given rise to the discourse that with the right model and enough data, any question can be answered with ML regardless of the premises of the question. This line of argument particularly presents a danger when long discredited pseudoscience (such as physiognomy) or highly contested questions are reduced to a matter of ``state-of-the-art'' model building and a question of hyperparameter tuning and the ``right'' dataset. 

As we have seen in Chapter~\ref{chp:values}, values such as accuracy, performance, and SOTA have become the most predominant values in ML. One of the implication of this is that as the race becomes to beat the SOTA benchmark, questioning priors is seen as something outside the purview of the field. In this regard, face/emotion recognition systems, for the most part, have bypassed the bigger question --- the very idea of inferring inner states from outer appearances threads between discredited pseudoscience and dubious or shaky enquiry --- and are seen as a question of accuracy and improvement. And when one is deep in the quest to improve the accuracy of, for example, an emotion recognition system, one has already adopted the fallacious premise that emotions are something that can be read off faces. The long history and critical work demonstrating the goundleness of inferring inner states from outer appearance is then bypassed and ignored entirely. This partly arises due to the field's tendency to discard critical historical work from adjacent fields and to ignore its dark history. The next section briefly looks this.

\section{Collective Amnesia of the Dark Past}

Ruha Benjamin noted in a keynote\footnote{https://venturebeat.com/2020/04/29/ruha-benjamin-on-deep-learning-computational-depth-without-sociological-depth-is-superficial-learning/} at ICLR (2020) that computational depth without historic or sociological depth is ``superficial learning''. Nothing any researcher produces is ever completely new but instead is always contingent upon past work. Yet, current research culture, from paper review processes to funding structures, incentivises values such as novelty, accuracy and performance. The attitude towards older work appears in the reluctance to read and cite it. This is not an issue only for the AI/ML community but academic research in general, although it seems especially prevalent in AI/ML. As we lose connection to past work, we are prone to repeat past mistakes and reinvent the wheel. The dangers of this lack of awareness are most obvious and alarming in areas of study that involve quantification, measurement and modeling of human characteristics such as intelligence and emotion, and endeavours that generally attempt to infer inner behaviour from outer characteristics, given their dark history. 

Computational and cognitive sciences --- fields that both rely on computational methods to carry out research as well as engage in research of computation itself --- are built on a foundation of racism, sexism, colonialism, Anglo- and Euro-centrism, white supremacy, and all intersections thereof~\cite{Lugones2016, Crenshaw1990}.
This is very apparent when one examines the history of fields such as genetics, statistics, and psychology, which were historically engaged in refining and enacting eugenics~\cite{Winston2020, Syed2020, Saini2019, Roberts2020, Cave2020}.
``Great'' scientists were eugenicists, e.g., Alexander Graham Bell,
Cyril Burt,
Francis Galton,
Ronald Fisher,
Gregory Foster,
Karl Pearson,
Flinders Petrie,
and Marie Stopes
\citep{BernalLlanos2020}. Over the last one hundred years, while some progress has been made --- such as recently acknowledging the past in the case of University College Londons's legacy of eugenics~\cite{BernalLlanos2020} --- these fields maintain their historically oppressive characteristics.


Due to computational sciences' history --- especially the lack of institutional self-awareness, which protects hegemonic interests --- white and male supremacy continues to sneak (back) into even ostensibly sensible research areas.
For example, under the guise of a seemingly scientific endeavour, so-called ``race science'' or ``race realism'' conceals much of the last two centuries' white supremacy, racism, and eugenics~\citep{Saini2019}.
Despite a wealth of evidence directly discrediting this racist pseudoscience, race realism  --- the eugenic belief that human races have a biologically based hierarchy that supports racist claims of racial inferiority or superiority --- is currently experiencing a rebirth, chiefly aided by AI and ML \citep[e.g.,]{Arcas2017}. This manifests in computer vision datasets (racist and misogynist taxonomies~\cite{birhane2021large}) as well as models. OpenAI's CLIP suffers from this, for example: as outlined by the authors themselve, images belonging to the 'Black' racial designation face an approximately 14\% chance of being mis-categorized as \texttt{[‘animal’, ‘gorilla’, ‘chimpanzee’, ‘orangutan’, ‘thief’, ‘criminal’ and ‘suspicious person’]} in their FairFace dataset experiment~\cite{clipradford2021learning}.

\begin{figure}[ht!]
\setlength{\tabcolsep}{2pt}
\renewcommand{\arraystretch}{1}
\centering
\newcommand\x{0.315}

\begin{tabular}{lll}
\textsf{\textbf{a}) ground truth \vspace{1mm}} 
&   \textsf{\textbf{b}) blurred input} 
&  \textsf{\textbf{c}) output} \\

\includegraphics[width=\x\linewidth]{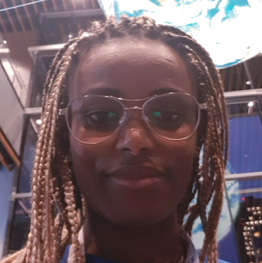}
 &  \includegraphics[width=\x\linewidth]{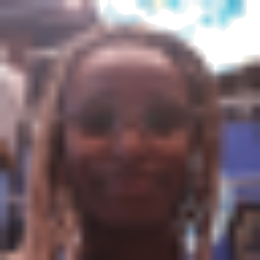}
 &  \includegraphics[width=\x\linewidth]{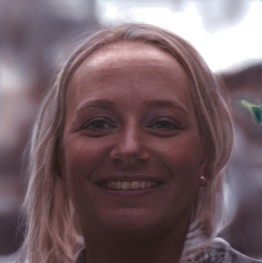} \\

  \includegraphics[width=\x\linewidth]{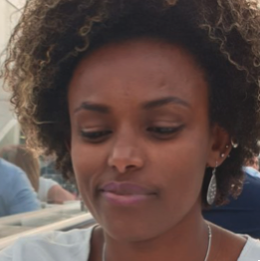}
 &  \includegraphics[width=\x\linewidth]{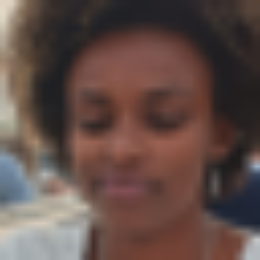}
 &  \includegraphics[width=\x\linewidth]{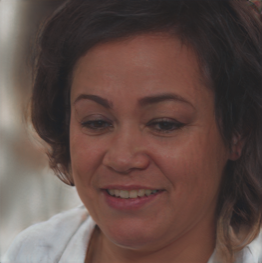}
 \\

  \includegraphics[width=\x\linewidth]{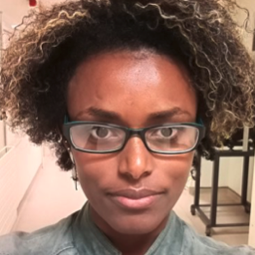}
 &  \includegraphics[width=\x\linewidth]{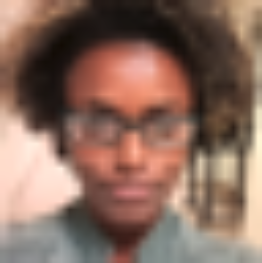}
 &  \includegraphics[width=\x\linewidth]{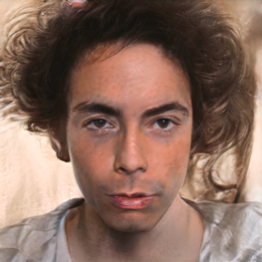}
 \\
\end{tabular}
\caption{Three examples of Abeba Birhane's face (column a) run through a depixeliser \citep{Menon2020}: input is column b and output is column c.
}
\label{fig:whiteface}
\end{figure}

Most current attempts to build intelligent systems hardly examine the history of intelligence research and the legacies of research on IQ and on race studies from the fields of statistics, genetics, and psychology \citep[e.g.,][]{BernalLlanos2020, Cell2020, Laland2020, Winston2020, Syed2020, prabhu2020large, Cave2020}.
Junk ``science'' from areas such as face research is revived and imbued with ``state-of-the-art'' machine learning models. 
This results in (at least partially) masquerading pseudoscience as science by use of vacuous and over-hyped technical jargon. This takes various forms and is hardly explicitly but manifests in often nuanced forms. Take a seemingly banal tool that depixelises images, for example (Figure~\ref{fig:whiteface}). When confronted with a Black woman's face, it ``corrects'' the Blackness and femininity. Here, the tool made the skin lighter, the nose slimmer, the lips thinner, and ``corrected'' the curly hair to straight hair. 

The downstream negative impact of such work is rarely considered and thus digitized pseudoscience is often packaged and deployed into high-stake decision-making processes, disproportionately impacting individuals and communities at the margins of society \citep[]{buolamwini2018gender}.
To wit, AI and ML are best seen as forces that wield power where it already exists, perpetuating harm and oppression \citep{kalluri2020don}, with both critical engagement with the past and ethical concerns largely missing (see Chapter \ref{chp:values} for empirical analysis of how such values are missing from most influential ML papers). 

In the present, white supremacism, racism, and colonialism are promoted through (increasingly) covert means and without the explicit consent of most research practitioners nor human participants.
In a similar manner, colonialism in academia does not take on the form of physical invasion through brute force~\cite{George2002, Birhane2019}. Instead we are left with the remnants of colonial era mentality: coloniality~\citep{mohamed2020decolonial}. 
White supremacist ideological inheritances, for example, are found in subtle forms in modern academic psychological, social, and cognitive sciences~\cite{Winston2020, Syed2020, Roberts2020} but more recently and much more explicitly in research practices such as computer vision (see, Chapter~\ref{chapter:computer vision}).

\begin{figure}[p]
    \begin{subfigure}[b]{\textwidth}
         \centering
         \includegraphics[height=0.45\textheight, trim={0 0 0 1cm}, clip]{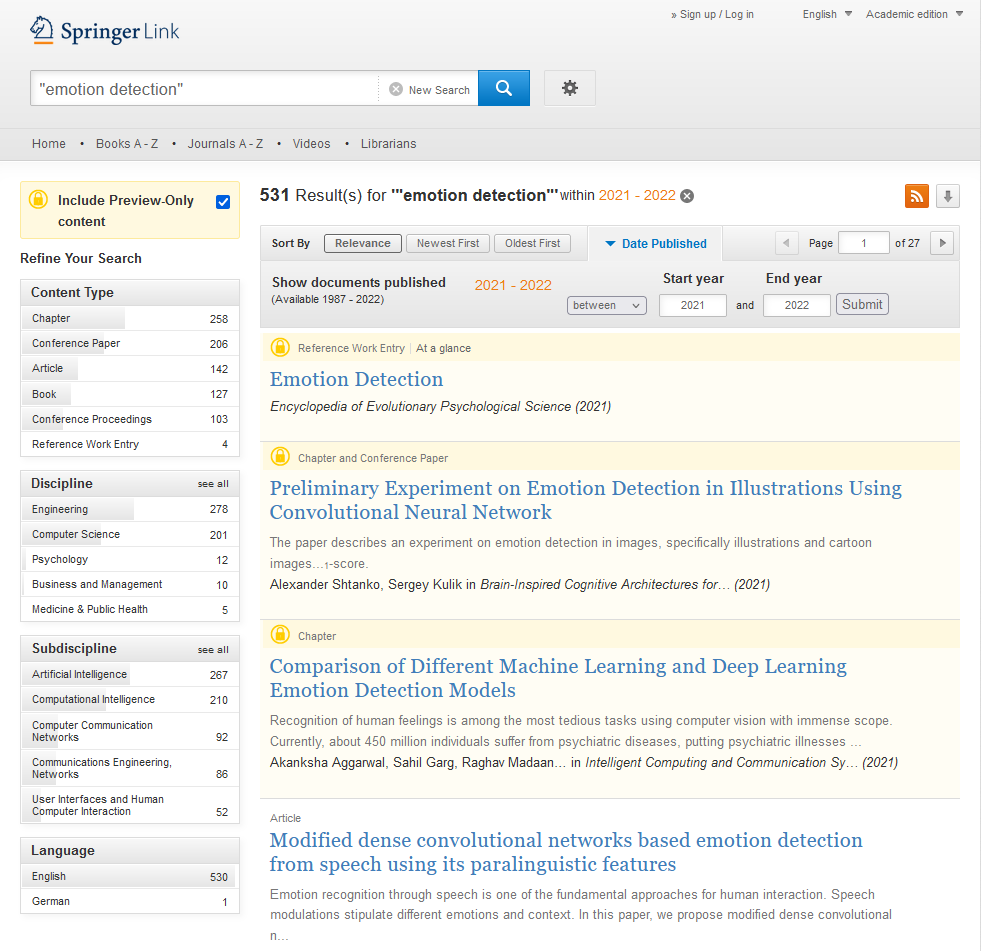}
         \caption{Search result for ``emotion detection'' in Springer Link}
     \end{subfigure}
         
    \begin{subfigure}[b]{\textwidth}
         \centering
         \includegraphics[height=0.45\textheight, trim={0 0 0 1cm}, clip]{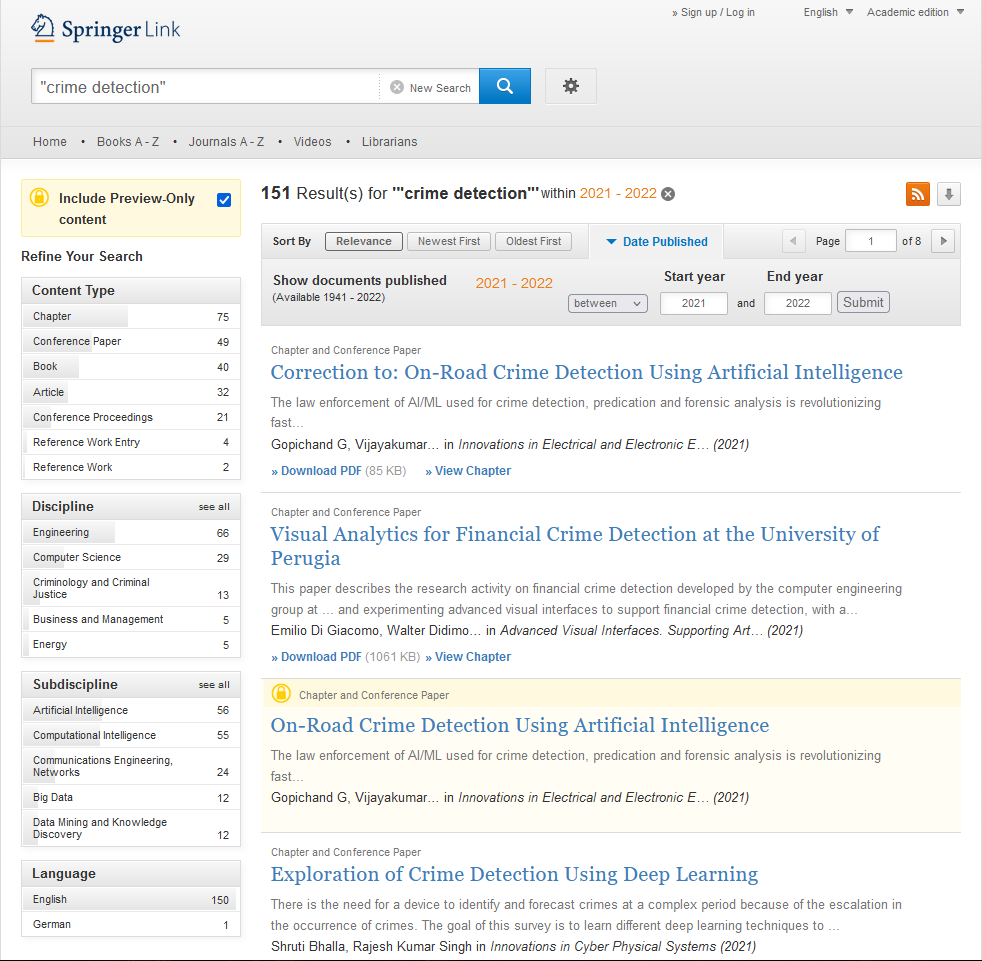}
         \caption{Search result for ``crime detection'' in Springer Link}
     \end{subfigure}
\caption{Search results from Springer Link showing publication of work on ``emotion detection'' (a) and ``crime detection'' (b).}
\label{fig:springer}
\end{figure}

Currently, harmful discredited pseudoscientific practices and theories like eugenics, phrenology, and physiognomy, even when explicitly promoted, face little to no pushback \citep{Saini2019, chinoy2019racist, stark2018facial}.
Springer, for example, was recently pressured to halt publication of a physiognomist book chapter.
Scholars and activists wrote an extensive rebuttal which was then signed by over two thousand 
experts from a variety of fields \citep{cct2020}.
No official statement was provided condemning such work by the editors or publishers, despite being explicitly called on to condemn this type of pseudoscience. Regardless, Springer continues to publish pseudoscience of similar magnitude. At the time of writing, for example, Springer has published 531 papers with ``emotion detection'' and 151 papers with ``crime detection'' (see Figures~\ref{fig:springer} (a) and (b) respectively) in their titles, since 2021 (within the past nine months alone). These publication come from predominantly engineering and computer science backgrounds where \textit{Engineering} as a discipline and \textit{Artificial Intelligence} as a sub-discipline by far produce most of the publications (see Figure \ref{fig:springer}). This reinforces the argument made earlier that modelling phenomena such as emotion and crime are increasingly seen as a matter of ``state-of-the-art'' model building rather than phenomena that require in-depth and rigorous \textit{understanding}. Although the content of the papers requires rigorous empirical investigation, approaching issues such as emotion, gender, and crime as primarily engineering problems (rather than social phenomena) are concerning, to say the least. 

While it is relatively easy to produce such models, to refute, oppose, or retract them requires a magnitude of resources, time and effort unmatched by the easy to produce them. 
In the rare cases where papers or datasets are retracted following outrage, it is the result of substantial effort often spearheaded by researchers who are junior, precarious, and/or of colour \citep[e.g.,][]{Gliske2020, Mead2020}.
A much higher energy barrier must be overcome to get such flawed work expunged from the academic record than to slip such work into the academic space in the first place, resembling a form of cheap talk.

\section{Cheap AI}
\label{sec:cheap}

\textit{Cheap talk (n): talk that results in harm to the marginalised but costs nothing for the speaker.}

As we have seen in Chapter~\ref{chp:automating}, insights from complex systems as well as embodied and enactive cognitive science illustrate that human behavior and social systems are contingent, dynamic, ambiguous, indeterminable and non-totalizable. This makes predictions difficult if not impossible. Even the best models we build are provisional, guided by the values, perspective, and choices of the modeller and can only tell a partial story, at best. Nonetheless, reductionist, misguided and pseudo-scientific models increasingly pervade the social world. Such models, when integrated into daily lives and used to aid decision making, become models that are not only limited, problematic or reductive but also \textit{dangerous}. These include tools that claim to predict complex social phenomena and largely inner states such as emotion, personality, and gender based on outer appearance such as face, gait, gesture, body recognition. I call such systems \textit{Cheap AI}. 

Cheap AI, a subset of Cheap Science, is produced when AI is inappropriately seen as a solution for challenges that it is not be able to solve. It is rooted in the faulty assumption that qualities such as trustworthiness, emotional state, and sexual preference are static characteristics with physical expression that can be read (for example) from our faces and bodies. Tools, that claims to detect dishonesty, 
criminality, 
emotional states, or systems that infer internal mental states from gait recognition —- all constitute \textit{Cheap AI}. Judgements made by these systems are inherently wholly misguided and fundamentally rooted in pseudoscience.

It has become commonplace to see academic papers or ``AI'' tools claiming to predict gender, criminality, emotion, personality, political orientation or another social attribute using machine learning. Whether such work indeed contain any element of AI is an important question worthy of investigation. Nonetheless, such question is out of the scope of this Section. 

Such work increasingly imposes a danger on society. Critics often label such work as pseudoscience, digital phrenology, physiognomy, AI snake oil, junk science, PhysiognomicAI, and bogus AI~\cite{narayanan2019, stark2018facial, crawford2021atlas,stark2021physiognomic}. These labels are fitting and valid. However, I identifying this work as \textit{Cheap AI} for it captures the fact that those producing it (usually a homogeneous group from privileged backgrounds, and predominantly powerful and big corporations and institutes) suffer little to no cost, while the people who serve as the testing grounds, frequently those at the margins of society, pay the heaviest price.

Cheapness emerges when a system makes it easy to talk at little or no cost to the speaker, while at the same time causing tangible harm to the most vulnerable, disenfranchised, and underserved. Within traditional sciences, cheapness manifests when racist, sexist, ableist, misogynist, transphobic and generally bigoted assumptions are re-packaged as scientific hypotheses, with the implication that the only viable way to reject them is to test them. Hisorically, much of the work from the intimately related fields of “race science” and IQ research constitutes Cheap Science. White supremacist ideological inheritances, for example, are found in subtle forms in modern academic psychological, social, and cognitive sciences~\citep{Winston2020, Syed2020, Roberts2020}. Currently, parts of vision research, particularly emotion, gender, and personality recognition tools are producing Cheap AI in masses. 


At the root of Cheap AI lies collective amnesia of past critical work demonstrating lack of scientific credibility of such work as well as historical and current prejudiced assumptions masquerading as objective enquiry. For instance, the very conception of racial categories derives from German doctor Johann Friedrich Blumenbach’s declaration in 1795 that there are ``five human varieties: Caucasians, Mongolians, Ethiopians, Americans, and Malays''~\cite{Saini2019}. This arbitrary classification placed white people of European descent at the top of the hierarchy and cleared the way for colonisation. Subsequently, seemingly scientific racial classifications have served as justifications for inhumane actions that include naturalised slavery, forced sterilization, and the genocidal Nazi attempt to exterminate the Jewish people, people with disabilities, and LGBTQ+ people; among other ``deviant'' classes. Today, racial classifications continue to justify discriminatory practices as a way to filter out ``inferior'' races~\cite{cook2014legacies} aided by computational work, predominantly from the domain of computer vision.

Like other pseudosciences, Cheap Science is built around oversimplifications and misinterpretations. The underlying objective of “race science”, for example, is to get at biological or cognitive differences in supposed capabilities between different races, ethnicities, and genders, on the presumption that there exists a hierarchy of inherent differences to be found between groups. IQ research, for instance, has asserted the existence of race-based differences, by reducing intelligence to a single number and by framing it as dependent upon race. Similarly, Cheap AI arrogates complex and contingent human behaviour to a face, gait, or body language. Tools based on these presumptions are then produced en masse to classify, sort, and “predict” human behaviour and action.

Cheap AI presents itself as something that ought to be tested, validated, or refuted in the marketplace of ideas; the question of whether a tool for predicting criminality as a matter of bigger and/or better data then becomes a fundamentally misguided question. This grants bigoted claims the status of scientific hypotheses, and frames the proponents and critics of Cheap AI as two sides of equal merit, with equally valid intent and equal power. This equivalence is false. While those creating or propagating Cheap AI may face criticism, or reputational harm (if they face these things at all), marginalised people risk discrimination, inhumane treatment, or even death as a result.

Time and time again, attempts to find meaningful biological differences between racial groups have been proven futile, laden with error (there exist more average differences within groups than between groups~\cite{lewontin1996biology}), and rooted in racist motivations. Yet, the same speculations persist today, differently framed and fuelled by facial recognition techniques. 
Despite decades of work warning against the dangers of a reductionist approach, so-called “emotion detection systems” continue to spread~\cite{stark2021ethics}. Affect recognition, an industry which is predicted to now be worth more than seventeen billion US dollars, is playing a role in reintroducing discredited practices of the past~\cite{crawford2021atlas}. Though criminality is a largely complex social phenomenon, claims are still made that AI systems can detect it based on images of faces. Although lies and deception are complex behaviours that defy quantification and measurement, assertions are still made that they can be identified from analysis of video-feeds of gaits and gestures. Alarmingly, this and similar work is fast becoming mainstream, increasingly appearing in prestigious academic venues and journals such as Springer (see Figure~\ref{fig:springer}).


Like segregation, much of Cheap AI is built on a logic of punishment. These systems embed and perpetuate stereotypes. From “deception detection” to “emotion recognition” systems, Cheap AI serves as a tool that “catches” and punishes those deemed to be outliers, problematic, or otherwise unconventional.

The seemingly logical and open-minded course of action—to withhold judgement on these systems on the premise that their merits lie in how “well” or “accurately” they work—lends them a false sense of reasonableness, giving the impression that the “self-correcting” nature of science will eliminate bad tools. It also creates the illusion of there being “two sides”. In reality, criticisms, objections, and calls for accountability drown in the sea of Cheap AI that is flooding day-to-day life. Cheap AI is produced at an unprecedented rate and huge amounts of money go into producing it~\footnote{DARPA has, for example, spent one million US dollars towards an app that supposedly can predict and decode emotions of allies as well as enemies in order to aid its military decisions (see https://www.forbes.com/sites/thomasbrewster/2020/07/15/the-pentagons-1-million-question-can-ai-predict-an-enemys-emotions/).}. By contrast, those working to reveal it as scientifically unfounded and ethically dangerous are scholars and activists working under precarious positions with little to no support, who are likely to suffer negative consequences.

By suspending judgement until wrong is proved, an ecosystem has been created where anyone can claim to have created obviously absurd and impossible tools (some of which are nonetheless taken up and applied) without facing any consequences for engaging in Cheap AI. Such creators and deployers may risk their reputations when their models are proven to be “inaccurate”. However, for those who face the burn of being measured by this tech, it can be a matter of life and death~\cite{obermeyer2019dissecting, banerjee2021reading}, resulting in years lost trying to prove innocence~\cite{hill2020accused,angwin2016machine}, and other grave forms of suffering.

\section{Recommendation}
The widespread application of Cheap AI has catastrophic consequences -- including entrenching historical injustice, exacerbating discriminatory practice, and threatening the right to privacy -- especially for individuals and groups at the margins of society. 
These mounting concerns have sparked conversations between researchers, regulators, and policy makers. Given that the very practice of building such models stands on questionable scientific grounds, and the negative consequences outweigh the benefits (if any benefits at all), one cautious way forward is a blanket ban on Cheap AI, more specifically systems that detect, classify, categorize, predict, and infer inner behaviour from outer appearance or characteristics. In a similar vein, \citet{stark2021physiognomic}, for example, have proposed the abolition of physiognomic AI and called for ``legislative action to forestall and roll back [its] proliferation'', more specifically,  ``that lawmakers should enact or expand biometric privacy laws to prohibit physiognomic AI. [...] in places of public accommodation.''~\citealp[p.1]{stark2021physiognomic}. Comparatively, applications such as face detection, verification and identification -- relatively less ambiguity and inference -- might require less strict measures. Nonetheless, they remain riddled with errors~\cite{buolamwini2018gender} and still infringe on rights to privacy when used in publicly accessible spaces~\cite{leslie2020understanding}. Thus, it is imperative that there are legal frameworks in place that guide and detail their use. 

In recognition of the threats posed by the use of facial and biometric recognition technologies a global call -- from entities including Access Now, Amnesty International, European Digital Rights (EDRi),
Human Rights Watch, Internet Freedom Foundation (IFF), and Instituto Brasileiro de Defesa do
Consumidor (IDEC) -- for a ban on the use biometric technology has emerged~\footnote{https://www.accessnow.org/cms/assets/uploads/2021/11/joint-statement-EU-AIA.pdf}. This global call emphasizes that the widespread use of these technologies exacerbates structural imbalances of power, and undermines human rights, including rights to privacy, data protection, and rights to equality and non-discrimination. The capacities of biometric technologies to identify, single out and track people subsequently harm marginalized communities given uneven power structures as the call highlights. Similarly, following a joint investigation with the Office of the Australian Information Commissioner (OAIC), UK's Information Commissioner’s Office (ICO) has found that Clearview AI Inc. fails to have lawful reasons for collecting data, to use data in a way that is fair and transparent, as well as to meet UK's biometric data protection standards. The UK's ICO has issued Clearview AI \pounds 7.5m and ordered the company to stop using publicly available personal data of UK residents~\cite{ICO2022}. In summary, and in light of these and other similar legal initiatives, a ban on Cheap AI (those including face, gender, and emotion classification and prediction) and/or strict legal safeguards (applications such as face identification and verification) are essential to protect rights to privacy, rights to equality and non-discrimination as well as to envision technological futures that are just and equitable. 





\section{Conclusion}
\label{sec:cheap_conc}

There is no quick solution to ending Cheap AI. Many factors contribute to the ecology in which it thrives. This includes blind faith in AI — or even just the idea of AI, the illusion of objectivity that comes with the field’s association with mathematics, Cheap AI’s creators and deployers’ limited knowledge of history and other relevant fields, a lack of diversity and inclusion, the privilege hazard (a field run by a group of mostly white, privileged men who are unaffected by Cheap AI’s harms), the tendency to ignore and dismiss critical voices, and a lack of accountability. We must recognise Cheap AI as a problem in the ecosystem. All of these factors and more need to be recognised and challenged so that Cheap AI is seen for what it is, and those producing it are held accountable.

Unfortunately, the retraction of a few papers or datasets, in an academic culture that fails to see the inherently racist, sexist, and white supremacist foundations of such work serves only as a band-aid on a bullet wound. 
The system itself needs to be rethought --- scholars should not, as a norm, need to form grassroots initiatives to instigate retractions and clean up the literature. Rather, the onus should fall on those producing, editing, reviewing, curating, managing, and funding (pseudo)scientific work. Strict and clear peer review guidelines and IRBs, for example, might provide a means to filter racist pseudoscience out \cite{boyd2020racism}. However, ultimately, it is the peer review and publishing system, and the broader academic ecosystem that need to be re-examined and reimagined in a manner that explicitly excludes harmful pseudoscience and subtly repackaged white supremacism from its model. 

It took Nazi-era atrocities, forced sterilizations, and other inhumane tortures for phrenology, eugenics, and other pseudosciences to be relegated from science’s mainstream to its fringe. It should not take mass injustice for Cheap AI to be recognised as similarly harmful and categorically rejected as ``oppressive and unjust'' \cite{stark2021physiognomic}. In addition to strict legal regulation and the enforcement of academic standards, we ourselves also bear a responsibility to call out and denounce Cheap AI, and those who produce it. 


    \part{}
    \label{part:iii}
    \chapter{Relational Ethics}
\label{chp:relational}




A persistent and recurrent trend within the AI ethics and fairness literature indicates that society's most vulnerable are disproportionally impacted. Yet, when algorithmic injustice and harm are brought to the fore, most of the solutions currently on offer (1) revolve around technical ``fixes'' and (2) do not center disproportionally impacted communities. This chapter proposes a fundamental shift from rationality to relationality in thinking about personhood, data, justice, and everything in between, and places ethics as something that goes above and beyond technical solutions. Outlining the idea of ethics built on the foundations of relationality, this chapter calls for a rethinking of justice and ethics as a set of broad, contingent, and fluid concepts and down-to-earth practices that are best viewed as a habit and not a mere methodology or an abstract principle for algorithmic systems. 

\section{Introduction}
\label{introduction}
Algorithmic decision-making increasingly pervades the social sphere. From allocating medical care \cite{obermeyer2019dissecting}, to predicting crimes \cite{lum2016predict}, selecting social welfare beneficiaries \cite{eubanks2018automating}, and identifying suitable job candidates \cite{raghavan2020mitigating,ajunwa2016hiring}; 
complex social issues are increasingly automated. AI and ML tools have become the hammer every messy social challenge is bashed with. 
High efficiency and seemingly neat shortcuts to complex and nuanced problems make algorithmic decision making attractive. However, automated and standardized solutions to complex and contingent social issues often contribute more harm than good --- they often fail to grasp complex problems and provide a false sense of solution and safety. Complex social issues require historical, political, and critical awareness and structural change. Any modeller working to automate issues of a social nature, in effect, is engaged in making moral and ethical decisions --- they are not simply dealing with purely technical work but with the \textit{making} of a social reality and a practice that actively impacts individual people. 

As social processes are increasingly automated and algorithmic decision-making integrated in various social spheres, socially and politically contested matters that were traditionally debated in the open are now reduced to mathematical problems with a technical solution \cite{mcquillan2020non}. The mathematization and formalization of social issues brings with it a \textit{veneer of objectivity} and positions its operations as value-free, neutral, and a-moral. The intrinsically political tasks of categorizing and predicting things like ``acceptable'' behaviour, ``ill'' health, and ``normal'' body type, then pass as apolitical technical sorting and categorizing tasks \cite{bowker2000sorting}. Unjust and harmful outcomes, as a result, are treated as side-effects that can be treated with technical solutions such as ``debiasing'' datasets \cite{gonen2019lipstick} rather than problems that have deep roots in the mathematization of ambiguous and contingent issues, historical inequalities and asymmetrical power hierarchies or unexamined problematic assumptions that infiltrate data practices. 


The growing body of work exposing algorithmic injustice has indeed 
brought forth increased awareness of these problems, subsequently spurring the development of various techniques and tactics to mitigate bias, discrimination, and harms. However, many of the ``solutions'' put forward 1) revolve around technical fixes and 2) do not centre individuals and communities that are disproportionally impacted. Relational ethics, at its core, is an attempt to unravel our assumptions and presuppositions and to rethink ethics in a broader manner via engaged epistemology in a way that puts the needs and welfare of the most impacted and marginalized at the centre. 

In the move to rethink ethics, concrete knowledge of the lived experience of marginalized communities is central. This begins with awareness and acknowledgement of historical injustices and the currently tangible impact of AI systems on vulnerable communities. The core of this framework is grounding ethics as a \textit{practice} that results in improved material conditions for individuals and communities while moving away from ethics as abstract contemplations or seemingly apolitical concepts such as ``fair'' and ``good''. Relational ethics, then, is a framework that necessitates we re-examine our underlying working assumptions, compels us to interrogate hierarchical power asymmetries, and stimulates us to consider the broader, contingent, and interconnected background that algorithmic systems emerge from (and are integrated into) in the process of prioritizing the welfare of the most vulnerable. 

Through the lens of \textit{relational ethics}, I explore the social, political, historical and dynamic nature of data science, machine learning, and AI and the need to rethink ethics in broader terms. This chapter primarily offers a critical analysis and encourages the practitioner/modeller to cultivate critical engagement. It departs from traditional scholarship within the data and AI ethics space that offer technical solutions or implementable remedies that attempt to mitigate problems of social, political, and justice nature.

The rest of the chapter is structured as follows. Section \ref{the roots} fleshes out the roots of relational ethics and provides comparisons of relationality with the dominant orthodoxy, rationality. Section \ref{ethics built on} then lays out the four tenets of relational ethics followed by a brief conclusion in Section \ref{closing}.

\section{Relational Ethics: The Roots}
\label{the roots}
Before delving into the roots and central tenets of relational ethics, it makes sense to make visible the dominant school of thought - rationality. Relationality exists both as a push back against rationality, but also on its own right, for example, in the case of ubuntu\footnote{A brief web search for ``ubuntu'' brings up information on a Linux operating system that has been around since 2004 usurping the original meaning of the word that has existed for centuries within sub-Saharan Africa. The appropriation of the word with its rich culture and history to a shallow tech sloganeering is not only wrongful but also symptomatic of the Western tech world's inability to centre non-Western perspectives while stripping them off their rich culture, history, and meaning.}, as a philosophy, ethics, and way of life \cite{menkiti1984person}. At the heart of relational ethics is the need to ground key concepts such as ethics, justice, knowledge, bias, and fairness in \textit{context}, \textit{history}, and an active, dynamic and \textit{engaging epistemology}. Fundamental to this is the need to shift from prioritizing rationality as of primary importance over to the supremacy of relationality.



\subsection{Rationality: The Dominant Orthodoxy}
\label{rationality}

\begin{displayquote}
``Renaissance thinkers like Montaigne acknowledged that universal, foundational principles cannot be applied to such practical matters as law, medicine and ethics; the role that context and history play in those areas prevents it.'' Alicia Juarrero \cite{juarrero2000dynamics} 
\end{displayquote}

The rational view serves as the backbone for much of Western science and philosophy permeating most fields of enquiry (and social and institutional practices) from the life sciences, to the physical sciences, the arts and humanities, and to the relatively recent field of computer science \cite{fromrationality2020, winograd1986understanding}. The rational worldview, the quintessential orthodoxy for Western thought, can be exemplified by the deep contention that reason and logical coherence are superior for knowledge production (in understanding the world) above and beyond relational and embodied becoming. The privileging of reason as the ultimate criterion makes knowing an abstract, passive and distant act. The deep quest for the rational worldview is certainty, stability, and order, and thus isolation, separation, and clear binaries form the foundations in place of connectedness, interdependence, and dynamic relations \cite{prigogine1984order}. Since the rational worldview has come to be seen as the standard, anything outside of this is viewed as an outlier. Spelling out what this worldview entails, what its underlying assumptions are, and the consequences for a subject of enquiry which inherits this worldview, therefore, is an important step towards providing context for the relational worldview.  


Although the rationalist worldview is a result from the accumulation of countless influences from pivotal thinkers, its lineage can be traced through Western intellectuals such as Newton and Descartes all the way back to Plato. René Descartes, the quintessential rationalist, attempted to establish secure foundations from which knowledge can be built based solely on reason and rational thought. In this quest, Descartes attempted to rid us of unreliable, changeable, and fallible human intuitions, senses, and emotions in favour of reason and timeless crystalline logic \cite{descartes1984philosophical}. At the heart of his quest was to uncover the permanent structures beneath the changeable and fluctuating phenomena of nature on which he could build the edifice of unshakable foundations of knowledge. Anything that can be doubted is eliminated. Subsequently, discussions and understanding of concepts such as knowledge and ethics tend to be abstract, gender-less, context-less, and race-less. Knowledge, according to this worldview, is rooted in the ideal rational, static, self-contained, and self-sufficient subject that contemplates the external world from afar in a ``purely cognitive'' manner as a disembodied and disinterested observer \cite{gardiner1998incomparable}. In the desire to establish timeless and absolute knowledge, abstract and context-less reasoning is prioritized over concrete lived experience submersed in co-relations, inter-dependence, fluidity, and connectedness \cite{merleau1968visible}. More fundamentally, as Ahmed \cite{ahmed2007phenomenology} contend, all bodies inherit history and the inheritance of Cartesianism is grounded in a white straight ontology. The reality of the Western straight white male masquerades as the invisible background that is taken as the ``normal'', ``standard'', or ``universal'' position. Anything outside of it is often cast as ``dubious'' or an ``outlier''.  


In a similar vein, and with a similar fundamental influence as Cartesianism, the Newtonian worldview aspired to pave the path for universal knowledge in a supposedly observer-free and totally ``objective'' manner. This thoroughly individualistic worldview sees the world as containing discrete, independent, and isolated atoms. Neat explanations and certainty in the face of ambiguity provide a sense of comfort. Within the physical world, Newtonian mechanistic descriptions allowed precise predictions of systems at any particular moment in the future, given knowledge of the current position of a system. This view fared poorly, however, when it came to capturing the messy, interactive, fluid, and ambiguous world of the living who are inherently context bound, socially embedded, and in continual flux. Living systems, such as social systems are, in complexity terms, non-compressible; meaning that there is no algorithm or model simpler than the system itself that can represent it in its entirety without leaving some things out \cite{cilliers2002complexity, richardson2001complexity}. People and social systems, as complex adaptive systems, are open and non-totalisable which means that a single, final and universal knowledge of such systems is impossible. 
We exist in a web of relation with others and embedded in our environment and in a continual dynamical interaction. Such interactions are nonlinear, which makes neat cause and effect relations difficult to establish. In a worldview that aspires for objective, universal, and timeless knowledge, the very idea of complex and changing interdependence and co-relations --- the very essence of being insofar as there can be any --- are not tolerated. Despite the inadequacy of the billiard ball model of Newtonian science in approaching complex adaptive systems such as human affairs, its residue prevails today, directly or indirectly \cite{juarrero2000dynamics,cilliers2002complexity} within the data sciences and the human sciences in general. 

The historic Bayesian framework of prediction \cite{doi:10.1098/rstl.1763.0053} has played a central role in establishing a normative explanation of behaviours \cite{hahn2014bayesian}. Bayes' approach, which is increasingly used in various areas including data science, machine learning, and cognitive science \cite{seth2014cybernetic,jones2011bayesian}, played a pivotal role in establishing the cultural privilege associated with statistical inference and set the ``neutrality'' of mathematical predictions. Price, who published the papers after Bayes' death, noted that Bayes' methods of prediction ``shows us, with distinctness and precision, in every case of any particular order or recurrency of events, what reason there is to think that such recurrency or order is derived from stable causes or regulations in nature, and not from any irregularities of chance'' \citep[p.374]{doi:10.1098/rstl.1763.0053}. However, despite the association of Bayes with rational predictions, Bayesian models are prone to spurious relationship and amplification of socially held stereotypes \cite{pager2009bayesian} without close scrutiny of priors. Horgan \cite{Bayes2016} notes, ``Embedded in Bayes’ theorem is a moral message: If you aren’t scrupulous in seeking alternative explanations for your evidence, the evidence will just confirm what you already believe.'' 

Dichotomous thinking — such as, subject versus object, emotion versus reason – persists within this tradition. Ethical and moral values and questions are often treated as clearly separable (and separate) from ``scientific work'' and as something that the scientist need not contaminate their ``objective'' work with rather than an inherent dimension of grasping, understanding and modelling the world. In its desire for absolute rationality, Western thought wishes to cleave thought from emotion, cultural influence, and ethical dimensions. Abstract and intellectual thinking are regarded as the most trustworthy forms of understanding and rationality is fetishized. 

Data science, and the wider discipline of computer science, have implicitly or explicitly inherited this worldview \cite{fromrationality2020}. These fields, by and large, operate with rationalist assumptions in the background. The view of the modeller is mistaken as ``the view from nowhere'' --- the ``neutral'' view. Misconceptions such as a universal, relatively static, and objective knowledge that can emerge from data are persistent \cite{gitelman2013raw}. Data science and data practices reincarnate rationalism in many forms, including in the manner in which messiness, ambiguity, and uncertainty are not tolerated; in the pervasive binary thinking (such as emotion vs reason, where the former is assumed to have no place in data science); the way in which data are often severed from the person (with emotions, hopes, and fears) that they are rooted in and the context in which they emerge; the manner in which the dominant view is taken as the ``God's eye view''; and the way questions of privilege and oppression are viewed as issues the data sciences need not concern themselves with. Not only does the inheritance of rationality to data sciences and computation make these fields inadequate to deal with complex and inherently indeterminable phenomena, Mhlambi \cite{fromrationality2020} has further argued that the AI industry, grounded in rationality, reproduces harmful and discriminatory outcomes.

\subsection{Relationality}
\label{relationality}
 
Contrary to the rationalist and individualist worldview, relational perspectives view existence as \textit{fundamentally co-existence in a web of relations}. Various schools of thought can be grouped under the umbrella of the relational framework with a core commonality of \textit{interdependence}, \textit{relationships}, and \textit{connectedness}. Relational-centred approaches include\footnote{This is not an exhaustive list of all approaches that could be identified as relational. The focus on these specific schools of thought and approaches, as opposed to others that might fall under relational approaches, is heavily influenced by the author's background and academic training.} Black feminist (Afro-feminist) epistemologies, embodied and enactive approaches to cognitive science, Bakhtinian dialogism, ubuntu (the Sub-Saharan African philosophy), and complexity science. Although these schools of thought vary in their subjects of enquiry, aims, objectives, and methods, they have relationality in common.

Relational frameworks emphasize the primacy of relations and dependencies. These accounts take their starting point in reciprocal co-relations. Kyselo \cite{kyselo2014body} for example, contends that the self is social through and through --- it is co-generated in interactions and relations with others. We achieve and sustain ourselves together with others. Similarly, according to the Sub-Saharan tradition of ubuntu as encapsulated by Mbiti's \cite{mbiti1969african} phrase \textit{`I am because we are, and since we are, therefore I am'}, a person comes into being through the web of relations. In a similar vein, Bakhtin \cite{bakhtin1984problems} emphasized that nothing is simply itself outside the matrix of relations in which it exists. It is only through an encounter with others that we come to know and appreciate our own perspectives and form a coherent image of ourselves as a whole entity. By \textit{`looking through the screen of the other’s soul,'} he wrote, \textit{`I vivify my exterior'}. Selfhood and knowledge are evolving and dynamic; the self is never finished – it is an open book \cite{birhane2017descartes}. 

Relational ethics takes its roots from these overlapping frameworks. The rest of this section delves into Afro-feminist thought and the enactive approach to cognitive science with the aim of providing an in-depth understanding of the roots of the relational worldview. 

\subsubsection{Afro-feminism}
\label{Afro-feminism}
\begin{displayquote}
``Knowledge without wisdom is adequate for the powerful, but wisdom is essential for the survival of the subordinate.'' Patricia Hill Collins \cite{collins2002black}
\end{displayquote}
 
Pushing back against the dominant Western orthodoxy, Afro-feminist epistemology grounds knowing in an active and engaged practice inseparable from the knower. The most reliable form of knowledge, especially \textit{concerning social and historical injustice}, is grounded in lived experience. One of the most prominent advocates of Afro-feminist epistemology, Patricia Hill Collins \cite{collins2002black}, emphasizes that people are not passive cognizers that contemplate and grasp the world in abstract forms from a distance but instead knowledge and understanding emerge from concrete lived experiences. 
At the heart of it, the Afro-feminist approach to knowing contends that concrete experiences are primary, and abstract reasoning secondary. Knowing and being are active processes that are necessarily political and ethical. Drawing core differences between the dominant Western tradition and the Afro-feminist perspective, Collins identifies two types of knowing: \textit{knowledge} and \textit{wisdom}. Knowledge is closely tied to what Collins calls ``book learning'' – learning that emerges from reasoning about the world from a distance in a rational way. This form of knowledge aspires to arrive at ``the objective truth'' that transcends context, time, specific and particular conditions, and lived experiences. Wisdom, on the other hand, is grounded in concrete lived experience. Formal education, according to Collins, is not the only route to such forms of knowledge and wisdom holds high credence in assessing knowledge claims. Distant statistics or theoretical accuracies do not take precedence over the actual experience of a person. Knowledge claims are not worked out in isolation from others but are developed in dialogue with the community. It is taken for granted that there exists an inherent connection between what one does and how one thinks. This is especially the case when the type of knowledge in question concerns \textit{oppression}, \textit{structural discrimination}, and \textit{racism}. Wisdom, and not ``book learning'', enables one to identify and resist oppression. From the core arguments of Afro-feminist epistemology, it follows that concepts such as ethics and justice need to be grounded in concrete events informed by lived experience of the most marginalized, individuals and communities that pay the highest price for algorithmic harm and injustice. 

Current data practices, for the most part, follow the rational model of thinking where data are assumed to represent the world ``out there'' in a ``neutral'' way. Yet, not only is it fallacious to assume complex social reality can be fully represented by data, the process of data collection, analysis and interpretation of results is a value-laden endeavour. In the process of data collection, for example, the data scientist decides what is worth measuring (making some things visible and others invisible by default) and how. In the process of data cleaning, rich information that provides context about which data are collected and how datasets are structured is stripped away. Emphasizing the importance of contexts for datasets, Loukissas \cite{loukissas2019all} has proposed a shift into thinking in terms of \textit{data settings} instead of \textit{datasets}. 

The rational worldview that aspires to an ``objective'' knowledge from a ``God’s eye view'' has resulted in the treatment of the researcher as invisible, their interests, values, and background as inconsequential. In contrast, for Afro-feminist thought, the researcher is an important participant in the knowledge production process \cite{nnaemeka2004nego}. For Sarojini Nadar \cite{nadar2014stories}, coming to know is an active and participatory endeavour with the power to transform. Consequently, data and our models portray and represent certain mode of reality while leaving out others.   

\subsubsection{Enactive Cognitive Science }
\label{enaction}
\begin{displayquote}
``Loving involves knowing, and [...] knowing involves loving. Loving and knowing, for human beings, entail each other. To understand knowing only ``coldly,'' abstractly, objectively is either not to see the loving involved, or not to know fully.'' Hanne De Jaegher \cite{de2019loving} 
\end{displayquote}

In a similar vein to Afro-feminist thought, the enactive cognitive science theory of participatory sense-making \cite{de2007participatory} advocates for an active and engaged knowing rooted in our relating. A proponent of this position, Hanne De Jaegher \cite{de2019loving}, contends that our most sophisticated human knowing lies in how we engage with each other. In \textit{`Loving and knowing: Reflections for an engaged epistemology'}, De Jaegher \cite{de2019loving} emphasizes that discrete, rational knowing comes at the detriment of \textit{Knowing-in-connection}. Far from a distant and ``objective'' discretising logic, knowing is an activity that happens in the relationship between the knower and the known. Proposing an understanding of human knowing in analogy with loving, De Jaegher argues that in knowing, like loving, what happens is not neutral, general, or universal. Knowers, like lovers, are not abstract subjects but are particular and concrete. ``\textit{Who loves matters}.'' And both loving and knowing take place in the relation between them \cite{de2019loving}. 

Human knowing is based not on purely rational logic, as the rational worldview assumes, but on living and connected know-hows. ``Our most sophisticated knowing'', according to De Jaegher, ``\textit{is full of uncertainty, inconsistencies, and ambiguities}.'' One of the consequences of prioritizing reason is that knowledge of the world and of other people becomes something that is rooted in the individual person’s rational reasoning – in direct contrast to engaged, active, involved, and implicated knowing. Humans are inherently historical, social, cultural, gendered, politicized, and contextualized organisms. Accordingly, their knowing and understanding of the world around them necessarily takes place through their respective lenses.

People are not solo cognizers that manipulate symbols in their heads and perceive their environment in a passive way, as the rationalist view would suggest, but they actively engage with the world around them in a meaningful and unpredictable way. Living bodies, according to Di Paolo, Cuffari, and De Jaegher \cite{di2018linguistic}, are processes, practices, and networks of relations which have ``more in common with hurricanes than with statues''. They are unfinished and always becoming, marked by\textit{``innumerable relational possibilities, potentialities and virtualities''} and not calculable entities whose behaviour can neatly be categorized and predicted in a precise way. Bodies: ``...grow, develop, and die in ongoing attunement to their circumstances... Human bodies are path-dependent, plastic, nonergodic, in short, historical. There is no true averaging of them.'' \citep[p.97]{di2018linguistic}. What might a version of ethics --- in the context of data practices and algorithmic systems --- that takes the core values of enactive cognitive science and Afro-feminist epistemology (described in sections \ref{Afro-feminism} and \ref{enaction}) as its foundations look like? Section \ref{ethics built on} details that. 


Before we delve into that, it is worth reemphasising that while the rational worldview tends to see knowledge, people, and reality in general as stable, for relational perspectives, we are fluid, active, and continually becoming. Nonetheless, the relational vs rational divide is not something that can be clearly demarcated but overlaps with fuzzy boundaries. Some approaches might prove difficult to fit in either category while others serve to bridge the gap -- Harding's \cite{harding1992rethinking} \textit{Strong Objectivity} is one such example that links relational and rational approaches. Furthermore, the relational and rational traditions exist in tension with a continual push and pull. For example, complexity science is a school of thought that emerged from this tension.   

\section{Ethics Built on the Foundations of Relationality}
\label{ethics built on}
\begin{displayquote}
``Ethics is a matter of practice, of down-to-earth problems and not a matter of those categories and taxonomies that serve to fascinate the academic clubs and their specialists.'' Heinz von Foerster in \cite{von2002metaphysics}
\end{displayquote}

What does the idea of \textit{ethics} -- within the context of data practices and algorithmic systems -- built on the foundations of relationality look like? This section seeks to elucidate. What follows is not a set of general guidelines, or principles, or a set of out-of-the-box tools that can be implemented to supposedly cleanse datasets of bias or to make a set of existing models ``ethical'' for the problems we are trying to grasp are deeply rooted, fluid, contingent, and complex. Neither is it a rationally and logically constructed ``theory of ethics'' which hypothesizes about morality in abstract terms. Rather, the following are the central tenets, informed by Afro-feminist and enactivist perspectives outlined in the previous section, that should aid in shifting toward a more humble and modest understanding complex systems such as people, knowledge and social systems. This is also a call for rethinking concepts such as data, ethics, models, matrices of oppression, and structural inequalities as inherently interlinked and processual.   


\subsection{Knowing that Centres Human Relations}
\label{knowing that centres}
Since knowing is a relational affair, who enters into the knower-known relations matters. Within the fields of computing and data sciences, the \textit{knower} is heavily dominated by privileged groups of mainly elite, Western, cis-gendered, and able-bodied white men \cite{broussard2018artificial}. Given that knower and known are closely tied, this means that most of the knowledge that such fields produce is reduced to the perspective, interest, and concerns of such dominant group. Subsequently, not only are the most privileged among us restricted to producing partial knowledge that fits a limited worldview (while such knowledge, tools, models, and technologies they produce are forced onto all groups, often disproportionately onto marginalized people), they are also poorly equipped to recognize injustice and oppression \cite{Berenstain2016}. D'Ignazio and F. Klein \cite{d2020data} call this phenomenon \textit{the privilege hazard}. This means that minoritized populations 1) experience harm disproportionally and 2) are better suited to recognize harm due to their epistemic privilege \cite{on1993marginality} while the reverse holds for those building and deploying models.  


\textbf{Centring the Disproportionally Impacted:}
\label{disproportionally}
The harm, bias, and injustice that emerge from algorithmic systems varies and is dependent on the training and validation dataset, the underlying taken for granted assumptions of the model, and the specific context the system is deployed in, amongst other factors. However, one thing remains constant: individuals and communities that are at the margins of society are disproportionally impacted. Some examples include object detection \cite{wilson2019predictive}; search engine results \cite{noble2018algorithms}; recidivism \cite{angwin2016machine}; gender recognition \cite{buolamwini2018gender}; gender classification \cite{hamidi2018gender,barlassee2020}; and medicine \cite{obermeyer2019dissecting}. Wilson et.al.'s findings in \cite{wilson2019predictive}, for instance, demonstrate that object detection systems designed to predict pedestrians display higher error rates identifying dark skin pedestrians while light-skinned pedestrians are identified with higher precision. The use of such systems situates the recognition of subjectivity with skin tone where whiteness is taken as ideal mode of being. Furthermore, gender classification systems often operate under essentialist assumptions and operationalize gender in a trans-exclusive way resulting in disproportionate harm to trans people \cite{keyes2018misgendering,hamidi2018gender}.

Given that harm is distributed disproportionately and that the most marginalized hold the epistemic privilege to recognize harm and injustice, relational ethics asks that for any solution that we seek, the starting point be the individuals and groups that are impacted the most. This means we seek to centre the needs and welfare of those that are disproportionally impacted and not solutions that benefit the majority. Most of the time this means not simply creating a fairness metric for an existing system but rather questioning what the system is doing, particularly examining its consequences on minoritized and vulnerable groups. This requires us to zoom out and draw the bigger picture. A shift from asking narrow questions such as \textit{how can we make a certain dataset representative?} to examining larger issues such as \textit{what is the product or tool being used for? Who benefits? Who is harmed? What are the factors that our model has taken into consideration (and what factors are left out as irrelevant). And are the factors we failed to consider or deemed irrelevant indeed so?} 

To some extent, the idea of \textit{centring the disproportionally impacted} shares some commonalities with aspects of \textit{participatory design}, where design is treated as a fundamentally participatory act \cite{slavin2016design} and even aspects human-centered design \cite{irani2010postcolonial} where individuals or groups whom technology is supposed to serve are placed at the centre. However, the idea of \textit{centring the disproportionally impacted} goes further than human-centered or participatory design as broadly construed. While the latter approaches can neglect those at the margins \cite{harrington2020forgotten}, shy away from power asymmetries and structural inequalities that permeate the social world, and ``mirror individualism and capitalism by catering to consumer's purchasing power at the expense of obscuring the hidden labor that is necessary for creating such system'' \cite{Lioyd2020} for the former, acknowledging these deeply ingrained structural hierarchies and hidden labour is a central starting point. In this regard, with a great emphasis on asymmetrical power relations, works such as Costanza-Chock \cite{costanza2018design}'s \textit{Design Justice} and Harrington \cite{harrington2020forgotten}'s \textit{The Forgotten Margins} are examples that provide insights into how centring the disproportionately impacted might be realized through design led by marginalized communities. 


The central implication of this in the context of a justice centred data practice is that minoritized populations that experience harm disproportionately hold the epistemic authority to recognize injustice and harm given their lived experience. Understandings these concepts and building just technologies therefore, needs to proceed from the experience and testimony of the disproportionately harmed. The starting point towards efforts such as ethical practice in machine learning or theories of ethics, fairness, or discrimination needs to centre the material condition and the concrete consequences an algorithmic tool is likely to bring on the historically marginalized. Having said that, these are efforts with extreme nuances and magnitudes of complexity in reality. For example, questions such as `how might a data worker engage vulnerable communities in ways that surface harms, when it is often the case that algorithmic harms may be secondary effects, invisible to designers and communities alike? What questions might be asked to help anticipate these harms?', `how do we make frictions, often the site of power struggles, visible?' are difficult questions but questions that need to be negotiated and reiterated by communities, data workers and model developers.

\subsection{Bias is not a Deviation from the ``Correct'' Description}
\label{bias is not a deviation}
One of the characteristics of a rational worldview is the tendency to perceive things as relatively static. In a supposedly objective worldview, bias, injustice, and discrimination are (mis)conceived as being able to be \textit{permanently corrected}. The common phrase ``bias in, bias out'' captures this deeply ingrained reductive thinking. Although datasets are often part of the problem, this commonly held belief relegates deeply rooted societal and historical injustices, nuanced power asymmetries, and structural inequalities to mere datasets. The implication is that if one can ``fix'' a certain dataset, then the deeper problems disappear. When we see bias and discrimination, what we see is problems that have surfaced as a result of a field that has thoughtlessly inherited deeply rooted unjust, racist, and white supremacist histories and practices~\cite{birhane2021algorithmic}. As D'Ignazio and F. Klein \cite{d2020data} contend, \textit{``addressing bias in a dataset is a tiny technological Band-Aid for a much larger problem''}. Furthermore, underlying the idea of ``fixing'' bias is the assumption that there exists a single \textit{correct description} of reality where a deviation from it has resulted in bias. As we have seen in Section \ref{rationality}, the idea of a single correct description, theory, or approach is reminiscent of the rationalist tradition where \textit{the correct way} is often synonymous with the \textit{status quo}. The idea of bias as something that can be eliminated, so to speak, once and for all, is misleading and problematic. Even if one can suppose that bias in a dataset can be ``fixed'', what exactly are we fixing? What is the supposedly bias free tool being applied to? Is it going to result in net benefit or harm to marginalized communities? Is the supposedly ``bias free'' tool used to punish, surveil, and harm anyway? And in Kalluri \cite{kalluri2020don}'s words, ``how is AI shifting power'' from the most to the least privileged? Looking beyond biased datasets and into deeper structural issues, historical antecedents, and power asymmetries is imperative. 

The rationalist worldview and its underlying assumptions are pervasive and take various nuanced forms. Within the computation and data sciences, the propensity to view things as relatively static manifests itself in the tendency to formulate subjects of study (people, ethics, and complex social problems in general) in terms of problem$\,\to\,$solution. Not only are subjects of study that do not lend themselves to this formulation discarded but also, this tradition rests on a misconception that injustice, ethics, and bias are relatively static things that we can \textit{solve once and for all}. Concepts such as bias, fairness, and justice, however, are moving targets. As we have discussed in section 2.2, neither people nor the environment and context they are embedded in are static. What society deems fair and ethical changes over time and with context and culture. The concepts of fairness, justice, and ethical practice are continually shifting. It is possible that what is considered ethical currently and within certain domains for certain societies will not be perceived similarly at a different time, in another domain, or by a different society. This, however, is not a call to relativism but rather an objection to static and final answers in the face of fluid reality. Adopting relational ethics means that we view our understandings, proposed solutions, and definitions of bias, fairness, and ethics as partially-open. This partial openness allows for revision and reiteration in accordance with the dynamic development of such challenges. This also means that this work is never \textit{done}.


\subsection{Prioritizing of Understanding over Prediction}
\label{prioritizing understaning}
\begin{displayquote}
``I have never been impressed with claims that structural linguistics, computer engineering or some other advanced form of thought is going to enable us to understand men without knowing them.'' Clifford Geertz \cite{geertz1973interpretation}
\end{displayquote}


The rationalist tradition's tendency toward timeless and generalizable knowledge aspires to establish timeless laws and generalizable theories. This pipeline takes observed commonalities, recurring similarities, and repeated patterns among past events or particular behaviours and abstracts them into generalizations that can be applied toward forecasting the future. Because the rationalist's focus is to uncover what remains constant regardless of context, culture, and time, the rationalist view embraces abstraction, generalization, and universal principles at the expense of concrete, particular, and contextual understanding — that is, knowledge grounded in active, concrete, and reciprocal relationships. According to Geertz \cite{geertz1973interpretation}, the desire to formulate general theories is in an \textit{irremovable tension} with the need to gain deep understanding of particular and contextual events and behaviors. The further theory goes, the deeper the tension. Theories and generalizations inevitably lack deep and contextual understanding of human thought. Theoretical disquisitions stand far from the immediacies of social life. Any generalization or theory constructed in the absence of deep understanding, not grounded in the concrete and particular, is vacuous.

On a similar note, the Russian philosopher Mikhail Bakhtin refers to the manner in which abstract general rules are derived from concrete human actions and behaviours as \textit{theoretism}. Bakhtin argues such attempts to abstract general rules from particulars ``loses the most essential thing about human activity, the very thing in which the soul of morality is to be found'' which Bakhtin calls, the \textit{``eventness''} of the event \cite{morson1989rethinking}. \textit{Eventness} is always a particular, and never exhaustively describable in terms of rules. In order to understand people, we must take into account \textit{``unrepeatable contextual meaning''}. Likewise, the historian of science Lorraine Daston contends that the strive for a universal law is a predicament that does not stand against unanticipated particulars since no universal ever fits the particulars \cite{daston2018calculation}. Commenting on current machine learning practices Daston \cite{Gross2020} explains: ``machine learning presents an extreme case of a very human predicament, which is that the only way we can generalize is on the basis of past experience. And yet we know from history —-- and I know from my lifetime —-- that our deepest intuitions about all sorts of things, and in particular justice and injustice, can change dramatically.''

While the rationalist tradition tends to aspire to produce generalizable knowledge disentangled from historical baggage, context, and human relations, relationalist perspectives strive for concrete, contextual, and relational understanding of knowledge, human affairs, and reality in general. Data science and machine learning systems sit firmly within the rationalist tradition. The core of what machine learning systems do can be exemplified as clustering similarities and differences, abstracting commonalities, and detecting patterns. Machine learning systems ``work'' by identifying patterns in vast amounts of data. Given immense, messy, and complex data, a machine learning system can sort, classify, and cluster similarities based on seemingly shared features. Feed a neural network labelled images of faces and it will learn to discern faces from not-faces. Not only do machine learning systems detect patterns and cluster similarities, they also make predictions based on the observed patterns \cite{o2013doing}. Machine learning, at its core, is a tool that predicts. It reveals statistical correlations with no understanding of causal mechanisms. 

Relational ethics, in this regard, entails moving away from building predictive tools (with no underlying understanding) to valuing and prioritizing in-depth and contextual understanding of the phenomena that we are building predictive models for. This means we examine the patterns we find and ask why we are finding such patterns. This in turn calls for interrogating contextual and historical norms and structures that might give rise to such patterns instead of using the findings as input towards building predictive systems and repeating existing structural inequalities and historical oppression. 

If we go back to the Bayesian models of inference mentioned in Section \ref{rationality}, we find that such models are prone to amplification of socially held stereotypes. Repeating Horgan \cite{Bayes2016}'s point: ``Embedded in Bayes’ theorem is a moral message: If you aren’t scrupulous in seeking alternative explanations for your evidence, the evidence will just confirm what you already believe.'' A data practice that prioritizes understanding over prediction is one that interrogates prior beliefs instead of using the evidence to confirm such belief and one that seeks alternative explanations by placing the evidence in a social, historical, and cultural context. In doing so, we ask challenging but important questions such as `to what extent do our initial beliefs originate in stereotypically held intuitions about groups or cultures?', `why are we finding the ``evidence'' (patterns) that we are finding?', and `how can we leverage data practices in order to gain an in depth understanding of certain problems as situated in structural inequalities and oppression?'



\subsection{ML as a Practice that Alters the Social Fabric}
\label{social fabric}
\begin{displayquote}
``Technology is not the design of physical things. It is the design of practices and possibilities.'' Lucy Suchman \cite{suchman2007human}
\end{displayquote}

Machine classification and prediction are practices that act directly upon the world and result in tangible impact \cite{mcquillan2018data}. Various companies, institutes, and governments use machine learning systems across a variety of areas. These systems process people's behaviours, actions, and the social world, at large. The machine-detected patterns often provide ``answers'' to fuzzy, contingent, and open-ended questions. These ``answers'' neither reveal any causal relations nor provide explanation on \textit{why} or \textit{how} \cite{pasquale2015black}. Crucially, the more socially complex a problem is, the less capable machine learning systems are of accurately or reliably classifying or predicting \cite{salganik2020measuring}. 
Yet, analytics companies boast their ability to provide insight into the human psyche and predict human behaviour \cite{Qualtrics}. Some even go so far as to claim to have built AI systems that are able to map and predict ``human states'' based on speech analysis, images of faces, and other data \cite{Affectiva}.

Thinking in relational terms about ethics begins with 
reconceptualizing data science and machine learning as practices that create, sustain, and alter the social world. 
The very declaration of a taxonomy brings some things into existence while rendering others invisible \cite{bowker2000sorting}. For any individual person, community, or situation, algorithmic classifications and predictions give either an advantage or they hinder. Certain patterns are made visible and types of being objectified while other types are erased. Some identities (and not others) are recognised as a pedestrian \cite{wilson2019predictive}, or fit for a STEM career \cite{lambrecht2019algorithmic}, or in need of medical care \cite{obermeyer2019dissecting}. Some are ignored and made invisible altogether. 

Categories cut and demarcate boundaries. They simplify and freeze nuanced and complex narratives obscuring political and moral reasoning behind a category. Over time, messy and contingent histories and political and moral stories hidden behind a category are forgotten and trivialized \cite{star2007enacting}. The process of categorizing, sorting, and generalizing, therefore, is far from a mere technical task. While seemingly invisible in our daily lives, categorization and prediction bring forth some behaviours and ways of being  as ``legitimate'', ``standard'', or ``normal'' while casting others as ``deviant''\cite{star2007enacting}. Seemingly banal tasks such as identifying and predicting ``employable'' or ``criminal'' characteristics carry grave consequences for those that do not conform to the status quo. 

Relational ethics encourages us to view data science in general, and the tasks of developing and deploying algorithmic tools that cluster and predict, as part of the practice of creating and reinforcing existing and historical inequalities and structural injustices. Therefore, in treating data science as a practice that alters the fabric of society, the data practitioner is encouraged to zoom out and ask such questions as `how might the deployment of a specific tool enable or constrain certain behaviours and actions?', `does the deployment of such a tool enable or limit possibilities, and for whom?' and `in the process of enabling some behaviours while constraining others, how might such a tool be encouraging/discouraging certain social discourse and norms?'


\section{Conclusion}
\label{closing}
Rethinking ethics is about undoing previous and current injustices to society’s most minoritized and empowering the underserved and systematically disadvantaged. This entails not devising ways to ``debias'' datasets or derive abstract ``fairness'' metrics but zooming out and looking at the bigger picture. Relational ethics encourages us to examine fundamental questions and unstated assumptions. This includes interrogating asymmetrical and hierarchical power dynamics, deeply ingrained social and structural inequalities and assumptions regarding knowledge, justice, and technology itself. 

Ethical practice, especially with regards to algorithmic predictions of social outcomes, requires a fundamental rethinking of justice, fairness, and ethics, above and beyond technical solutions. Ethics in this regard is not merely a methodology, a tool, or simply a matter of constructing a philosophically coherent theory but a down to earth practice that is best viewed as a habit --- a practice that alters the way we do data science. Relational ethics is a process that emerges through the re-examination of the nature of existence, knowledge, oppression, and injustice. Algorithmic systems never emerge in a social, historical, and political vacuum, and to divorce them from the contingent background that they are embedded in is erroneous. Relational ethics provides the framework to rethink the nature of data science through a relational understanding of being and knowing.

    \chapter{Conclusion}
\label{chapter:conclusion}

No work is ever complete. We live in a dynamic, fluid, and ever-becoming world, not a static one that can be pinned down and finalized once and for all. No dataset or model can capture a complex phenomenon in its entirety and provide complete knowledge of such phenomena. Additionally, nothing is ever \textit{purely} novel or springs from a vacuum. Anything we produce is always in relation to something within a certain context and exists in a web of relations; it is either (partly) based on what has been done, said, and/or in anticipation of a response. In a similar vein, this thesis is neither fully complete nor purely novel. Although this thesis brings various disciplines, methods, and schools of thought together in a unique way to narrate the story of complex human behaviour, the impossibility of machine prediction, and ways to approach problematic consequences of ML tools, it stands on the shoulders of giants and a great deal of previous work. More particularly, work that has often been outcast to the margins as an outlier from each respective field: critical complexity, Black feminist scholarship, decolonial, critical race and data studies. This chapter provides some closing remarks by summarizing the thesis' contributions and proposing some possible directions for future work.

\section{Summary and Contributions of the Thesis}

Using insights from post-Cartesian perspectives (complex systems and embodied and enactive approaches to cognitive science), this thesis has put forward a theoretical argument on why machine prediction of complex behaviour is impossible in principle. The more a phenomenon is socially, culturally, and historically intertwined, the more unreliable, problematic, and in some cases, dangerous our predictive models can be. More than anything else, predictive tools \textit{pick up} historical inequalities, social stereotypes, and normative patterns. This thesis illustrated these points by delving into computer vision, a sub-field of AI, and examining large scale datasets and application of vision research, such as ``emotion recognition''. 

The dataset audit works on both ImageNet and Tiny Images, described in Chapter~\ref{chapter:computer vision}, provided a fine grained analysis of the core concerns that were outlined theoretically in Chapter~\ref{chp:automating}. Additionally, this groundbreaking work has had a wide reaching impact within the computer vision community, including the retraction of the Tiny Images dataset. This thesis has also critically examined scientifically erroneous and ethically dubious applications of computer vision, such as ``emotion recognition'' systems and put forward an argument for discarding such tools outright. 

No work emerges in a vacuum and all scientific endeavours are embedded in value systems, current norms, and trends. ML is no different. There are a multitude of forces which give rise to the prominence of certain values in the ML community: the direct or indirect influence of other work in the field, feedback from reviewers (or attempts to preempt such feedback), corporate review boards which circumscribe what can and cannot be said, community norms (both implicit and explicit), as well as the broader academic tradition that views publications as the highest currency. By examining recent influential ML research, this thesis has illustrated the underlying values of ML research. Although ML research is seemingly technical and neutral, the underlying values are actually politically loaded. This work also unveiled values that are both most prominent and also \textit{absent}, and the increasingly predominant corporate presence within the field. This work was the first of its kind whereby qualitative and quantitative evidence were presented, and put forward a novel methodology for advancing such work. 

Finally, using insights from post-Cartesian approaches as well as Black feminist and critical studies, this thesis has put forward an ethical framework that goes beyond narrow technical solutions and fairness metrics; a relational ethics approach. As we have seen, AI and ML systems are best viewed as an ecology rather than a technical tool. As an ecology, they encompass internet sourced datasets, invisible labour (Ghost work)~\cite{gray2019ghost}, natural resources, and more. These tools also require societal uptake, and are value-laden, and politically and socially loaded. A relational ethics approach allows a rethinking of ethics in a manner that accounts for this ecology. 

\section{Possible Directions for Future Work}
At the center of most scientific endeavors stands the desire to produce generalizable knowledge based on abstracted similarities, patterns, and commonalities which then allow for prediction. This relies on supposedly disentangling a complex phenomenon from its historical and contextual contingencies; stabilizing dynamic and continual processes; finalizing open and non-totalizable phenomena; and bringing the phenomenon down to its abstract ``essense'' or to a single representation. By supposedly disentangling contingencies and boiling a phenomenon down to its abstract ``essense'', scientists claim to get at free-standing knowledge or even a truth that applies regardless of time or context. This desire for free-standing knowledge ``uncontaminated'' by a particular culture, history, and context is largely inherited from the Newtonian worldview that dominates much of Western (both natural and human) science. Such worldviews stand at odds with the reality of how complex phenomena unfold but provide a way (albeit illusory) to control, manipulate, and predict complex behaviour. 

This aspiration exists in tension with gaining in-depth and contextual understanding of people and society. While most domains of enquiry that deal with people and society tread some balance between prediction and understanding, data science and machine learning stand firmly on the former. Much of current statistical and mathematical modelling operates on de-contextualization and abstraction which then allow generalizations and predictions. Predictions are often taken as probabilistic truths and yet are detached from the specific humans and societies that they are made on. With such data and modelling practices, bigger and more data is seen as a substitute for causal understanding. As discussed in Section~\ref{prioritizing understaning}, algorithmic injustice suffered by minoritized communities is partly rooted in the fields’ fetishization of prediction with limited understanding of people, society, histories, and structural injustices. One promising direction for future work is to lean on insights from fields where \textit{understanding} a phenomenon is seen as foundational, before driving generalizations or building predictive tools. In this regard cultural anthropology, specifically Geertz's method~\cite{geertz1973interpretation} of \textit{thick description} -- describing a phenomenon with as much detail as possible with the aim of understanding -- stands as a means to equitable ML models that primarily provide understanding of a subject of study.

\subsection{Modelling Complexity and the Future}

Knowledge of our self and of the world is fluid, dynamic, and continually moving. Any understanding of the person-society-technology relationship can thus only be a moving target. Our data and models can only capture a snapshot of this moving target and they need to remain partially open. This accommodates the uncertainties of ongoing change and provides room for dialogue, negotiation, reiteration, and revision of any claims and positions. This means that how algorithmic systems are designed and implemented requires continual negotiation between the different stakeholders, including the often ignored stakeholders such as data subjects and disproportionately negatively impacted groups. In the case of computer vision, for example, this includes people whose images have been sourced and used without their awareness or consent. In fact, the views and input of vulnerable communities that are disproportionately negatively impacted by algorithmic decision making need to be central at all stages of the design, development, and deployment process. This shift in perspective is central to moving towards a just society.  

Furthermore, as we have seen, ML systems simplify, abstract, classify and predict by reducing complex, messy, dynamic and ambiguous phenomena to abstract (and oftentimes inaccurate) representations. Another proposed way of moving forward to a just society is to envision a fundamentally different kind of technology that is grounded in ambiguity, fluidity, and diversity of experience. In their vision of \textit{diversity computing}, Fletcher et al.~\citep*{fletcher2018diversity} envisage fundamentally new kinds of computing devices that reflect, promote, and embrace differences rather than eliminating them. This requires acknowledgement that our static models are not capable of capturing dynamic human behaviour which, in turn, means both being modest about our models and also striving for models built on the fundamental understanding of the inherent diversity and indeterminability of the world.  

As we have seen, complex phenomena are unfinalizible and inexhaustible, which means that we can never capture human behaviour or social phenomena in their entirety with models. Complex systems are incompressible, meaning there is no single accurate representation of the system that is simpler than the system itself. However, the implication is not that we should stop striving to capture complexity in our models, or that we should stop building models, but rather that we should acknowledge our model’s limitations and pitfalls and be modest about our claims of a model's knowledge~\cite{cilliers2002complexity}. 

Finally, the main points that I have discussed in this section are crucial for a radical transformation of AI that embraces uncertainties and indeterminabilities and serves the most disadvantaged. Although acknowledging responsibility and accountability is an important first step, pleading the powerful to take responsibility and be considerate to the vulnerable is simply not sufficient. Just like science has found ways to evade ethical responsibility by means of systematic separation of ``objective science'' and ethics, the field of AI will do the same if allowed. In some respects, it has already been doing so with impunity. In fact, global technology giants spend millions~\citep{molla2019} actively lobbying to influence legislation in their favour, resulting in less agency and more harm to the masses. The capitalist ecosystem in which ML systems are built and deployed presents one of the greatest challenges. Even for the most well meaning technologists, the incentive structures pressure individuals to develop technology that maintains the status quo and existing power structures, and produces maximum profit. Technology that envisages radical shift in power (from the most to the least powerful) stands in stark opposition to current technology that maximizes profit and efficiency. It is an illusion to expect technology giants to develop AI that centres the interests of the marginalized. Strict regulations, social pressure through organized movements, strong reward systems for technology that empowers the least privileged, and a completely new application of technologies -- which require vision, imagination, and creativity -- all pave the way to a technologically just future. 



\subsection{Justice and the Future}

\begin{quote}

``What does it mean when the tools of a racist patriarchy are used to examine the fruits of that same patriarchy? It means that only the most narrow parameters of change are possible and allowable.''

Lorde~\cite{Lorde1984}
\end{quote}

This thesis is first and foremost an endeavour to critically examine both the scientific basis and ethical dimensions of current ML/AI. Thus, it primarily lays bare the limitations, failings, and problems surrounding AI systems broadly construed. But this is not to imply that AI systems cannot be beneficial (to marginalized communities, AI is already beneficial to the wealthy and powerful), or cannot function in a way that serves the most disenfranchised or minoritized. Far from it. It is possible and important to envision such a future. 

It is impossible to predict the future, but that does not mean that we can't envision it. In fact, envisioning the type of future we want is a crucial first step to working towards making a better future happen. As Lewontin~\cite{lewontin1996biology} sharply put it: ``skepticism should not be confused with a cynicism or defeatism for while the former can lead to action, the latter to passivity.'' In order to envision a future, it is important to know what the current state of the field is and what prominent values underlie ML. In this regard, this thesis (Chapter~\ref{chp:values}) has laid the crucial ground work. The most prominent values underlying ML research currently include values such as ``performance'', ``accuracy'' and ``SOTA'' while values such as ``fairness'' and ``justice'' are almost entirely missing. 

Yet, the current values and the present state of the field are all contingent; it could be otherwise. 
The emergence of certain values into prominence and the perception of these values as purely technical and intrinsic to ML, go hand in hand, all contributing towards the thinking that alternatives are beyond the scope of ML. Thus, characterization of the way the field is now is useful for understanding, shaping, dismantling, or transforming what is, and for articulating and bringing about alternative visions. This encourages the reader to remember that, what felt intrinsic and like it defined the realm of possibility not so long ago has frequently been transformed.

Finally, because of the rise of corporate influence in ML research, we see datasets getting ever larger with insufficient documentation of their contents, and little to no caution towards how they impinge on privacy or how they bake-in stereotypes. Cheap AI is projected to be a multi-billion dollar industry. The massive natural and ecological costs of producing AI, the hidden labour -- Ghost work -- behind AI, as well as the uneven benefit and harm distribution, all indicate that AI needs to be seen as a social, structural, ecological, cultural, and historical phenomenon, not just ones and zeros on a machine. This means that a truly just and beneficial (to the disproportionately harmed) AI requires a radical rethinking of social structures, asymmetries of power, uneven benefit-harm distributions, and oppressive systems, tacking problems from their roots and not just a tweaking of parameters or devising mathematical formulae for systems that enforce the status quo.

    \newpage
    \bibliographystyle{unsrtnat}
    \renewcommand{\bibname}{References}
    \addcontentsline{toc}{chapter}{References}
    {\small \bibliography{references}}

\end{document}